\documentclass[aps,prd,twocolumn,superscriptaddress,showpacs,amsmath,amssymb]{revtex4}

\usepackage{graphicx}
\usepackage{dcolumn}
\usepackage{bm}

\def\Pomeron{I\hspace{-1.2mm}P}

\begin{document}

\title{Observation of Exclusive Dijet Production at the Fermilab Tevatron $\bar pp$ Collider}
\affiliation{Institute of Physics, Academia Sinica, Taipei, Taiwan 11529, Republic of China} 
\affiliation{Argonne National Laboratory, Argonne, Illinois 60439} 
\affiliation{Institut de Fisica d'Altes Energies, Universitat Autonoma de Barcelona, E-08193, Bellaterra (Barcelona), Spain} 
\affiliation{Baylor University, Waco, Texas  76798} 
\affiliation{Istituto Nazionale di Fisica Nucleare, University of Bologna, I-40127 Bologna, Italy} 
\affiliation{Brandeis University, Waltham, Massachusetts 02254} 
\affiliation{University of California, Davis, Davis, California  95616} 
\affiliation{University of California, Los Angeles, Los Angeles, California  90024} 
\affiliation{University of California, San Diego, La Jolla, California  92093} 
\affiliation{University of California, Santa Barbara, Santa Barbara, California 93106} 
\affiliation{Instituto de Fisica de Cantabria, CSIC-University of Cantabria, 39005 Santander, Spain} 
\affiliation{Carnegie Mellon University, Pittsburgh, PA  15213} 
\affiliation{Enrico Fermi Institute, University of Chicago, Chicago, Illinois 60637} 
\affiliation{Comenius University, 842 48 Bratislava, Slovakia; Institute of Experimental Physics, 040 01 Kosice, Slovakia} 
\affiliation{Joint Institute for Nuclear Research, RU-141980 Dubna, Russia} 
\affiliation{Duke University, Durham, North Carolina  27708} 
\affiliation{Fermi National Accelerator Laboratory, Batavia, Illinois 60510} 
\affiliation{University of Florida, Gainesville, Florida  32611} 
\affiliation{Laboratori Nazionali di Frascati, Istituto Nazionale di Fisica Nucleare, I-00044 Frascati, Italy} 
\affiliation{University of Geneva, CH-1211 Geneva 4, Switzerland} 
\affiliation{Glasgow University, Glasgow G12 8QQ, United Kingdom} 
\affiliation{Harvard University, Cambridge, Massachusetts 02138} 
\affiliation{Division of High Energy Physics, Department of Physics, University of Helsinki and Helsinki Institute of Physics, FIN-00014, Helsinki, Finland} 
\affiliation{University of Illinois, Urbana, Illinois 61801} 
\affiliation{The Johns Hopkins University, Baltimore, Maryland 21218} 
\affiliation{Institut f\"{u}r Experimentelle Kernphysik, Universit\"{a}t Karlsruhe, 76128 Karlsruhe, Germany} 
\affiliation{Center for High Energy Physics: Kyungpook National University, Daegu 702-701, Korea; Seoul National University, Seoul 151-742, Korea; Sungkyunkwan University, Suwon 440-746, Korea; Korea Institute of Science and Technology Information, Daejeon, 305-806, Korea; Chonnam National University, Gwangju, 500-757, Korea} 
\affiliation{Ernest Orlando Lawrence Berkeley National Laboratory, Berkeley, California 94720} 
\affiliation{University of Liverpool, Liverpool L69 7ZE, United Kingdom} 
\affiliation{University College London, London WC1E 6BT, United Kingdom} 
\affiliation{Centro de Investigaciones Energeticas Medioambientales y Tecnologicas, E-28040 Madrid, Spain} 
\affiliation{Massachusetts Institute of Technology, Cambridge, Massachusetts  02139} 
\affiliation{Institute of Particle Physics: McGill University, Montr\'{e}al, Canada H3A~2T8; and University of Toronto, Toronto, Canada M5S~1A7} 
\affiliation{University of Michigan, Ann Arbor, Michigan 48109} 
\affiliation{Michigan State University, East Lansing, Michigan  48824} 
\affiliation{University of New Mexico, Albuquerque, New Mexico 87131} 
\affiliation{Northwestern University, Evanston, Illinois  60208} 
\affiliation{The Ohio State University, Columbus, Ohio  43210} 
\affiliation{Okayama University, Okayama 700-8530, Japan} 
\affiliation{Osaka City University, Osaka 588, Japan} 
\affiliation{University of Oxford, Oxford OX1 3RH, United Kingdom} 
\affiliation{University of Padova, Istituto Nazionale di Fisica Nucleare, Sezione di Padova-Trento, I-35131 Padova, Italy} 
\affiliation{LPNHE, Universite Pierre et Marie Curie/IN2P3-CNRS, UMR7585, Paris, F-75252 France} 
\affiliation{University of Pennsylvania, Philadelphia, Pennsylvania 19104} 
\affiliation{Istituto Nazionale di Fisica Nucleare Pisa, Universities of Pisa, Siena and Scuola Normale Superiore, I-56127 Pisa, Italy} 
\affiliation{University of Pittsburgh, Pittsburgh, Pennsylvania 15260} 
\affiliation{Purdue University, West Lafayette, Indiana 47907} 
\affiliation{University of Rochester, Rochester, New York 14627} 
\affiliation{The Rockefeller University, New York, New York 10021} 
\affiliation{Istituto Nazionale di Fisica Nucleare, Sezione di Roma 1, University of Rome ``La Sapienza," I-00185 Roma, Italy} 
\affiliation{Rutgers University, Piscataway, New Jersey 08855} 
\affiliation{Texas A\&M University, College Station, Texas 77843} 
\affiliation{Istituto Nazionale di Fisica Nucleare, University of Trieste/\ Udine, Italy} 
\affiliation{University of Tsukuba, Tsukuba, Ibaraki 305, Japan} 
\affiliation{Tufts University, Medford, Massachusetts 02155} 
\affiliation{Waseda University, Tokyo 169, Japan} 
\affiliation{Wayne State University, Detroit, Michigan  48201} 
\affiliation{University of Wisconsin, Madison, Wisconsin 53706} 
\affiliation{Yale University, New Haven, Connecticut 06520} 
\author{T.~Aaltonen}
\affiliation{Division of High Energy Physics, Department of Physics, University of Helsinki and Helsinki Institute of Physics, FIN-00014, Helsinki, Finland}
\author{J.~Adelman}
\affiliation{Enrico Fermi Institute, University of Chicago, Chicago, Illinois 60637}
\author{T.~Akimoto}
\affiliation{University of Tsukuba, Tsukuba, Ibaraki 305, Japan}
\author{M.G.~Albrow}
\affiliation{Fermi National Accelerator Laboratory, Batavia, Illinois 60510}
\author{B.~\'{A}lvarez~Gonz\'{a}lez}
\affiliation{Instituto de Fisica de Cantabria, CSIC-University of Cantabria, 39005 Santander, Spain}
\author{S.~Amerio}
\affiliation{University of Padova, Istituto Nazionale di Fisica Nucleare, Sezione di Padova-Trento, I-35131 Padova, Italy}
\author{D.~Amidei}
\affiliation{University of Michigan, Ann Arbor, Michigan 48109}
\author{A.~Anastassov}
\affiliation{Rutgers University, Piscataway, New Jersey 08855}
\author{A.~Annovi}
\affiliation{Laboratori Nazionali di Frascati, Istituto Nazionale di Fisica Nucleare, I-00044 Frascati, Italy}
\author{J.~Antos}
\affiliation{Comenius University, 842 48 Bratislava, Slovakia; Institute of Experimental Physics, 040 01 Kosice, Slovakia}
% ...
\collaboration{CDF Collaboration\footnote{With visitors from $^a$University of Athens, 15784 Athens, Greece, 
$^b$Chinese Academy of Sciences, Beijing 100864, China, 
$^c$University of Bristol, Bristol BS8 1TL, United Kingdom, 
$^d$University Libre de Bruxelles, B-1050 Brussels, Belgium, 
$^e$University of California Irvine, Irvine, CA  92697, 
$^f$University of California Santa Cruz, Santa Cruz, CA  95064, 
$^g$Cornell University, Ithaca, NY  14853, 
$^h$University of Cyprus, Nicosia CY-1678, Cyprus, 
$^i$University College Dublin, Dublin 4, Ireland, 
$^j$University of Edinburgh, Edinburgh EH9 3JZ, United Kingdom, 
$^k$University of Heidelberg, D-69120 Heidelberg, Germany, 
$^l$Universidad Iberoamericana, Mexico D.F., Mexico, 
$^m$University of Manchester, Manchester M13 9PL, England, 
$^n$Nagasaki Institute of Applied Science, Nagasaki, Japan, 
$^o$University de Oviedo, E-33007 Oviedo, Spain, 
$^p$Queen Mary, University of London, London, E1 4NS, England, 
$^q$Texas Tech University, Lubbock, TX  79409, 
$^r$IFIC(CSIC-Universitat de Valencia), 46071 Valencia, Spain, 
}}
\noaffiliation

\date{\today}

\begin{abstract}
We present the first observation and cross section measurement of exclusive dijet production in $\bar pp$ interactions, $\bar pp\rightarrow \bar p+\mbox{dijet}+p$. Using a data sample of 310 pb$^{-1}$ collected by the Run II Collider Detector at Fermilab at $\sqrt s$=1.96 TeV, exclusive cross sections for events with two jets of transverse energy $E_T^{jet}\geq 10$ GeV have been measured as a function of minimum $E_T^{jet}$. The exclusive signal is extracted from fits to data distributions based on Monte Carlo simulations of expected dijet signal and background shapes. The simulated background distribution shapes  are checked in a study of a largely independent data sample of 200 pb$^{-1}$ of $b$-tagged jet events, where exclusive dijet production is expected to be suppressed by the $J_z=0$ total angular momentum selection rule. Results obtained are compared with theoretical expectations, and implications for exclusive Higgs boson production at the $pp$ Large Hadron Collider at $\sqrt s=$14~TeV are discussed. 
\end{abstract}


\maketitle

\section{Introduction}\label{sec:Intro}
Exclusive dijet production in $\bar pp$ collisions is a process in which both the antiproton and proton escape the interaction point intact and a two-jet system is centrally produced:
\begin{equation}
\bar{p} + p \rightarrow  \bar{p}' + (jet_1+jet_2) + p'.
\label{eq:ex2jets}
\end{equation}

This process is a particular case of dijet production in double Pomeron exchange (DPE), a diffractive process in which the antiproton and proton suffer a small fractional momentum loss, and a system $X$ containing the jets is produced, 
\begin{equation}
\bar{p} + p \rightarrow [\bar{p}' + \Pomeron_{\bar p}] + [p' + \Pomeron_p] \rightarrow \bar{p}' + X + p',
\label{eq:incl2jets}
\end{equation}
where $\Pomeron$ designates a Pomeron, defined as an exchange consisting of a colorless combination of gluons and/or quarks carrying the quantum numbers of the vacuum. 

\begin{figure}
 \begin{center}
\includegraphics[width=8cm]{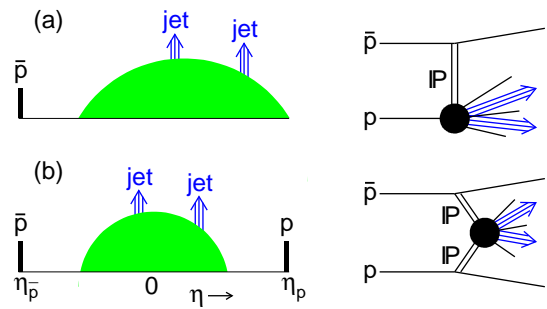}
\vglue -1.4in
  \caption{Illustration of event topologies in pseudorapidity, $\eta$, 
and associated Pomeron exchange 
diagrams for dijet production in (a) single diffraction and (b) double 
Pomeron exchange. The shaded areas on the left side represent ``underlying event'' particles not associated with the jets [from Ref.~\cite{RunI_DPE}]. 
\label{fig:SD_DPE}}
 \end{center}
\end{figure}

In a particle-like Pomeron picture (e.g. see~\cite{ref:IS}), the system $X$ may be thought of as being produced by the collision of two Pomerons, $\Pomeron_{\bar p}$ and $\Pomeron_p$,
\begin{equation}
\Pomeron_{\bar p} +\Pomeron_p \rightarrow X \Rightarrow Y_{\Pomeron/{\bar p}}+(jet_1+jet_2)+Y_{\Pomeron/p},
\label{eq:pompom}
\end{equation}
where in addition to the jets the final state generally contains Pomeron remnants designated by $Y_{\Pomeron/{\bar p}}$ and $Y_{\Pomeron/p}$. 
Dijet production in DPE is a sub-process to dijet production in single diffraction (SD) dissociation, where only the antiproton (proton) survives while the proton (antiproton) dissociates. Schematic diagrams for SD and DPE dijet production  are shown in Fig.~\ref{fig:SD_DPE} along with  event topologies in pseudorapidity space (from Ref.~\cite{RunI_DPE}).   
In SD, the escaping $\bar p$ is adjacent to a rapidity gap, defined as a region of pseudorapidity devoid of particles~\cite{rapidity}. 
A rapidity gap arises because the Pomeron exchanged in a diffractive process is a colorless object of effective spin $J\ge 1$ and carries the quantum numbers of the vacuum. In DPE, two such rapidity gaps are present. 

Dijet production in DPE may occur as an exclusive process~\cite{excl_define} with only the jets in the final state and no Pomeron remnants, either due to a fluctuation of the Pomeron remnants down to zero or with a much higher cross section in models in which the Pomeron is treated as a parton and the dijet system is produced in a $2\rightarrow 2$ process analogous to  $\gamma\gamma\rightarrow jet+jet$~\cite{BL}. 

In a special case  exclusive dijets may be produced through an intermediate state of a Higgs boson decaying into $\bar bb$:
\begin{equation}
\Pomeron_{\bar p} +\Pomeron_p \rightarrow H^0 \rightarrow (\bar b\rightarrow jet_1)+(b\rightarrow jet_2).
\label{eq:pompomHiggs}
\end{equation}
\begin{figure}
 \begin{center}
  \includegraphics[width=3.5cm]{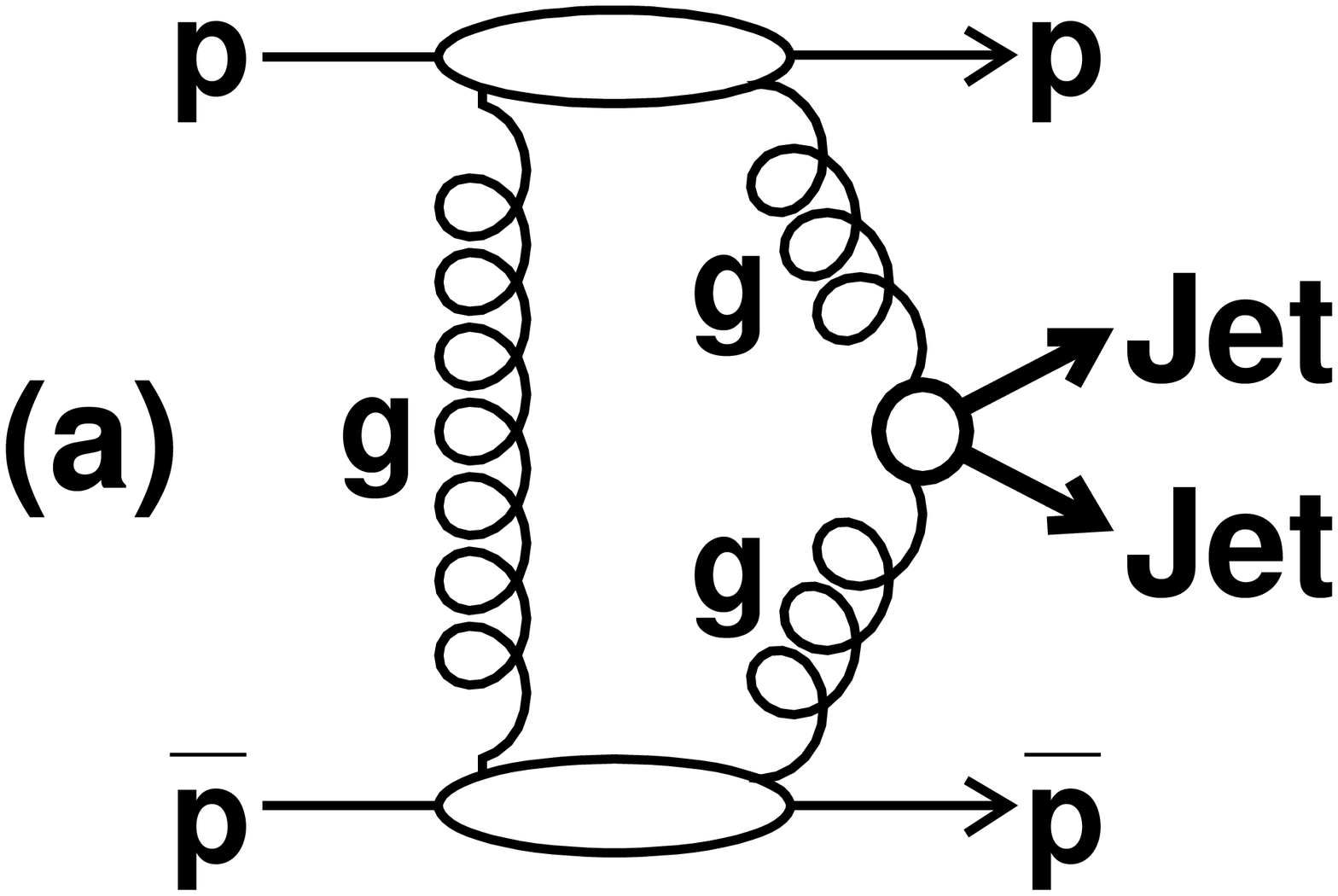}\hspace{0.5cm}
  \includegraphics[width=3.5cm]{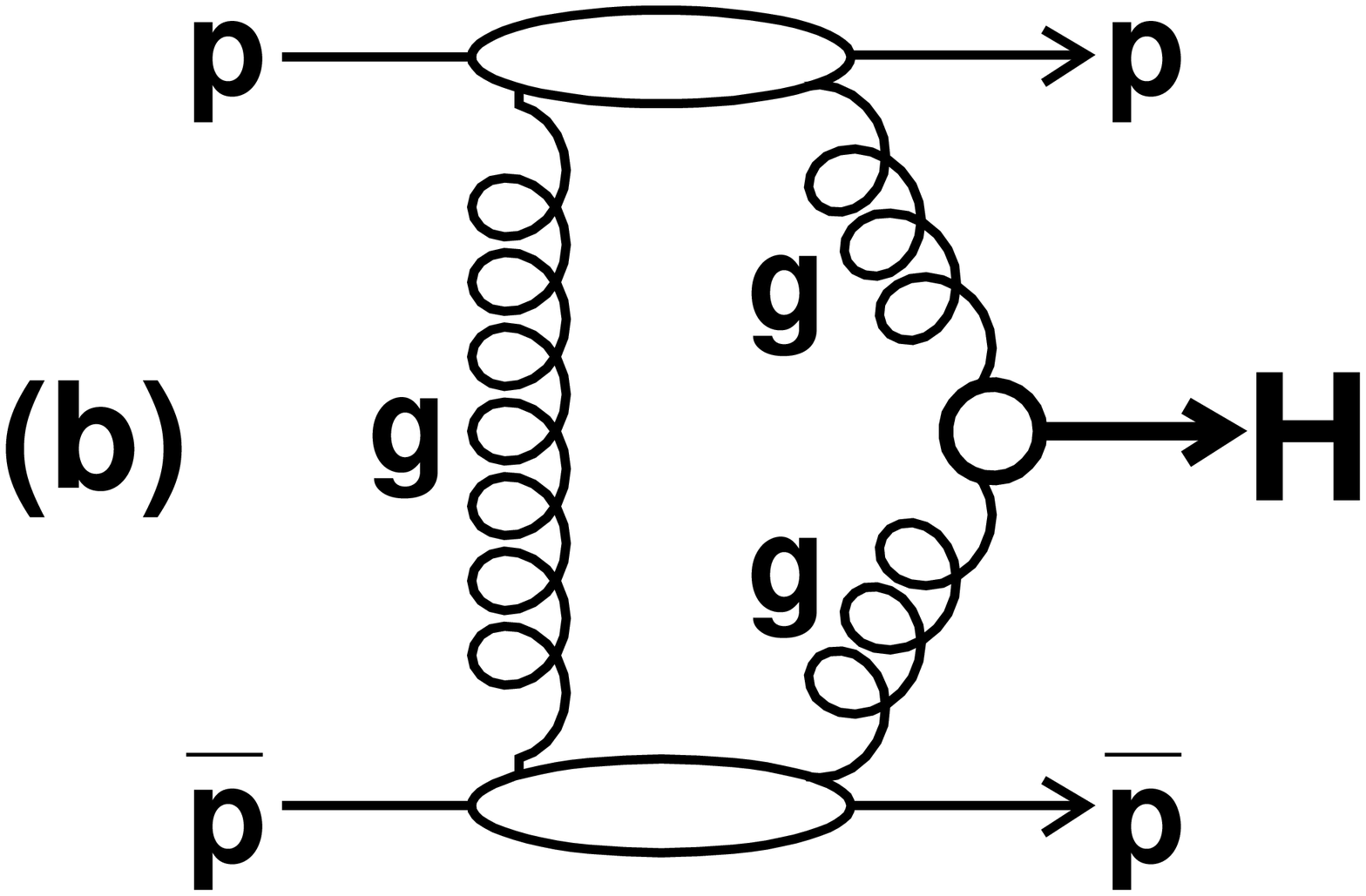}
  \caption{Leading order diagrams for (a) exclusive dijet  and (b) exclusive Higgs boson production in $\bar{p}p$ collisions. 
\label{fig:excl_diagram}}
 \end{center}
\end{figure}
 
Exclusive production may also occur through a $t$-channel color-singlet two gluon exchange at leading order (LO) in perturbative quantum chromo-dynamics (QCD), as shown schematically in Fig.~\ref{fig:excl_diagram}~(a), where one of the two gluons takes part in the hard scattering that produces the jets, while the other neutralizes the color flow~\cite{KMRmethod}. 
A similar diagram, Fig.~\ref{fig:excl_diagram}~(b), is used in~\cite{KMRmethod} to calculate exclusive Higgs boson production. 

Exclusive dijet production has never previously been observed in hadronic collisions. In addition to providing information on QCD aspects of vacuum quantum number exchange, there is currently intense interest in using measured exclusive dijet production cross sections to calibrate theoretical predictions for exclusive Higgs boson production at the Large Hadron Collider (LHC).
Such predictions are generally hampered by large uncertainties due to non-perturbative suppression effects associated with the rapidity gap survival probability. As these effects are common to exclusive dijet and Higgs boson production mechanisms, dijet production potentially provides a ``standard candle'' process against which to calibrate the theoretical models~\cite{KMRmethod,Royon}.     

In Run I (1992-96) of the Fermilab Tevatron $\bar pp$ collider operating at 1.8~TeV, the Collider Detector at Fermilab (CDF) collaboration made the first observation of dijet production by DPE)~\cite{RunI_DPE} using an inclusive sample of SD events, $\bar{p}p \rightarrow \bar{p}'X$, collected by triggering on a $\bar{p}$ detected in a forward Roman Pot Spectrometer (RPS). DPE dijet events were selected from this sample by requiring, in addition to the $\bar{p}$ detected by the RPS, the presence of two jets with transverse energy $E_T>7$~GeV and a rapidity gap in the outgoing proton direction in the range $2.4<\eta<5.9$~\cite{COORD}. In the resulting sample of 132 inclusive DPE dijet events, no evidence for exclusive dijet production was found, setting a 95~\% confidence level upper limit of 3.7 nb on the exclusive production cross section. At that time, theoretical estimates of this cross section ranged from $\sim 10^3$ larger~\cite{berera} to a few times smaller~\cite{KMRmethod} than our measured upper bound. More data were clearly needed to observe an exclusive dijet signal and test theoretical predictions of kinematical properties and production rates. 

In Run II-A (2001-06), with the Tevatron  providing $\bar pp$ collisions at $\sqrt{s}=1.96$ TeV, two high statistics data samples of DPE dijet events were collected by the upgraded CDF~II detector: one of inclusive dijets, and another largely independent sample of $b$-quark jets. The analysis of these data is the subject of this paper. 
The results obtained provide the first evidence for exclusive dijet production in $\bar pp$ collisions. 

The paper is organized as follows.
In Sec.~\ref{sec:Strategy}, we present the strategy employed to control the experimental issues involved in searching for an exclusive dijet signal.  We then describe the detector (Sec.~\ref{sec:detector}), the data samples and event selection (Sec.~\ref{sec:data}), the data analysis for inclusive DPE (Sec.~\ref{sec:xsec_incl}) and exclusive dijet production (Sec.~\ref{sec:excl_search}), results and comparisons with theoretical predictions (Sec.~\ref{sec:xsec}), and background shape studies using heavy flavor quark jets (Sec.~\ref{sec:hf}). Implications for exclusive Higgs boson production at the Large Hadron Collider are discussed  in Sec.~\ref{sec:higgs}, and  conclusions are presented in Sec.~\ref{sec:conclusion}. 

\begin{figure*}
 \begin{center} 
 \includegraphics[width=17.5cm]{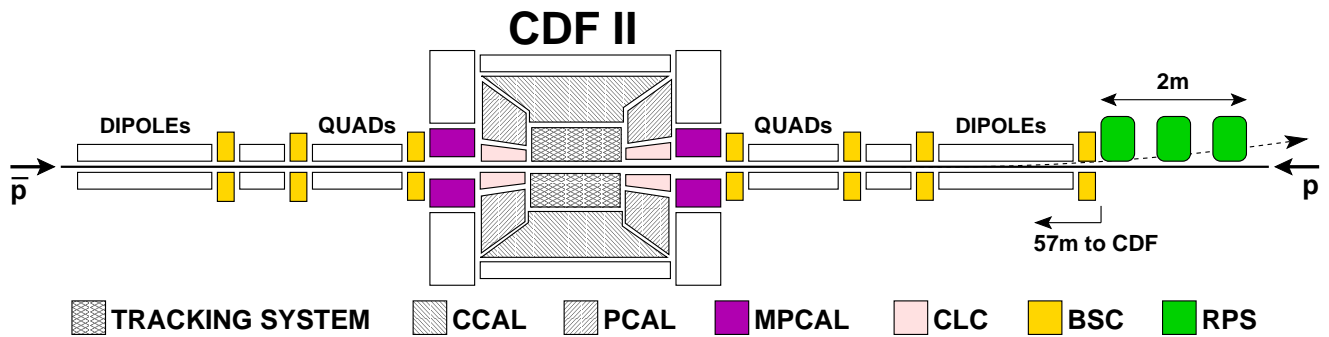}
 \caption{Schematic drawing (not to scale) of the CDF~II detector.\label{fig:CDFII_detector}}
 \end{center} 
\end{figure*}

\section{Strategy}\label{sec:Strategy}
Exclusive dijet production is characterized by two jets in the final state and no additional final state particles except for the escaping forward proton and antiproton. Therefore, searching for exclusive dijet production would ideally require a full acceptance detector in which all final state particles are detected, their vector momenta are measured, the correct particles are assigned to each jet, and ``exclusivity'' is certified by the absence of any additional final state particle(s). Assigning particles to a jet is a formidable challenge because the detector threshold settings used to reduce noise may inadvertently either eliminate particles with energies below threshold or else result in noise being counted as additional particles if the thresholds are set too low. To meet this challenge, we developed a strategy incorporating detector design, online triggers, data sets used for background estimates, and an analysis technique sensitive to an exclusive signal but relatively immune to the above  mentioned effects. 

The exclusive signal is extracted using the ``dijet mass fraction'' method developed in our Run~I data analysis. From the energies and momenta of the jets in an event, the ratio $R_{jj}\equiv M_{jj}/M_X$ of the dijet mass $M_{jj}$ to the total mass $M_X$ of the final state (excluding the $\bar p$ and $p$) is formed and used to discriminate between the signal of exclusive dijets, expected to appear at $R_{jj}=1$, and the background of inclusive DPE dijets, expected to have a continuous distribution concentrated at lower $R_{jj}$ values. Because of smearing effects in the measurement of $E_T^{jet}$ and $\eta^{jet}$ and gluon radiation from the jets the exclusive dijet peak is broadened and shifts to lower $R_{jj}$ values. The exclusive signal is therefore obtained by a fit of the $R_{jj}$ distribution to expected signal and background shapes generated by Monte Carlo (MC) simulations. The background shape used is checked  with an event sample of heavy quark flavor dijets, for which exclusive production is expected to be suppressed in LO QCD by the $J_z=0$ selection rule of the hard scattered di-gluon system, where $J_z$ is the projection of the total angular momentum of the system along the beam direction~\cite{KMR2001}.

\section{Detector}\label{sec:detector}
The CDF~II detector, shown schematically in Fig.~\ref{fig:CDFII_detector}, is described in detail elsewhere~\cite{CDFII}. The detector components most relevant
for this analysis are the charged particle tracking system, the central and plug calorimeters, and a set of detectors instrumented
in the forward pseudorapidity region. 
The CDF tracking system consists of a silicon vertex detector (SVX~II)~\cite{SVX},
 composed of double-sided microstrip silicon sensors arranged in five cylindrical shells of radii
between 2.5 and 10.6 cm, and an open-cell drift chamber~\cite{COT} of 96 layers organized in 8 superlayers
with alternating structures of axial and $\pm2^{\circ}$ stereo readout within a  radial range between
40 and 137 cm. Surrounding the tracking detectors is a superconducting solenoid, which provides a
magnetic field of 1.4~T. Calorimeters located outside the solenoid are physically divided into a 
central calorimeter (CCAL)~\cite{CEM, CHA}, covering the pseudorapidity range $|\eta|<1.1$, and a plug calorimeter 
(PCAL)~\cite{PCAL}, covering the region $1.1<|\eta|<3.6$. These calorimeters are segmented into projective towers of granularity $\Delta\eta\times\Delta\phi \approx 0.1\times15^{\circ}$.

\begin{figure}
 \begin{center} 
 \includegraphics[width=7.5cm]{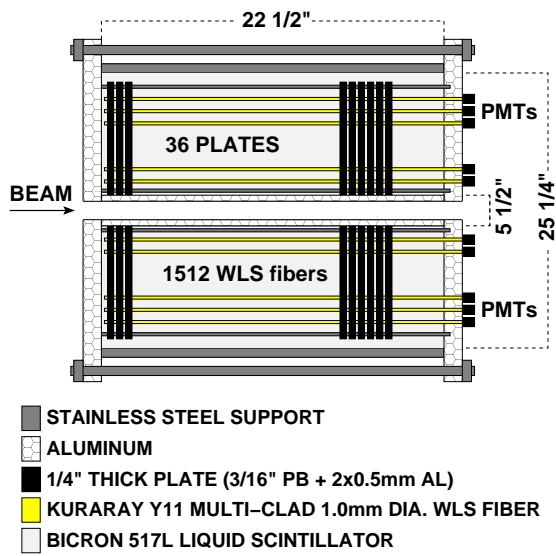}
 \caption{Schematic cross sectional view of one of the two forward MiniPlug Calorimeters installed in CDF~II.\label{fig:MiniPlug}}
 \end{center} 
\end{figure}

The forward detectors~\cite{PoScdf}, which extend the coverage into the $\eta$ region beyond 3.6, consist of the MiniPlug calorimeters (MPCAL)~\cite{MPCAL}, the Beam Shower Counters (BSC), a Roman Pot Spectrometer (RPS), and a system of Cerenkov Luminosity Counters (CLC).  The MiniPlug calorimeters, shown schematically in Fig.~\ref{fig:MiniPlug}, are designed to measure the energy and lateral position of particles in the region $3.6<|\eta|<5.2$. They  consist of alternating lead plates and liquid scintillator layers perpendicular to the beam, which are read out by wavelength shifting fibers that pass through holes drilled through the plates parallel to the beam direction. Each MiniPlug is 32 (1.3) radiation (interaction) lengths deep. The BSC are scintillation counters surrounding the beam pipe at three (four) different locations on the outgoing proton (antiproton) side of the CDF~II detector. Covering the range $5.4<|\eta|<5.9$ is the BSC1 system, which is closest to the interaction point (IP) and is used for measuring beam losses and for triggering on events with forward rapidity gaps. Lead plates of thickness 1.7 radiation lengths precede each BSC1 counter to convert $\gamma$ rays to $e^+e^-$ pairs to be detected by the scintillators. The RPS, located at $\sim57$ m downstream in the antiproton beam direction, consists of three Roman pot stations, each containing a scintillation counter used for triggering on the $\bar{p}$, and a scintillation fiber tracking detector for measuring the position and angle of the detected $\bar{p}$. The CLC~\cite{CLC}, covering the range $3.7<|\eta|<4.7$, which substantially overlaps the MiniPlug coverage, are normally used in CDF to measure the number of
inelastic $\bar{p}p$ collisions per bunch crossing and thereby the luminosity. In this analysis, they are also used to refine the rapidity gap definition by detecting charged particles that might penetrate a MiniPlug without interacting and thus produce too small a pulse height to be detected over the MiniPlug tower thresholds used. 

\section{Data Samples and Event Selection}\label{sec:data}
Three data samples are used in this analysis, referred to as the DPE, SD, and non-diffractive (ND) event samples. The exclusive signal is derived from the DPE event sample, while the SD and ND samples are used for evaluating backgrounds. The total integrated luminosity of
 the DPE sample is $312.5\pm18.7$ pb$^{-1}$.

The following trigger definitions are used:

{\bf J5:} a single CCAL or PCAL calorimeter trigger tower of $E_T>5$ GeV.

{\bf RPS:} a triple coincidence among the three RPS trigger counters in time with a $\bar{p}$ gate.

{$\overline{\mbox{\bf BSC1}}_{p}$:} a BSC1 veto on the outgoing proton side.

\noindent The three event samples were collected with the following triggers:\\   
{ND}$\equiv$J5, {SD}$\equiv$J5 $\cdot$ RPS, {DPE}$\equiv$J5 $\cdot$ RPS $\cdot\,\overline{{\mbox{BSC1}}_p}$.

The DPE events, from which cross sections are calculated, were sampled at a rate of one out of five events to accommodate the trigger bandwidth.
In the above sample definition, ND events include SD and DPE contributions, and SD events include DPE ones. This results in a ``contamination'' of background distributions by signal events, which is taken into account in the data analysis.

The selection cuts used in the data analysis include:

{\bf VTX cut} (ND, SD, and DPE): no more than one reconstructed primary vertex within $|z|<60$~cm, imposed to reduce the number of overlap events occurring during the same beam-beam crossing at the IP.

{\bf RPST cut} (SD and DPE): RPS trigger counter pulse height cut, imposed to reject ``splash'' triggers caused by particles hitting the beam pipe in the vicinity of the RPS and spraying the RPST counters with secondary particles.

{\bf JET cut} (ND, SD, and DPE): events are required to  have at least two jets with transverse energy $E_T^{jet}>10$~GeV within $|\eta|<2.5$. The transverse energy of a jet is defined as the sum $E_T^{jet}\equiv \Sigma_iE_i\sin(\theta_i)$ of all calorimeter towers at polar angles $\theta_i$ within the jet cone. Jets are reconstructed with the midpoint algorithm~\cite{MidPoint}, which is an improved iterative cone clustering algorithm, using a cone radius of 0.7 in $\eta$-$\phi$ space and based on calorimeter towers with $E_T$ above 100 MeV. The $E_T$ of a jet is defined as the sum of the $E_T$ values of the clustered calorimeter towers. The jet $E_T$ is corrected for the relative response of the calorimeters and for the absolute energy scale.

The above selection cuts define the DPE data sample (DPE) and are summarized below:
\begin{eqnarray}
&\mbox{\bf DPE sample:}\label{DPEsample}\label{dpesample:}
\\
&J5 \cdot RPS\cdot \overline{BSC1_p} \cdot VTX \cdot RPST\cdot JET.\nonumber
\end{eqnarray}
The DPE data sample  consists of 415~688 events. 

Backgrounds in the DPE event sample fall into two general categories: (i) SD dijet events, in which the  $\overline{BSC1_p}$ requirement is fulfilled by a downward $BSC1_p$ multiplicity fluctuation to zero, and (ii) overlaps between a ND J5 trigger and a RPS trigger provided by either a low mass soft SD event that has no reconstructed vertex or by a scattered beam halo or ND event particle. To reduce these backgrounds, two more requirements are imposed on the data: a large rapidity gap on the outgoing proton direction, {LRG}$_p$, and passing the $\xi_{\bar{p}}^X$ cut defined below.  
  
{\bf LRG{\boldmath $_p$}:}  this requirement is implemented by demanding zero multiplicities in MPCAL$_p$ and CLC$_p$, $N_{MPCAL}^p=N_{CLC}^p=0$, added to the trigger requirement of BSC1$_p=0$. The LRG$_p$ approximately covers the range of $3.6<\eta<5.9$. This selection cut enriches the DPE event sample in exclusive events by removing non-exclusive backgrounds.
Although there is a substantial overlap between the pseudorapidity regions covered by MPCAL$_p$ and CLC$_p$, the requirements of MPCAL$_p=0$ and  CLC$_p=0$ are nevertheless complementary, as the two systems detect hadrons and electromagnetic particles with different efficiencies. 

\begin{figure}
 \begin{center}
 \includegraphics[width=7.8cm]{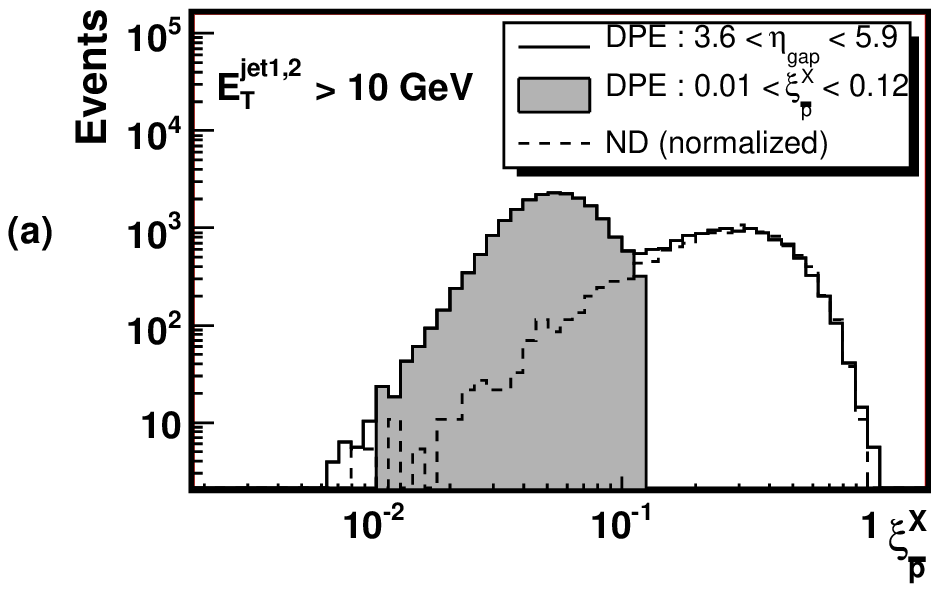}
 \includegraphics[width=7.8cm]{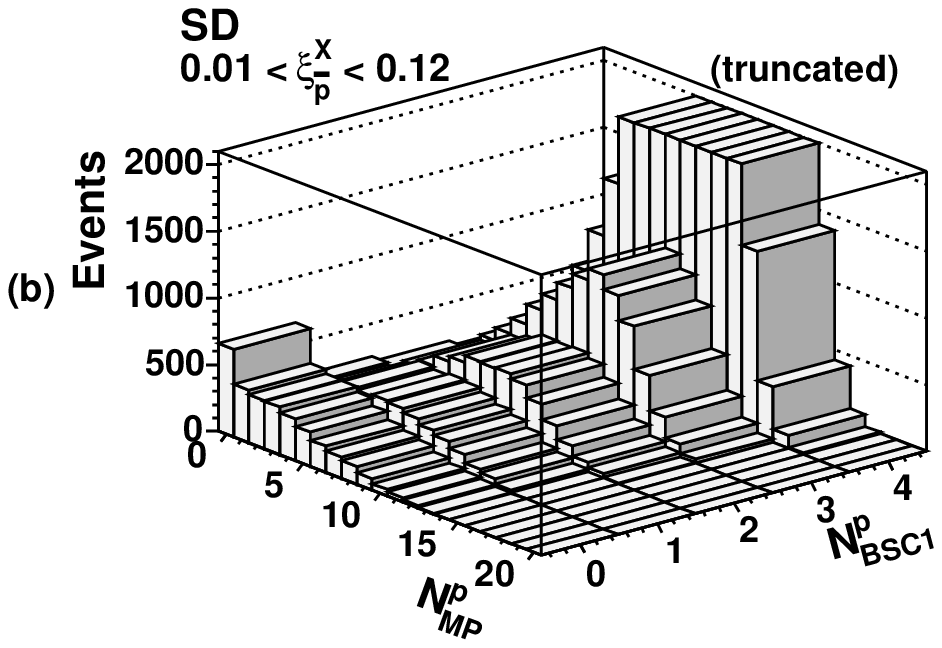}
 \caption{(a) $\xi_{\bar{p}}^X$ distribution of DPE events passing the ${\rm LRG}_p$ requirement (solid histogram), with the 
shaded area representing events in the region $0.01<\xi_{\bar{p}}^X<0.12$; the dashed histogram shows the $\xi_{\bar{p}}^X$ 
distribution for ND events passing the same ${\rm LRG}_p$ requirement and normalized to the solid histogram in the plateau region of $0.22<\xi_{\bar{p}}^X<0.50$; (b) MPCAL hit multiplicity, ${\rm N}_{MP}^{p}$, vs. ${\rm BSC1}_p$ hit counter multiplicity, ${\rm N}_{BSC1}^p$, in SD 
events with $0.01<\xi_{\bar{p}}^X<0.12$.
\label{fig:xi_pbar}}
 \end{center}
\end{figure}

{\boldmath $\xi_{\bar{p}}^X$} {\bf cut:} $0.01<\xi_{\bar{p}}^X<0.12$. 
In the high instantaneous luminosity environment of Run~II, multiple $\bar{p}p$ interactions occurring in the same beam-beam bunch crossing may result in overlap events consisting of a ND dijet event overlapped by a soft SD event with a leading $\bar{p}$ triggering by the RPS. These events, which are a background to both DPE and SD dijet events, can be well separated from diffractively produced dijet events using the variable $\xi_{\bar{p}}^X$, defined as  
\begin{equation}
 \xi_{\bar{p}}^X = \frac{1}{\sqrt{s}}\,\sum_{i=1}^{N_{tower}}(E_T^i\,e^{-\eta^i}),
\label{eq:xiX}
\end{equation}
where the sum is carried out over all calorimeter towers with $E_T>100$ MeV for CCAL and PCAL, and $E_T>20$ MeV for MPCAL. The tower $E_T$ and $\eta$ are measured with respect to the primary vertex position. The variable $\xi_{\bar{p}}^X$ represents the fractional longitudinal momentum loss of the $\bar p$ measured using calorimeter information. For events with a gap on the $\bar p$ side, $\xi_{\bar{p}}^X$ is calibrated by comparing data with Monte Carlo generated events. Calibrated $\xi_{\bar{p}}^{X}$ values were found to be in good agreement with values of $\xi_{\bar{p}}$ measured by the RPS, $\xi_{\bar{p}}^{\rm RPS}$. 
On the proton side where there is no RPS, $\xi_p^{X}$ is obtained from calorimeter information using Eq.~\ref{eq:xiX} in which $-\eta$ in the exponent is changed to $+\eta$ and is calibrated using the MC technique that was validated by the comparison with RPS data on the $\bar p$ side. Fig.~\ref{fig:xi_pbar}~(a) shows $\xi_{\bar{p}}^X$ distributions for events of the DPE event sample selected with the LRG$_p$ requirement. The events in the peak at 
$\xi_{\bar{p}}^X\sim0.05$ are dominated by DPE dijets, while the broad peak around $\xi_{\bar{p}}^X\sim0.3$ are residual overlap ND dijet events for which the LRG$_p$ is caused by downward multiplicity fluctuations. 

\begin{figure*}
\begin{center}
 \includegraphics[width=5.8cm]{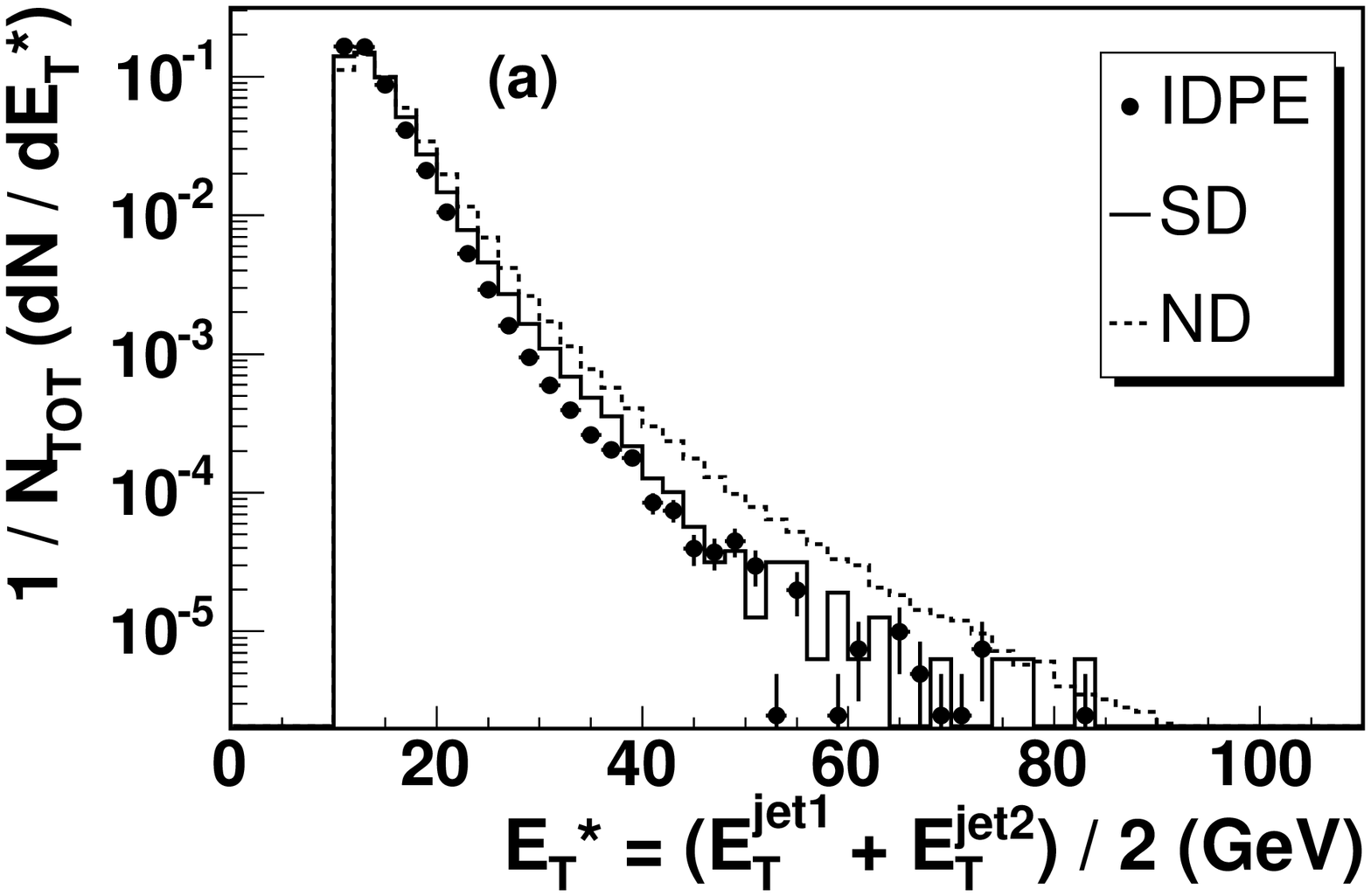}
 \includegraphics[width=5.8cm]{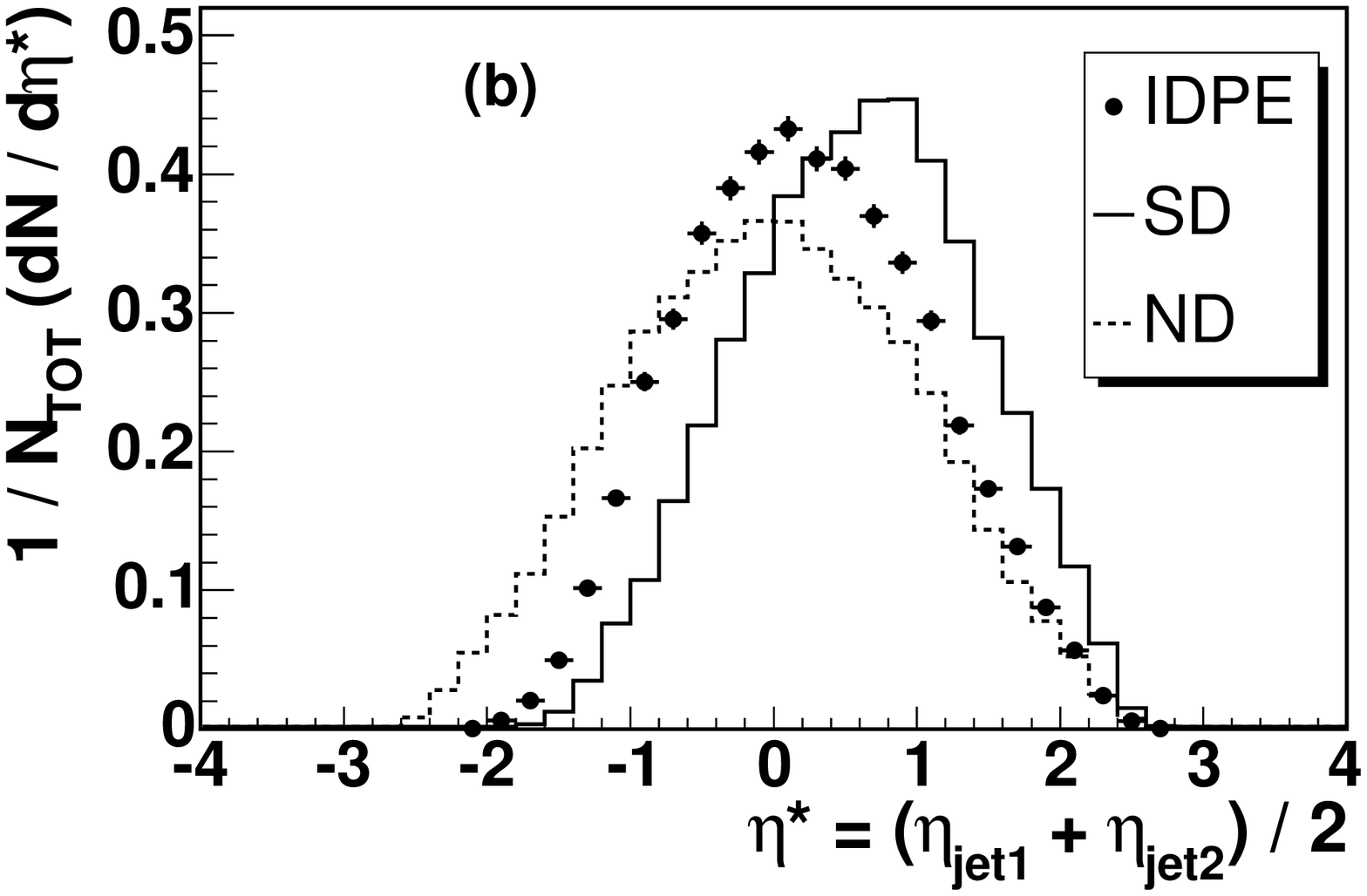}
 \includegraphics[width=5.8cm]{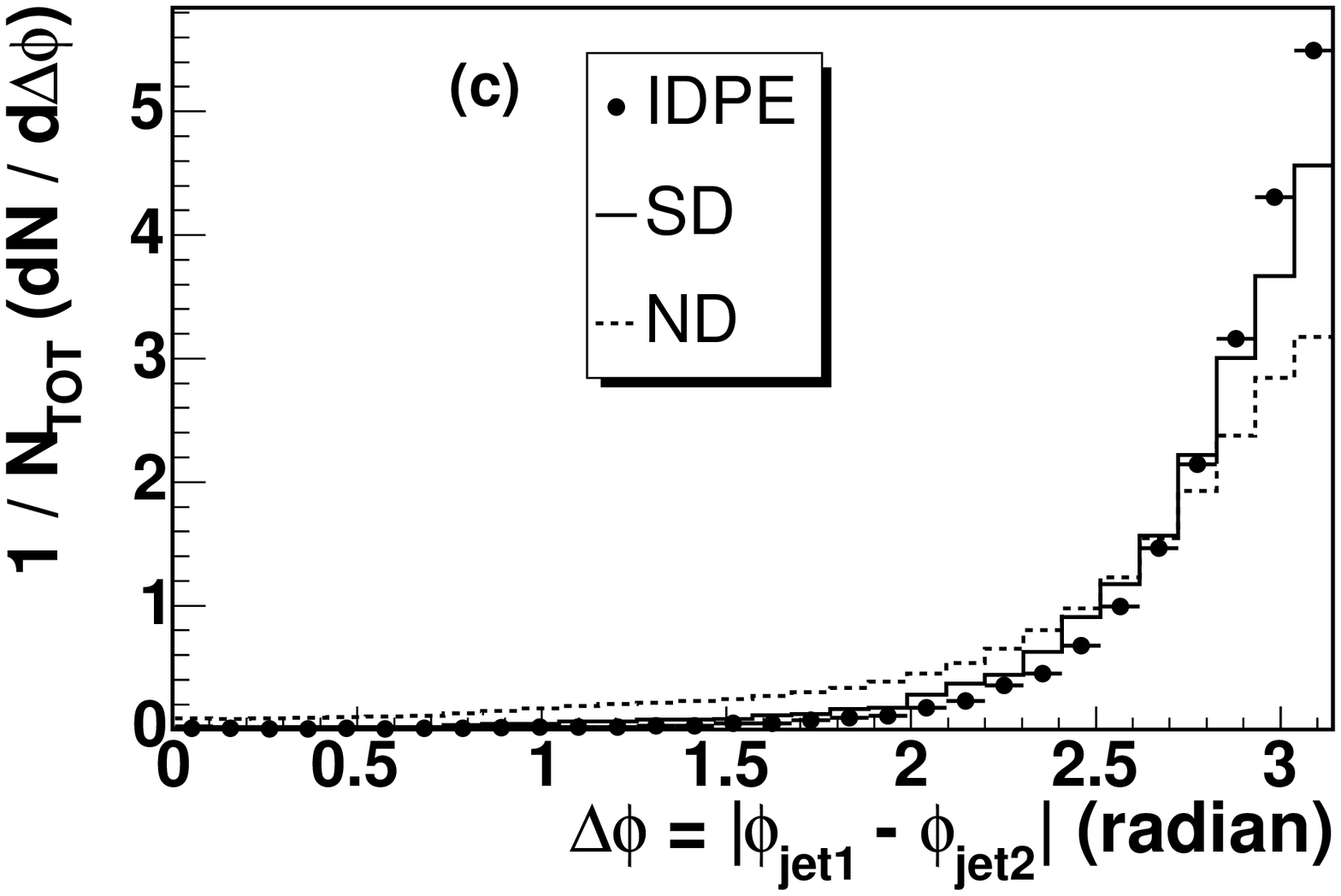}
 \caption{(a) Mean $E_T$, (b) mean $\eta$ of the two leading jets, and (c) azimuthal angle difference between the two leading jets of $E_T>10$ GeV in IDPE (circles), SD (solid histograms), and ND (dashed histograms) dijet events. 
\label{fig:jet_kin}}
 \end{center}
\end{figure*}

In this analysis, we use the DPE dominated events in the range $0.01<\xi_{\bar{p}}^X<0.12$. The same $\xi_{\bar{p}}^X$ requirement is used in selecting the SD event sample. Figure~\ref{fig:xi_pbar}~(b) shows the MPCAL hit multiplicity, $N_{MP}^p$, vs.
BSC1 hit counter multiplicity, $N_{BSC1}^p$, for the SD event sample.  The majority of the events have $5<N_{MP}^p<10$ and $N_{BSC1}^p\geq3$, but there are also some events with $N_{BSC1}^p=N_{MP}^p=0$, which are due to DPE events in the SD event sample efficiently passing the $\overline{BSC1_p}$ trigger requirement. 

The above trigger and offline selection requirements define  the inclusive DPE event sample (IDPE).\\
\begin{eqnarray}
&\mbox{\bf IDPE sample:}\label{IDPEsample}\\
&DPE\cdot LRG_p\cdot\xi_{\bar p}^X.\nonumber
\end{eqnarray}
The IDPE sample contains 20~414 events. 

In Fig.~\ref{fig:jet_kin}, we compare distributions of the mean dijet transverse energy, 
$E_T^*=(E_T^{jet1}+E_T^{jet2})/2$, mean pseudorapidity, $\eta^*=(\eta_{jet1}+\eta_{jet2})/2$, and 
azimuthal angle difference, $\Delta\phi=|\phi_{jet1}-\phi_{jet2}|$, for IDPE (points),
SD (solid histogram), and ND (dashed histogram) events. 
All distributions are normalized to unit area. The IDPE, SD and ND distributions exhibit the following features: (a) the $E_T^*$ distributions for IDPE, SD and ND events are similar at low $E_T^*$ , but reach larger $E_T^*$ values for SD and ND events due to  the higher c.m.s. energies of $\Pomeron$-$p$ and $\bar pp$ collisions relative to $\Pomeron$-$\Pomeron$ collisions; (b) the ND $\eta^*$ distribution is symmetric about $\eta^*=0$, as expected, and the DPE  distribution is approximately symmetric as the jets are produced in collisions between two Pomerons of approximately equal momentum (due to the approximately equal gap size on the $p$ and $\bar p$ sides), while the SD distribution is boosted toward positive $\eta^*$ (outgoing $p$ direction) due to the jets being produced in collisions between a proton carrying the beam momentum, $p_0$, and a Pomeron of much smaller momentum, $\xi_{\bar{p}}p_0$; and (c) the jets are more back-to-back in SD than in ND events, and even more so in IDPE events, due to less gluon radiation being emitted in events where colorless Pomerons are exchanged.

\section{Inclusive DPE Dijet Production}\label{sec:xsec_incl}
The cross section for inclusive DPE dijet production is obtained from the IDPE event sample using the expression
\begin{equation}
 \sigma_{DPE}^{incl} = \frac{N_{DPE}^{jj}(1-F_{BG})}{L \cdot \epsilon},
 \label{eq:xsec}
\end{equation}
where $N_{DPE}^{jj}$ is the number of DPE dijet events corrected for losses due to multiple interactions and for smearing effects on $E_T^{jet}$ due to the detector resolution, $F_{BG}$ is the non-DPE background fraction, $L$ is the integrated luminosity, and $\epsilon$ is the total event selection efficiency including detector acceptance. Details are provided below. 

\subsection{Non-DPE background events}\label{subsec:bg}
There are two sources of non-DPE background events in the IDPE event sample underneath the DPE peak at $0.01<\xi_{\bar{p}}^X<0.12$ shown in Fig.~\ref{fig:xi_pbar}~(a): one due to ND dijet events and the other due to SD ones. In both cases, the ${\rm LRG}_p$ requirement of ${\rm N}_{BSC1}^p={\rm N}_{MP}^p={\rm N}_{CLC}^p=0$ is satisfied by downward multiplicity fluctuations.

\paragraph*{\bf Non-diffractive background.} The ND background is caused by the RPS being triggered  either by a real antiproton from an overlapping soft SD event or by a particle originating in beam-pipe or beam-gas interactions.  This background is estimated from the 
$\xi_{\bar{p}}^X$ distribution of ND dijet events with ${\rm N}_{BSC1}^p={\rm N}_{MP}^p={\rm N}_{CLC}^p=0$ normalized to the $\xi_{\bar{p}}^X$ distribution of DPE events in the region $0.22<\xi_{\bar{p}}^X<0.50$, which is dominated by ND events. The DPE (normalized ND) $\xi_{\bar{p}}^X$ distribution is shown in Fig.~\ref{fig:xi_pbar}~(a) as a solid (dashed) histogram. 
Integrating the ND distribution over the range $0.01<\xi_{\bar{p}}^X<0.12$, we obtain the fraction of ND dijet background in the  IDPE event sample to be $F_{BG}^{ND}=13.3\pm0.2$~\%. 

\begin{figure}
 \begin{center}
 \includegraphics[width=4.2cm]{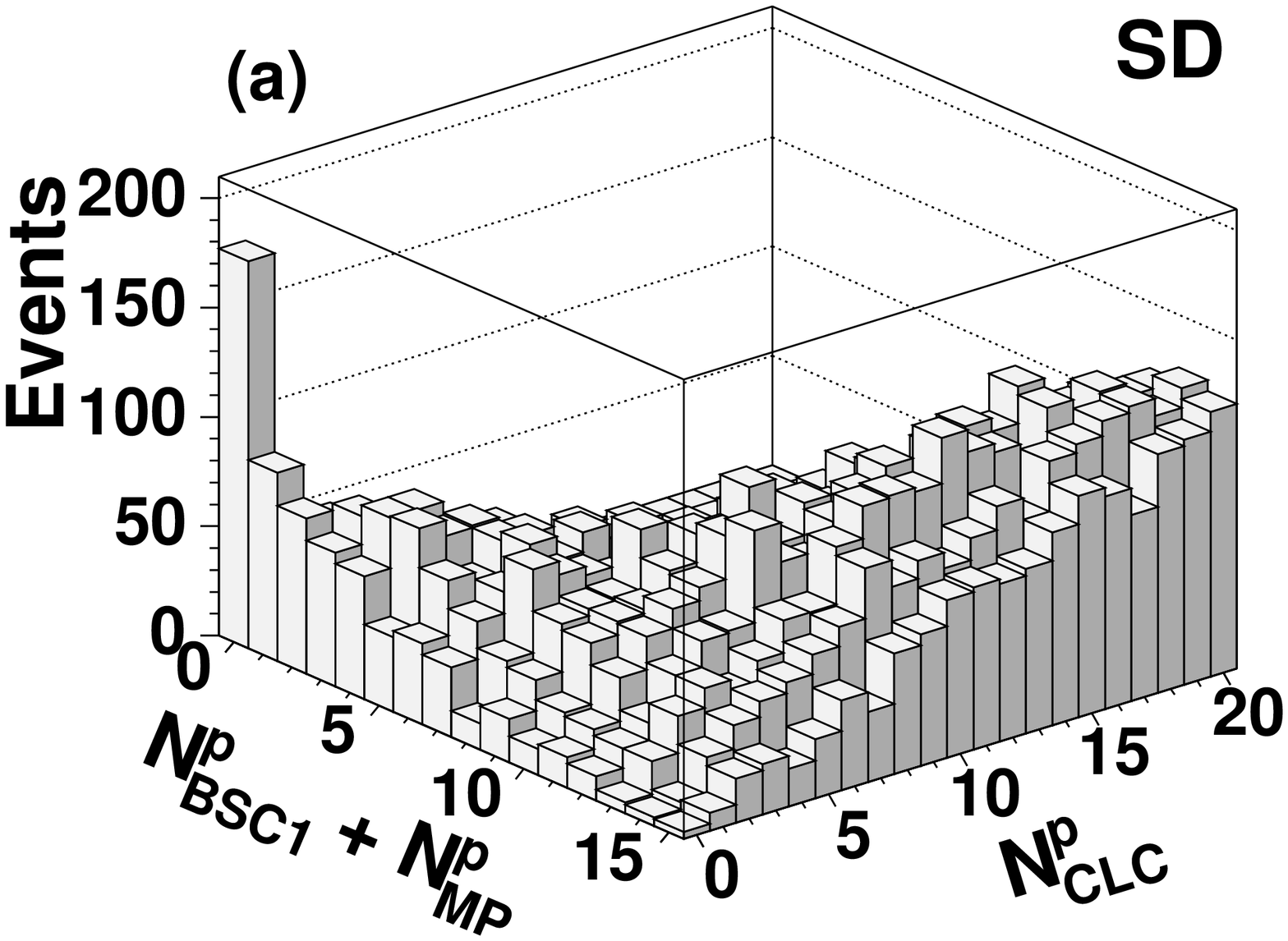}
 \includegraphics[width=4.2cm]{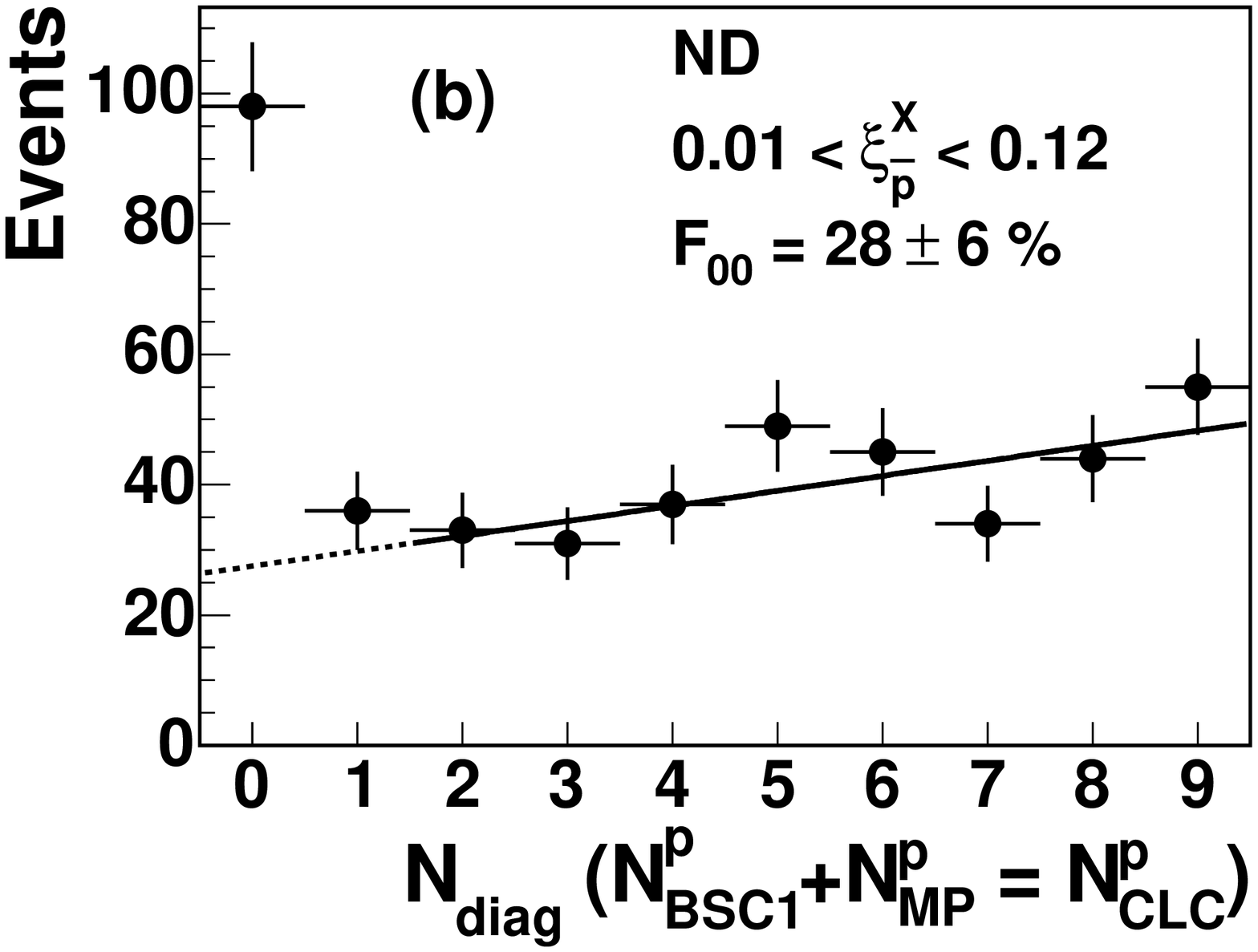}
 \caption{(a) Sum of the BSC1 hit counter multiplicity and MPCAL hit multiplicity, ${\rm N}_{BSC1}^p+{\rm N}_{MP}^p$, versus
CLC hit multiplicity, ${\rm N}_{CLC}^p$, in SD events with dijets of $E_T^{jet1,2}>10$ GeV and $0.01<\xi_{\bar{p}}^X<0.12$;
(b) multiplicity distribution along the diagonal line in the left plot, ${\rm N}_{diag}$, defined by ${\rm N}_{BSC1}^p+{\rm N}_{MP}^p$=${\rm N}_{CLC}^p$, with
the solid line representing a linear fit in the region $2\leq {\rm N}_{diag}\leq 9$, and the dashed line the extrapolation of the fit to
the ${\rm N}_{diag}=0$ bin.
\label{fig:sd_bg}}
 \end{center}
\end{figure}

\paragraph*{\bf Single diffractive  background.} The SD background is estimated by examining the correlation between
${\rm N}_{BSC1}^p+{\rm N}_{MP}^p$ and ${\rm N}_{CLC}^p$ in the SD data sample. Figure~\ref{fig:sd_bg}~(a) shows the distribution of 
${\rm N}_{BSC1}^p+{\rm N}_{MP}^p$ versus ${\rm N}_{CLC}^p$ for SD dijet events with $0.01<\xi_{\bar{p}}^X<0.12$.
The multiplicity along the diagonal, $N_{diag}$, defined by ${\rm N}_{BSC1}^p+{\rm N}_{MP}^p$=${\rm N}_{CLC}^p$, is well fitted with a linear function in the  region $2\leq N_{ diag}\leq 14$, as shown in  Fig.~\ref{fig:sd_bg}~(b).  The diagonal distribution is used because it monotonically decreases as $N_{diag}\rightarrow 0$ providing the least background under the peak. Extrapolating the fit to the bin with ${\rm N}_{BSC1}^p={\rm N}_{MP}^p={\rm N}_{CLC}^p=0$ yields a SD background fraction of $F_{00}=24\pm4$~\%. After correcting for a ND content of 42\% in the SD data, estimated by applying the method used in evaluating $F_{BG}^{ND}$, we obtain a single diffractive background fraction of $F^{SD}_{BG}=F_{00}\times(1-0.42)=14\pm3$~\%.

\subsection{Corrections for multiple interactions}\label{subsec:corr1}
Multiple interactions in the same beam-beam crossing may produce additional events which overlap the DPE event and cause it to fail the event selection requirements by contributing extra event vertices and/or by spoiling the rapidity gap on the proton side.    
 Corrections for DPE event losses due to multiple interactions are considered separately for overlapping events with one or more reconstructed vertices, and for overlapping events which do not have a vertex but nevertheless spoil the $LRG_p$. The latter also account for $LRG_p$ losses due to  beam background and/or detector noise.

\paragraph*{\bf Overlap events  with a reconstructed vertex.} The average number of inelastic $\bar{p}p$ interactions per bunch crossing
is given by $\bar{n}_i = {\cal L}_i\cdot\sigma_{inel}/f_0$, where ${\cal L}_i$ is the instantaneous luminosity, $\sigma_{inel}$ the inelastic interaction cross section, and $f_0$ the Tevatron bunch crossing frequency of 1.674 MHz. 
The average number of $\bar pp$ interactions which have a vertex, $\bar{n}_i^{vtx}$, is obtained by replacing $\sigma_{inel}$ with $\sigma_{inel}^{vtx}$, the cross section of $\bar{p}p$ interactions with a vertex. From Poisson statistics, the probability that no $\bar{p}p$ interaction producing a vertex occurs in a beam-beam bunch crossing is given by ${\rm P}(0)= \exp(-\bar{n}_i^{vtx})$.
The number of observed DPE events, $N_{DPE}$, corrected for losses due to multiple interactions  that yield overlap events with a vertex, 
${\rm N}^{corr}_{DPE}$, is obtained by weighting every event by ${\rm P}(0)^{-1}$ and summing up 
over all DPE events: ${\rm N}_{DPE}^{corr} = \sum_{i=1}^{{\rm N}_{DPE}}\exp(\bar{n}_i^{vtx})$.

The value of $\sigma_{inel}^{vtx}$, which is needed for evaluating $\bar n_i^{vtx}$, is obtained from an analysis of the fraction of events with one or more reconstructed vertices contained in a sample of zero-bias events collected by triggering on beam-beam crossings during the same time period in which the DPE sample was taken. 
The zero-bias sample is split into small sub-samples corresponding to different time slots of data taking to account for changes in beam and detector conditions, and the fraction of events with $\ge 1$ vertex as a function of instantaneous luminosity for each sub-sample is fit to the expected fraction given by $1-{\rm P}(0)=1-\exp(-{\cal L}_i\cdot\sigma_{inel}^{vtx}/f_0)$  with 
$\sigma_{inel}^{vtx}$ as a free parameter. The average value obtained from the fits is $\sigma_{inel}^{vtx}=30.3\pm1.5$~(syst) mb, where the uncertainty is evaluated from the variations observed among the different sub-samples.
Using this value, we obtain ${\rm N}_{DPE}^{corr}=189~317\pm1325$ events for the IDPE sample. The $\pm 1.5$~mb uncertainty in $\sigma_{inel}^{vtx}$ leads to an uncertainty of $\sim 2$~\% on ${\rm N}_{DPE}^{corr}$, 
which is negligibly small compared to other uncertainties discussed below.

\paragraph*{\bf Overlap events with no reconstructed vertex.} The rapidity gap of DPE events remaining after rejecting events with more than one vertex could be further spoiled by the presence of additional soft $\bar{p}p$ interactions with no reconstructed vertex, by beam background, or by detector noise. The correction for these effects is obtained from the same zero-bias samples by selecting events with no reconstructed vertex and evaluating the fraction ${\rm F}_{gap}$ of events with ${\rm LRG}_p$. The correction factor, ${\rm F}_{gap}^{-1}$, evaluated for bins of different instantaneous luminosity and data taking time, is then 
applied to ${\rm N}_{DPE}^{corr}$ for the same instantaneous luminosity and time bins.
Within the instantaneous luminosity range of
$10^{31}<{\cal L}_i<4\times10^{31}$ cm$^{-2}$s$^{-1}$ of our DPE data sample, ${\rm F}_{gap}$ varies between 70~\% and 30~\%. 

\subsection{Event selection efficiency}\label{subsec:eff}
\paragraph*{\bf Jet selection efficiency.} The trigger efficiency for jets with a calorimeter trigger tower of $E_T>5$~GeV is obtained from a sample of minimum-bias (MB) events triggered only on a CLC coincidence between the two sides of the detector. The $E^{jet}_T$ and $\eta^{jet}$ are reconstructed using the same algorithm as that used in the analysis of the IDPE dijet event sample. For MB events that contain a calorimeter trigger tower of $E^{tower}_T>5$ GeV, jets are selected if the trigger tower is contained within the $\Delta\eta$-$\Delta\phi$ cone of the jets. The trigger efficiency per jet is determined in bins of $E^{jet}_T$ and $\eta^{jet}$ as the ratio of the number of jets containing a trigger tower of $E^{tower}_T>5$ GeV to the total number of jets in all MB events. 
The single tower trigger efficiency for a given DPE dijet event, $\epsilon_{ST}$, is derived from the efficiency per jet, the number of jets in the event, and the $E_T$ and $\eta$ values of each jet. The DPE data are corrected for the trigger efficiency by assigning a weight of $\epsilon_{ST}^{-1}$ to each  event.

\paragraph*{\bf RPS trigger efficiency.} The efficiency of triggering on a leading antiproton in the  RPS trigger counters may be expressed as the product of the trigger   
counter acceptance, $A_{RPS}$, and the  
RPS detector efficiency, $\epsilon_{RPS}$. The latter
can be further factorized into two terms: the efficiency for finding the antiproton hits, $\epsilon_{RPSh}$,  and the efficiency  of the hit signals passing the trigger requirement, $\epsilon_{RPSt}$. 
From a study of trigger counter signals produced by particles reconstructed as single tracks using a zero-bias event 
sample, we obtain $\epsilon_{RPSh}=97\pm1$~\%. Using zero-bias events with signals in the trigger counters consistent with the response expected for minimum ionizing particles, $\epsilon_{RPSt}$ is found to be unity. The trigger counter acceptance is obtained from a simulation of SD events using the beam transport matrix to carry the recoil $\bar p$ from the IP to the RPS detectors. The total RPS acceptance for DPE dijet events 
is obtained from the expression
\begin{equation}
 A_{RPS}^{total} = \frac{N_{DPE}}{\displaystyle{\sum_{i=1}^{N_{DPE}}}A_{RPS}(\xi_{\bar{p}_i}^{RPS},|t_{\bar{p}_i}^{RPS}|)^{-1}},
\end{equation}
where $t_{\bar{p}}^{RPS}$ is the four momentum transfer squared measured by the RPS and ${\rm N}_{DPE}$ the total number of DPE dijet events. For the events collected in our data taking period we obtain $A_{RPS}^{total}=78.4\pm0.3$ (stat.)~\%.

\paragraph*{\boldmath{$\xi_{\bar{p}}^X$} \bf{cut efficiency.}} The requirement of $0.01<\xi_{\bar{p}}^X<0.12$ is used as a pre-selection cut to reduce ND dijet background due to superimposed $\bar{p}p$ interactions. However, this cut also removes some DPE events. The efficiency for DPE events retained by this requirement is obtained from the $\xi_{\bar{p}}^X$ distributions of the DPE and ND dijet events shown in Fig.~\ref{fig:xi_pbar}~(a) and used in Sec.~\ref{subsec:bg} to estimate the ND dijet background fraction in the IDPE data. Subtracting the normalized ND from the DPE events and evaluating the ratio of events within $0.01<\xi_{\bar{p}}^X<0.12$ to the total number of events yields an efficiency of $98.5\pm0.2$~\%. The deficit of this efficiency with respect to unity is due to fluctuations and calorimeter resolution effects causing a small fraction of events to migrate outside the selected $\xi_{\bar{p}}^X$ range.

\paragraph*{\bf Single vertex cut efficiency.} The single vertex requirement (VTX cut), which is imposed to reject events with multiple interactions, also rejects single
interaction events with extra misidentified vertices resulting from ambiguities in track reconstruction. Comparing the number of IDPE events  
before and after imposing this requirement, 
we obtain a single vertex cut efficiency of $\epsilon_{1vtx}=98\pm1$~\%. Using a similar method, the efficiency of the requirement of the vertex
position being within $|z|<60$~cm is determined to be $\epsilon_{zvtx}=92\pm2$~\%.

\paragraph*{\bf Jet reconstruction efficiency.} The results presented are based on events with at least two jets of $E_T^{jet}>10$~GeV. The reconstruction of such relatively low $E_T$ jets in the CDF~II calorimeters is prone to  inefficiencies associated with the calorimeter measurement of particle energies and the jet reconstruction algorithm used. Jet reconstruction efficiencies are studied using Monte Carlo dijet event samples generated with {\sc pythia} 6.216~\cite{PYTHIA} and processed through a {\sc geant}-based detector simulation~\cite{CDFSIM}. Simulated jets are reconstructed at both particle and calorimeter levels using the same jet reconstruction algorithm as that used in the data analysis. Then, events with matched pairs of particle and calorimeter level jets in $y$-$\phi$ space are selected satisfying the requirement of $\Delta R = [(y_{CAL}-y_{HAD})^2 + (\phi_{CAL}-\phi_{HAD})^2]^{1/2} \leq 0.7$, where $y_{CAL}$ ($y_{HAD}$) and $\phi_{CAL}$ ($\phi_{HAD}$) are the rapidity and azimuthal angle of a calorimeter (particle) level jet. If more than one calorimeter level jet matches a  hadron level jet, the closest matched calorimeter level jet is chosen. Using this method, the jet reconstruction efficiency $\epsilon_{jet}$, defined as the fraction of hadron level jets that have a matched calorimeter level jet, is obtained as a function of hadron level jet $E_T$ and $\eta$.  We find that the value of  $\epsilon_{jet}$ is $\sim 83$~\% at $E^{jet}_T=10$ GeV and reaches full efficiency at $E^{jet}_T\sim 25$ GeV. The dijet reconstruction efficiency for a given  DPE event, $\epsilon_{dijet}$, is determined from the jet reconstruction efficiencies for the $E_T$ and $\eta$ of the jets in the event. In evaluating cross sections, each DPE dijet event is assigned a weight of $\epsilon_{dijet}^{-1}$, and the number of DPE dijet events is recalculated.

\subsection{Jet $E_T$ energy smearing}\label{subsec:jetsmear_incl}
The reconstruction of low $E_T$ jets suffers from energy smearing effects due to large fluctuations in the calorimeter response to low $E_T$ particles. These  effects, convoluted with a steeply falling $E^{jet}_T$ spectrum, cause migration of jets into adjacent $E^{jet}_T$ bins. The smearing is unfolded as a function of $E^{jet}_T$ using correction factors derived from inclusive DPE dijet events generated with the {\sc pomwig} Monte Carlo simulation~\cite{POMWIG}, described in Sec.~\ref{subsec:pomwig}, followed by a simulation of the detector. The $E^{jet}_T$ spectra of the second highest $E_T$ jet at particle and calorimeter levels are then compared. No matching between particle and calorimeter level jets in $y$-$\phi$ space is performed. The second highest $E^{jet}_T$  is used in order to conform with the minimum $E^{jet}_T$ thresholds imposed on $E_T^{jet2}$ in our cross section measurements and in available theoretical predictions. Calorimeter level jets are corrected for the relative response of the calorimeters and for the absolute energy scale. The correction factors, obtained for each $E_T^{jet2}$ bin as the ratio of the number of particle level jets to the number of calorimeter level jets, vary from  $0.93\pm0.03$ to $1.03\pm0.03$ within the region  of $10<E_T^{jet2}<50$ GeV. This correction is applied to the measured inclusive DPE dijet cross section as a function of $E_T^{jet2}$.

\begin{table*}
 \caption{Diffractive/Pomeron structure function (DSF) used in the inclusive dijet {\sc pomwig} Monte Carlo simulations. 
 \label{tab:pdf}}
 \begin{ruledtabular}
 \begin{center}
  \begin{tabular}{ll}
   DSF       & Definition \\\hline
   CDF$\oplus$H1 & $F_{jj}^D(\beta,Q^2)$ of Eq.~(\ref{eq:FjjD}) for $\Pomeron_{p(\bar p)}$ and H1 LO QCD fit2 for $\Pomeron_{\bar p(p)}$~\cite{RunI_dijet, RunI_DPE} \\
   CDF        & $F_{jj}^D(\beta,Q^2)$ of Eq.~(\ref{eq:FjjD}) for both $\Pomeron_p$ and $\Pomeron_{\bar p}$~\cite{RunI_dijet} \\
   H1-fit2    & H1 LO QCD fit2 with extended $Q^2$ range~\cite{H1fit} \\
   ZEUS-LPS   & NLO QCD fit to ZEUS LPS data~\cite{ZEUS_LPS}\\
  \end{tabular}
 \end{center}
 \end{ruledtabular}
\end{table*}

\section{Exclusive Dijet Production}\label{sec:excl_search}
The  exclusive dijet signal contained in the IDPE data sample  
is evaluated from the distribution of the dijet mass fraction, $R_{jj}$ (=$M_{jj}/M_X$), by measuring the excess of events at high $R_{jj}$ over expectations from the {\sc pomwig} Monte Carlo DPE event generator~\cite{POMWIG}, which does not simulate the exclusive process. Below, in Sec.~\ref{subsec:pomwig}, we demonstrate that the IDPE dijet data are well described by a combination of an inclusive MC generated distribution plus a non-DPE background obtained from data, in Sec.~\ref{subsec:search} we present the search for an exclusive dijet signal at high $R_{jj}$, in Sec.~\ref{subsec:exc_mc} we discuss expectations from an exclusive dijet Monte Carlo simulation, and in Sec.~\ref{subsec:inc+exc} we compare the data with an appropriately normalized combination of inclusive plus  exclusive MC generated events. Cross section results for exclusive dijet production are presented in Sec.~\ref{sec:xsec}.    

\subsection{Inclusive {\sc pomwig} Monte Carlo Simulation}\label{subsec:pomwig}We first compare data distribution shapes with {\sc pomwig} predictions to verify that the data are well described by the MC simulation apart from deviations expected from the possible presence in the data of an exclusive dijet signal. The data used are the IDPE event sample defined in Eq.~(7) in Sec.~\ref{sec:data}, which contains  20~414 events. 
While this sample should contain a larger fraction of exclusive dijet events than the total DPE event sample defined in Eq.~(5), it is used because in searching for an exclusive signal, agreement between {\sc pomwig} predictions and data is more relevant if checked in a kinematic region as close as possible to that where the exclusive signal is expected. 

Dijet events are generated in {\sc pomwig}
using a $2\rightarrow2$ processes with Pomeron remnants (see  Eq.~\ref{eq:pompom}) and a minimum transverse momentum cut of $p^{min}_T=7$~GeV/c.
Each event is processed through the detector simulation and is required to pass the data analysis cuts.
In comparisons with IDPE data, the SD and ND backgrounds expected in the data are normalized to their respective 14.0~\%  and 13.3~\% values, estimated as described above in Sec.~\ref{sec:xsec_incl}, and are added to the {\sc pomwig} generated events. 
The MC distributions of {\sc pomwig} DPE plus SD and ND background events and the corresponding data distributions are normalized to the same area. 
Background SD distribution shapes are obtained from data satisfying the IDPE event sample requirements except for $\overline{{\rm BSC}_p}$, which is replaced by ${\rm N}_{BSC1}^p+{\rm N}_{MP}^p\leq1$ 
and ${\rm N}_{CLC}^p\leq1$ excluding events with ${\rm N}_{BSC1}^p={\rm N}_{ MP}^p={\rm N}_{CLC}^p=0$;
ND shapes are extracted from J5 data satisfying the ${\rm LRG}_p$ and $\xi_{\bar p}^X$ requirements.

As a diffractive/Pomeron structure function we use   
$F_{jj}^D(\beta,Q^2)\propto \beta^{-1}$~\cite{RunI_dijet}, where $\beta$ is the longitudinal momentum fraction of the parton in the Pomeron related to the $x$-Bjorken variable $x_{Bj}$ ($x$-value of the struck parton) by $\beta\equiv x_{Bj}/\xi$. The $Q^2$ dependence of $F_{jj}^D$ is implemented as a weight to $F_{jj}^D(\beta)$, 
determined from the CTEQ6L~\cite{CTEQ6L} proton parton distribution function (PDF) at the $Q^2$ scale of the event. The justification for using the proton PDF is based on a Run~II CDF measurement of a rather flat $Q^2$ dependence of the ratio of SD to ND structure functions, indicating that the Pomeron evolves with $Q^2$ similarly to the proton~\cite{KG_d2006}.
We assign 46~\% and 54~\% of the Pomeron momentum to quarks 
($u$, $d$, $\bar u$, $\bar d$) and gluons, respectively, as measured by CDF in Run~I from diffractive $W$, dijet, and 
$b$-quark production~\cite{RunI_Wjjb}. In view of the above considerations, the following structure function form is employed in the MC program, 
\begin{eqnarray}
 F_{jj}^D(\beta, Q^2) &=& 0.46 \cdot \frac14 \sum_{q=u,d,\bar{u},\bar{d}}\left[\frac{\omega_q(Q^2)}{\beta_q+a}\right]\nonumber\\
                      &&  + 0.54\cdot\frac{\omega_g(Q^2)}{\beta_g+a}, \label{eq:FjjD}
\end{eqnarray}
where $\omega_{q(g)}(Q^2)$ is a weight used to include the $Q^2$ dependence of the quark (gluon) PDF and 
$a=10^{-5}$ is an arbitrary parameter employed to avoid a divergence at $\beta=0$.

Diffractive/Pomeron structure functions (DSFs) are also provided in the {\sc pomwig} 
MC program, obtained from QCD analyses of H1 diffractive DIS data~\cite{H1fit}. Two of the H1 DSFs used in {\sc pomwig} are the H1 LO QCD fit2 ({H1-fit2}) with a $Q^2$ range extended to $10^5$~GeV$^2$ to cover the CDF 
range~\cite{H1Q2max}, and
the H1 NLO QCD fit3 ({H1-fit3}). 
Recently, QCD analyses of diffractive structure functions have also been performed by the ZEUS collaboration using diffractive DIS data
obtained with a Leading Proton Spectrometer (LPS)~\cite{ZEUS_LPS}, and also by the rapidity gap (or Mx) method~\cite{ZEUS_Mx}. 
We have implemented programs returning NLO QCD fits for {ZEUS-LPS} and {ZEUS-Mx} structure functions for use in {\sc pomwig}~(see Ref.~\cite{GLP_fit} for Mx data).
However, a more recent QCD analysis of diffractive DIS data performed by H1 using larger data samples and incorporating data from different final states~\cite{H12006} yields DSFs favoring the H1-fit2 DSF over the H1-fit3 DSF and in good shape agreement with the ZEUS-LPS DSF, while disfavoring the ZEUS-Mx DSF. Therefore, for consistency among measured DSFs at HERA, we exclude the H1-fit3 and ZEUS-Mx DSFs from this analysis.

Guided by our Run~I DPE dijet analysis results, in which $F_{jj}^D$ measured from the ratio of DPE to SD dijet events was found to agree 
in shape and normalization with H1-fit2, while the $F_{jj}^D$ of Eq.~(\ref{eq:FjjD}) measured from diffractive dijet production is suppressed by a factor of $\sim 10$ relative to that from H1-fit2, we use $F_{jj}^D$ of Eq.~(\ref{eq:FjjD}) for the Pomeron emitted by the $p\;(\bar p)$ and H1-fit2 for that emitted by the $\bar p\;(p)$~\cite{KG_d2006}. This combination, which will be referred to as
{CDF$\oplus$H1}, is used as the default DSF in the {\sc pomwig} event generation. 
The four diffractive/Pomeron structure functions used in the analysis are listed in Table~\ref{tab:pdf}. 

\begin{figure*}
 \begin{center} 
 \includegraphics[width=15cm]{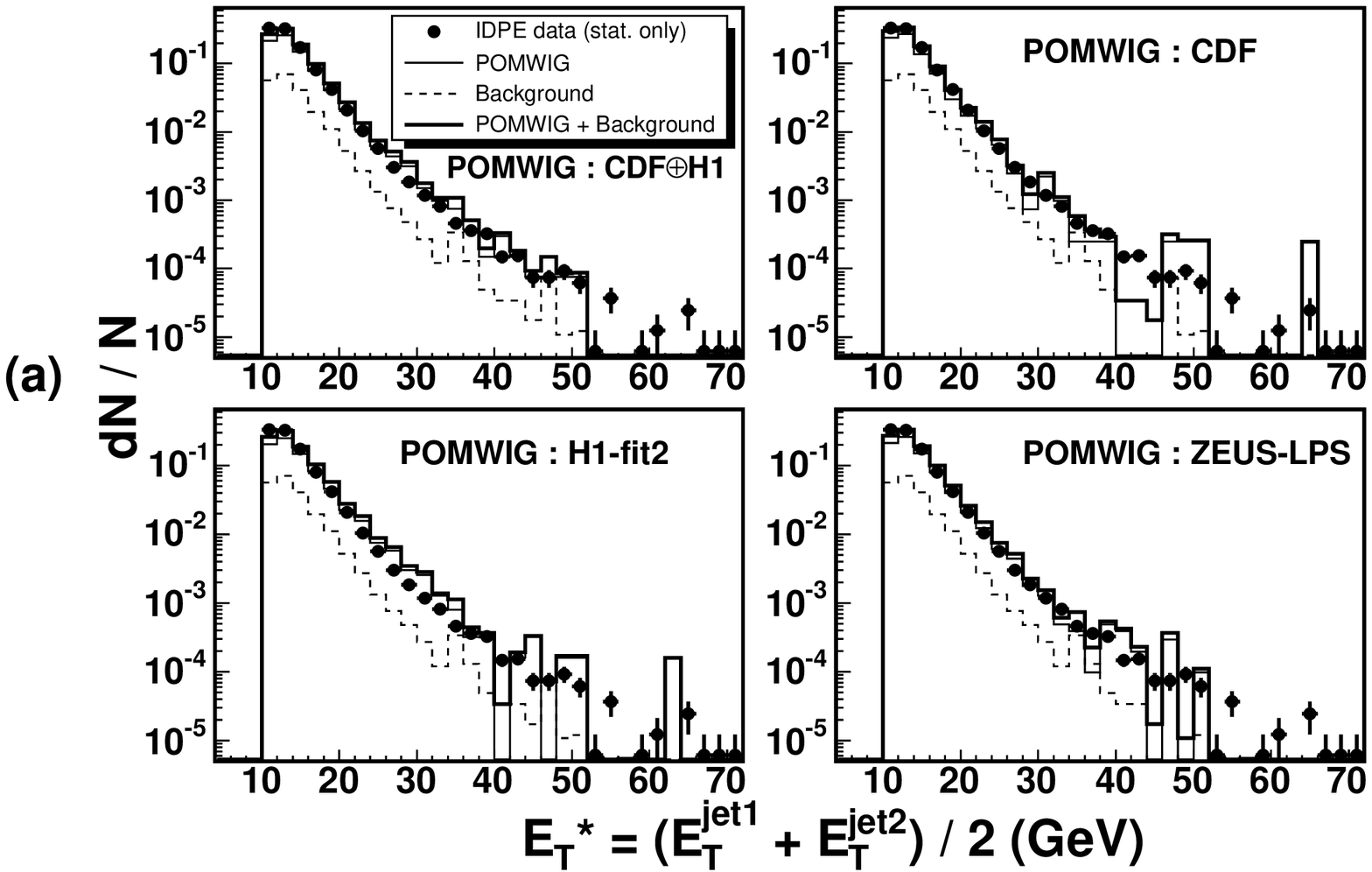}
 \includegraphics[width=15cm]{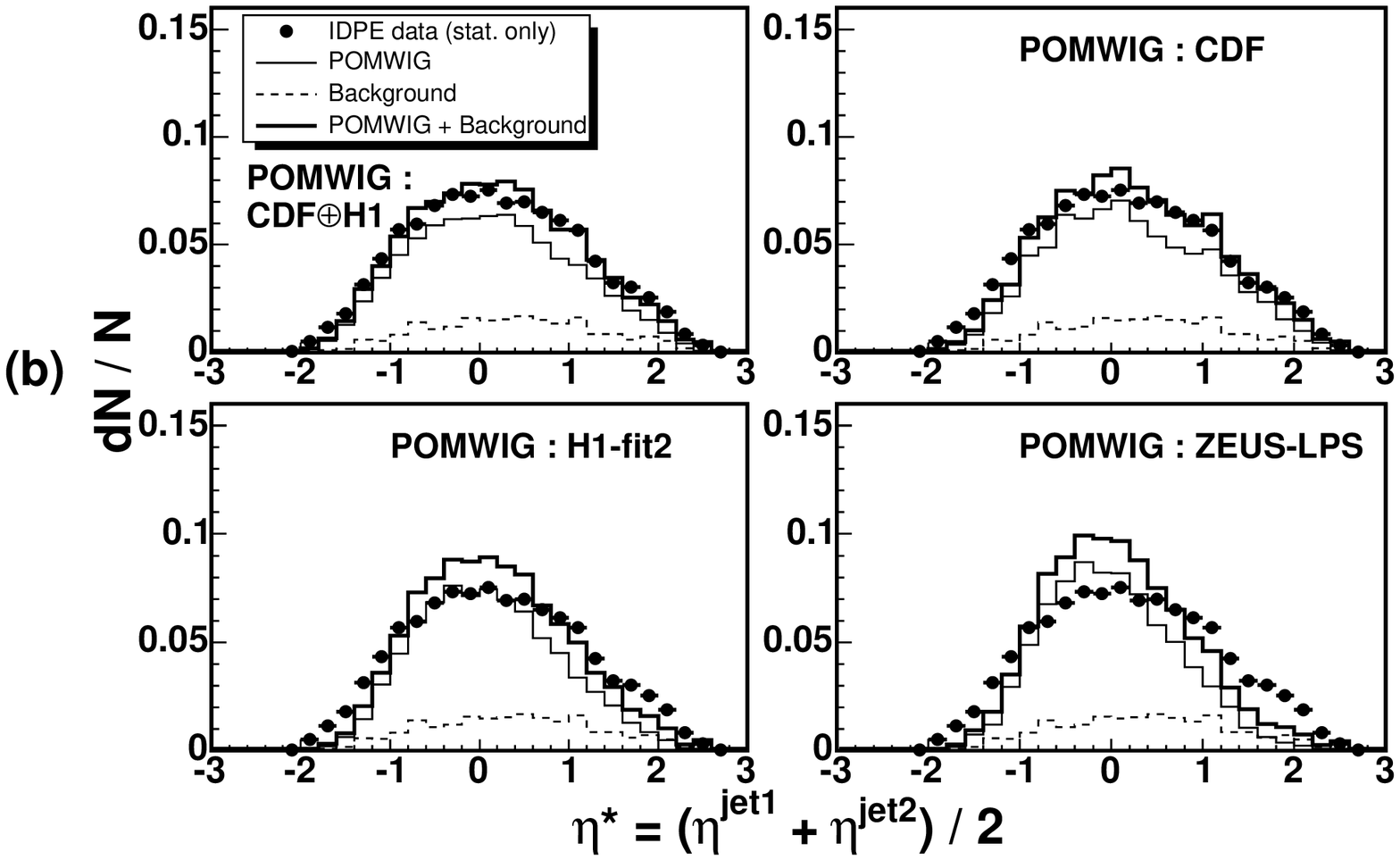}
 \caption{(a) Mean transverse energy $E_T^*$ and (b) mean pseudorapidity $\eta^*$ of the two highest $E_T$ jets 
in IDPE data (points) and {\sc pomwig} MC events (thick solid histograms) composed of {\sc pomwig} DPE signal (thin solid histograms) 
and the sum of SD and ND background events (dashed histograms). The data and {\sc pomwig}+background distributions are normalized to unit area. The {\sc pomwig} generated distributions in the plots (a) and (b) correspond to the four different diffractive/Pomeron structure functions used: 
CDF$\oplus$H1, CDF, H1-fit2, and ZEUS-LPS.
\label{fig:pomwig_et_eta}}
 \end{center}
\end{figure*}

\begin{figure*}
 \begin{center}
 \includegraphics[width=15cm]{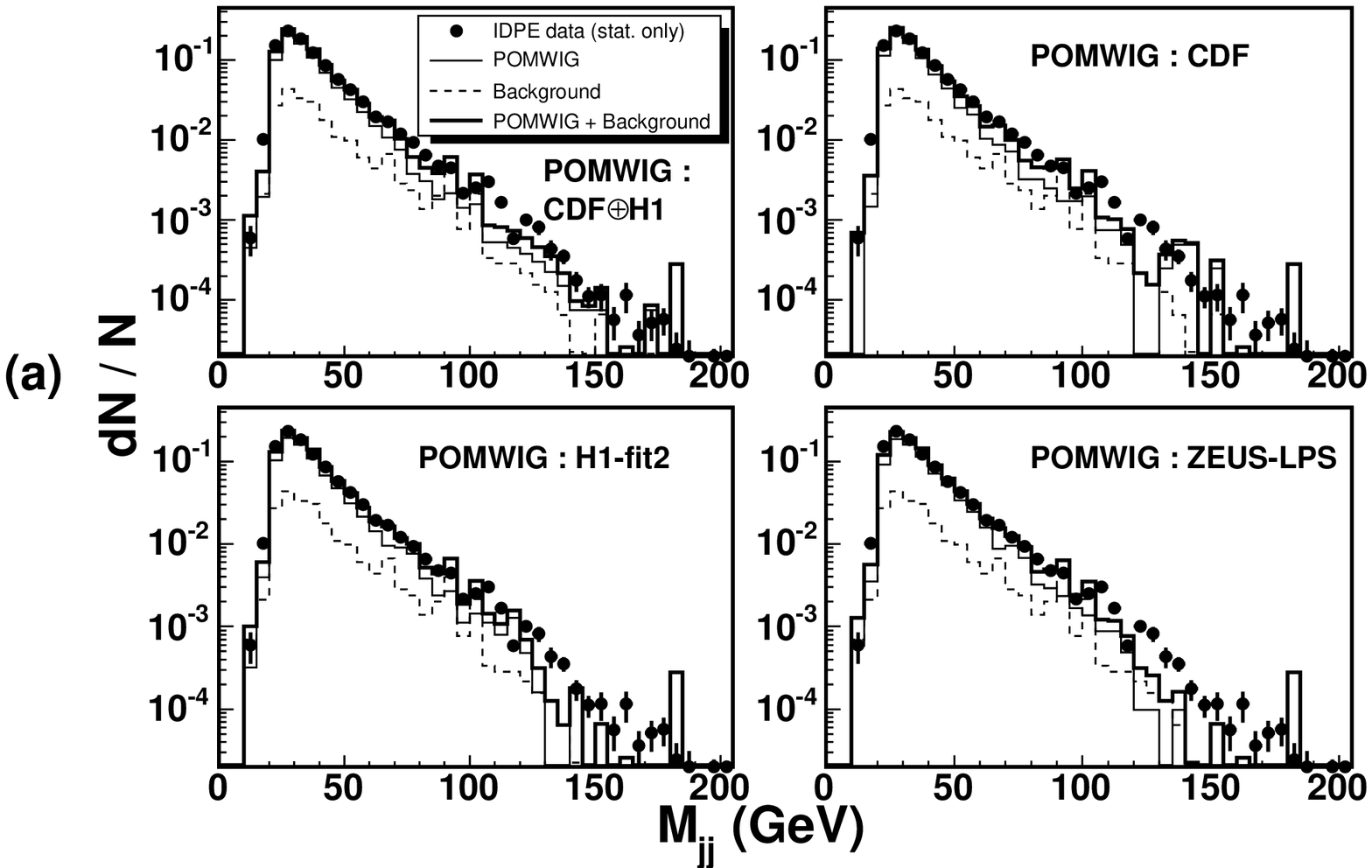}
 \includegraphics[width=15cm]{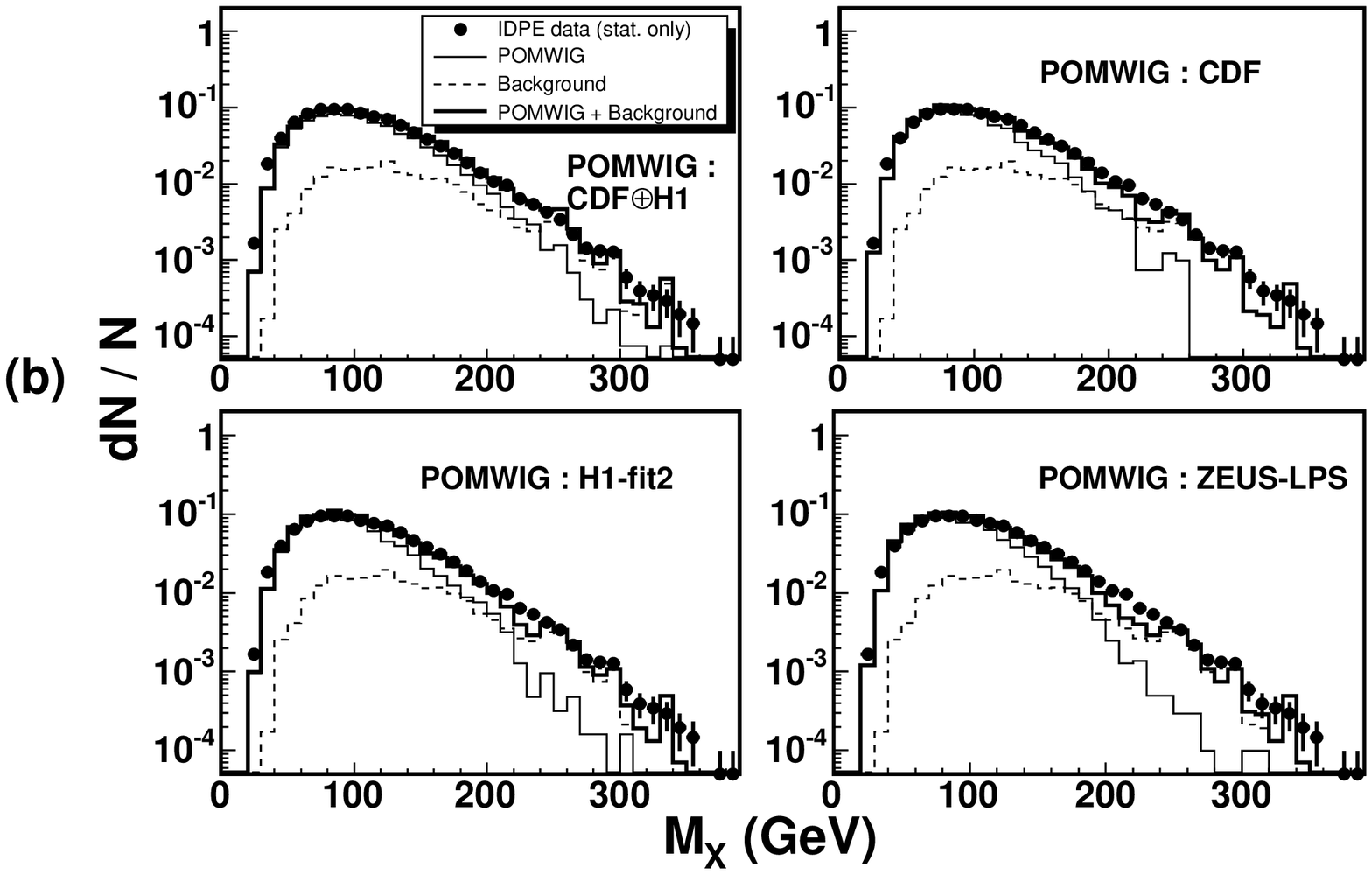}
 \caption{(a) Dijet mass $M_{jj}$ and (b) central hadronic system mass, $M_X$ 
in IDPE data (points) and {\sc pomwig} MC events (thick solid histograms) composed of {\sc pomwig} generated
DPE dijet events (thin solid histograms) and the sum of SD and ND generated background events (dashed histograms). 
The data and {\sc pomwig}+background distributions are normalized to unit area. The {\sc pomwig} DPE distributions in each set of four plots 
correspond to the four different diffractive/Pomeron structure functions used: CDF$\oplus$H1, CDF, H1-fit2, and ZEUS-LPS.
\label{fig:pomwig_mjj}}
 \end{center}
\end{figure*}

In Fig.~\ref{fig:pomwig_et_eta}, we compare (a) the average dijet $E_T^{jet}$ and (b) the average $\eta^{jet}$ distributions, $E_T^*$ and $\eta^*$, between IDPE data and {\sc pomwig} generated events using CDF$\oplus$H1, CDF, H1-fit2, and ZEUS-LPS diffractive/Pomeron structure functions. 
While all $E_T^*$ distributions have similar shapes, the data $\eta^*$ distribution is broader than all simulated ones. The larger width of the data $\eta^*$ distribution is due to the presence in the data of exclusive signal events concentrated in the pseudorapidity region around $\eta\sim -1$ (see Sec.~\ref{subsec:inc+exc}).
Figure~\ref{fig:pomwig_mjj} shows data and {\sc pomwig} distributions of the dijet invariant mass, $M_{jj}$, and mass of
the central hadronic system, $M_X$, where  
$M_X=[(\sum_{i=1}^{N_{tower}} E_i)^2 - (\sum_{i=1}^{N_{tower}} E_i\vec{n}_i)^2]^{1/2}$, where $E_i$ is the energy of a tower with $E_T>100$~MeV for CCAL or PCAL and $E_T>20$~MeV for MPCAL and $\vec{n}_i$ is a unit vector pointing to the center of the tower. Good agreement is observed between data and MC generated distributions for all four diffractive/Pomeron structure functions.

\subsection{Search for exclusive dijets}\label{subsec:search}
\begin{figure}[htp]
 \begin{center}
 \includegraphics[width=8.5cm]{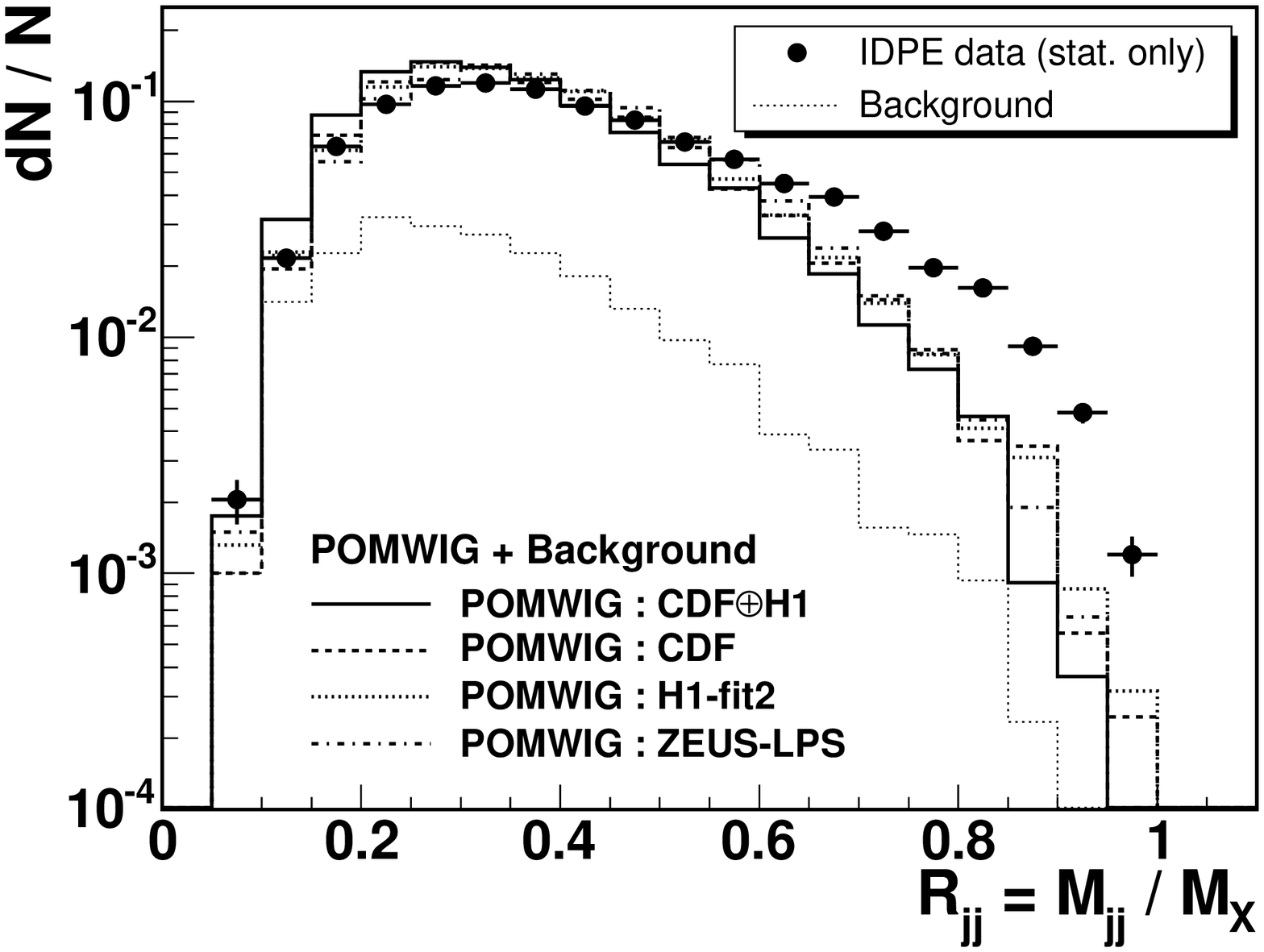}
 \caption{Dijet mass fraction, $R_{jj}$, in IDPE data (points) and {\sc pomwig} MC events (upper histograms), 
composed of {\sc pomwig} DPE signal and the sum of SD and ND background events (lower dashed histogram) normalized to the background fraction in the data. 
The upper four histograms correspond to the four
different diffractive/Pomeron structure functions used in {\sc pomwig}: CDF$\oplus$H1 (solid), CDF (dashed), H1-fit2 (dotted) and
ZEUS-LPS (dot-dashed histogram). These four histograms and the data distribution are normalized to unit area.   
\label{fig:pomwig_rjj}}
 \end{center}
\end{figure}

We search for exclusive dijet production by comparing data with {\sc pomwig} simulated  dijet mass fraction distributions, $R_{jj}=M_{jj}/M_X$, looking for an excess of data over simulation at high $R_{jj}$. Data and four {\sc pomwig} $R_{jj}$ distributions obtained with CDF$\oplus$H1, CDF, H1-fit2, and ZEUS-LPS DSFs are shown in Fig.~\ref{fig:pomwig_rjj}. All distributions are normalized to unit area.

\noindent An excess of data over simulated events at high $R_{jj}$ is observed for all four DSFs used in the simulation. This excess is examined for consistency with the presence in the data of an exclusive dijet signal by applying selection cuts expected to enhance the appearance of the signal. The following successive cuts have been studied:

{\bf (a) LRG}{\boldmath $_{\bar p}:$} this cut, which is the equivalent of 
 ${\rm LRG}_p$, retains events with ${\rm N}_{BSC1}^{\bar{p}}={\rm N}_{MP}^{\bar{p}}={\rm N}_{CLC}^{\bar{p}}=0$, enforcing a gap approximately covering the region $3.6<\eta<5.9$. 

{\bf (b) {\boldmath {$E_T^{jet3}<5$}}~GeV:} third jet veto. Applying a veto on events with three (or more) jets of $E_T^{jet3}\geq 5$~GeV further enhances the exclusive dijet signal, resulting  in a narrower exclusive signal peak in the $R_{jj}$ distribution. This requirement tends to shift events toward high $R_{jj}$ values by removing, for instance, exclusive dijet events which contain extra reconstructed jets originating from gluon radiation in parton showers. In evaluating exclusive cross sections, the loss of such events is accounted for by correcting the data for the exclusive signal acceptance obtained from exclusive dijet MC simulations.

{\bf (c) {\boldmath{$\eta^{jet}$}}-cut:} $\eta^{jet1}$ and/or $\eta^{jet2}$~$<-0.5$. This cut exploits correlations in the $\eta^{jet1}$ vs. $\eta^{jet2}$ distribution, which is more symmetric around $\eta^{jet1}=\eta^{jet2}=0$ for inclusive than for exclusive events. 

\begin{figure}
 \begin{center}
 \includegraphics[width=8.5cm]{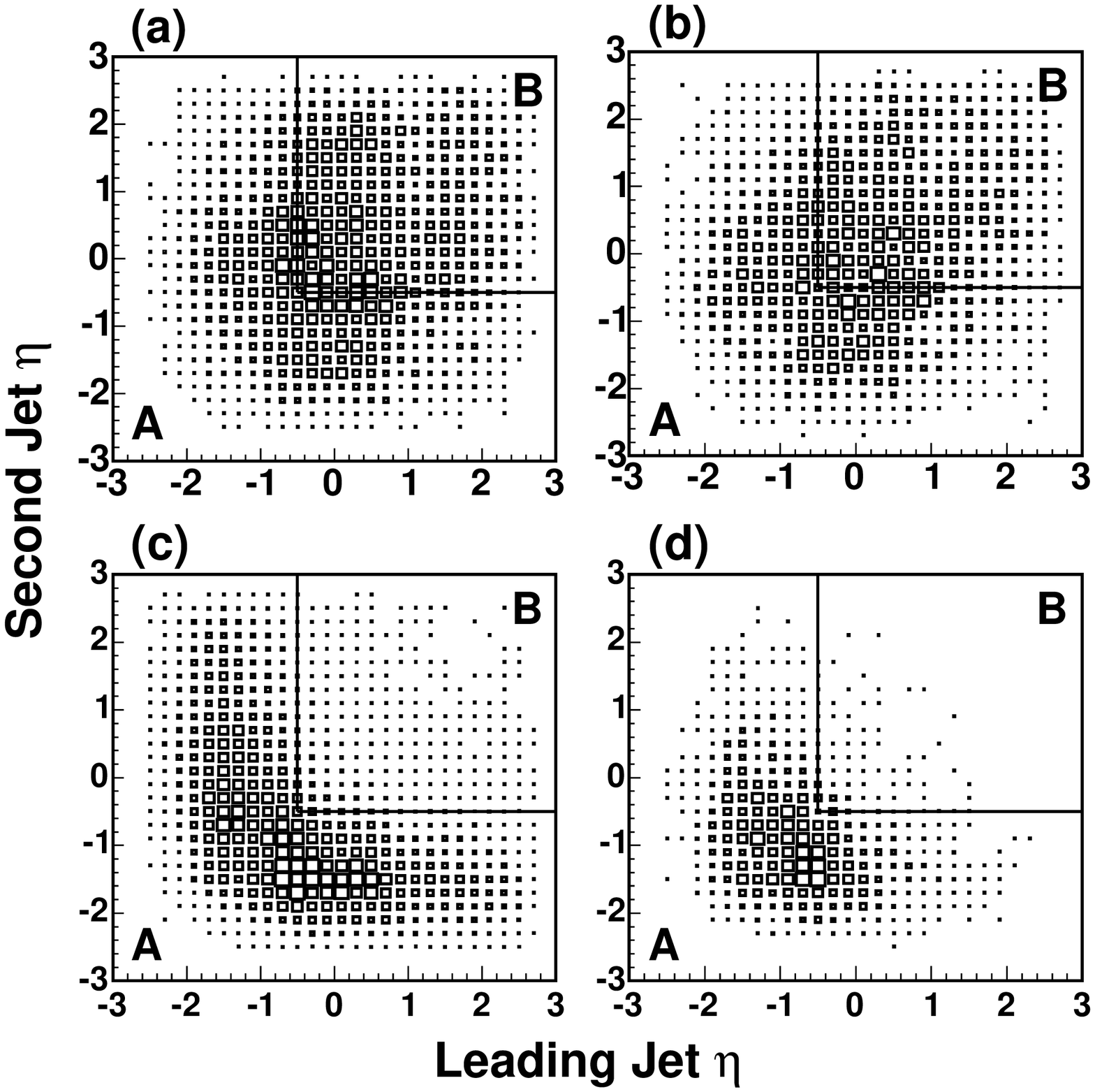}
 \caption{Second jet $\eta$ vs. leading jet $\eta$  for events with two jets of $E_T^{jet1,jet2}>10$~GeV satisfying all IDPE requirements plus the additional requirements of $LRG_{\bar p}$ and $E_T^{jet3}<5$~GeV (third jet veto): 
(a)  data, (b) {\sc pomwig} generated events, (c) exclusive {\sc ExclDPE} generated events, and (d) exclusive {\sc ExHuME} generated events. The solid lines represent the $\eta^{jet}$-cuts of $\eta^{jet1}$ and/or $\eta^{jet2}$~$<-0.5$ used to divide the $\eta^{jet1}$-$\eta^{jet2}$ space into the exclusive signal-enriched region A and the non-exclusive dominated region B. 
\label{fig:jet_eta_cuts}}
 \end{center}
\end{figure}

\begin{figure}
 \begin{center}
   \includegraphics[width=4.2cm]{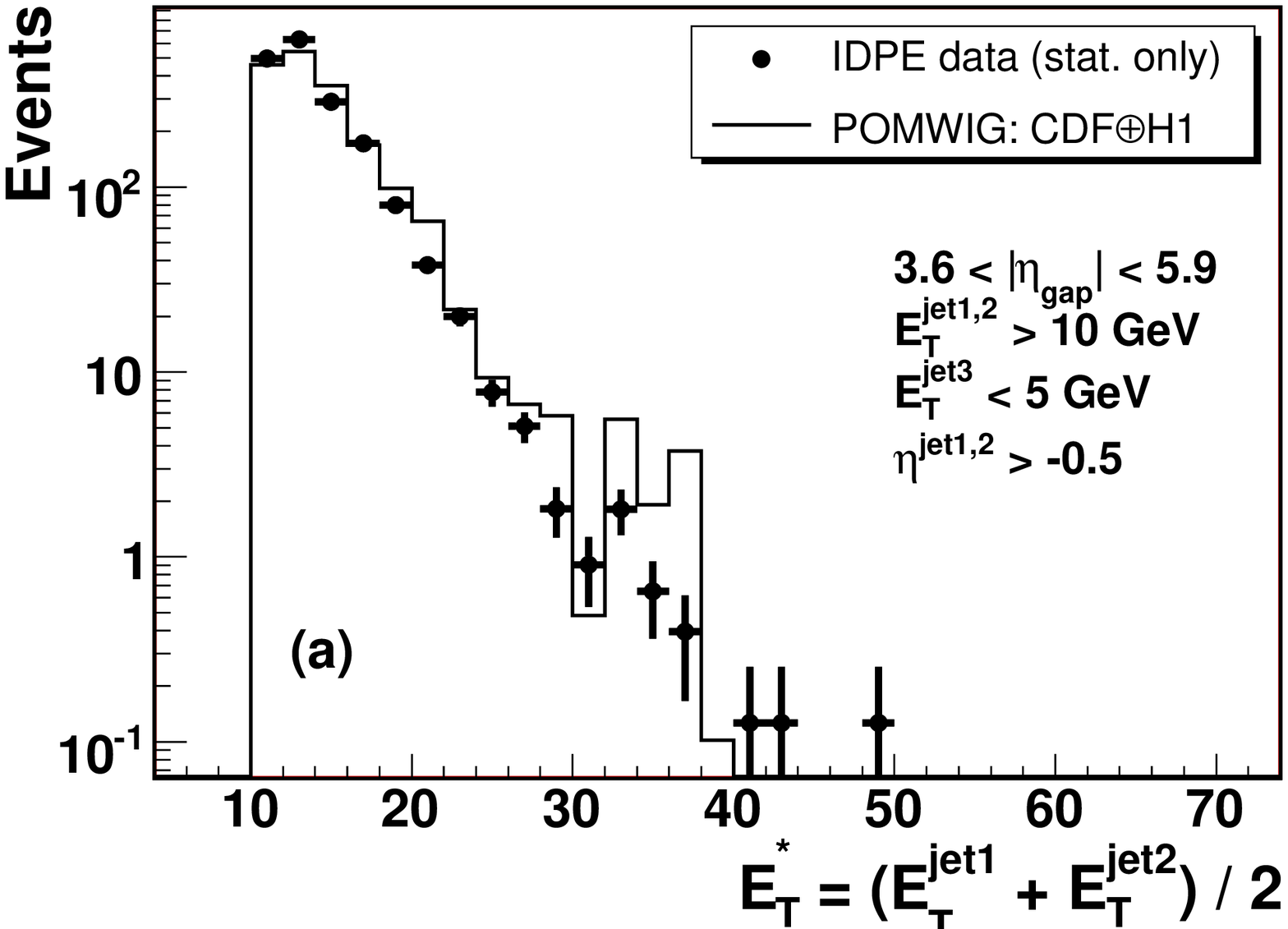}
   \includegraphics[width=4.2cm]{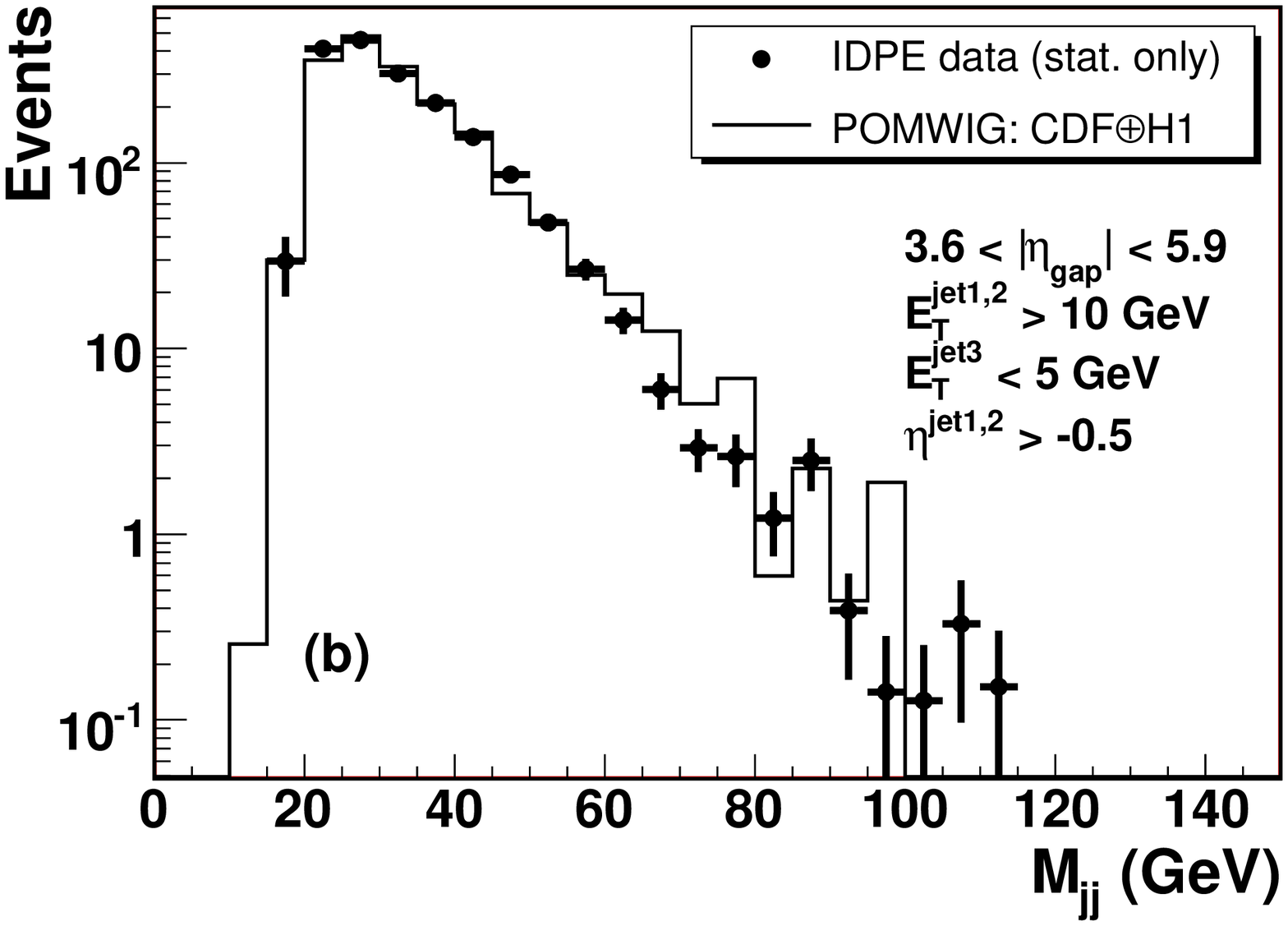} \\
   \includegraphics[width=4.2cm]{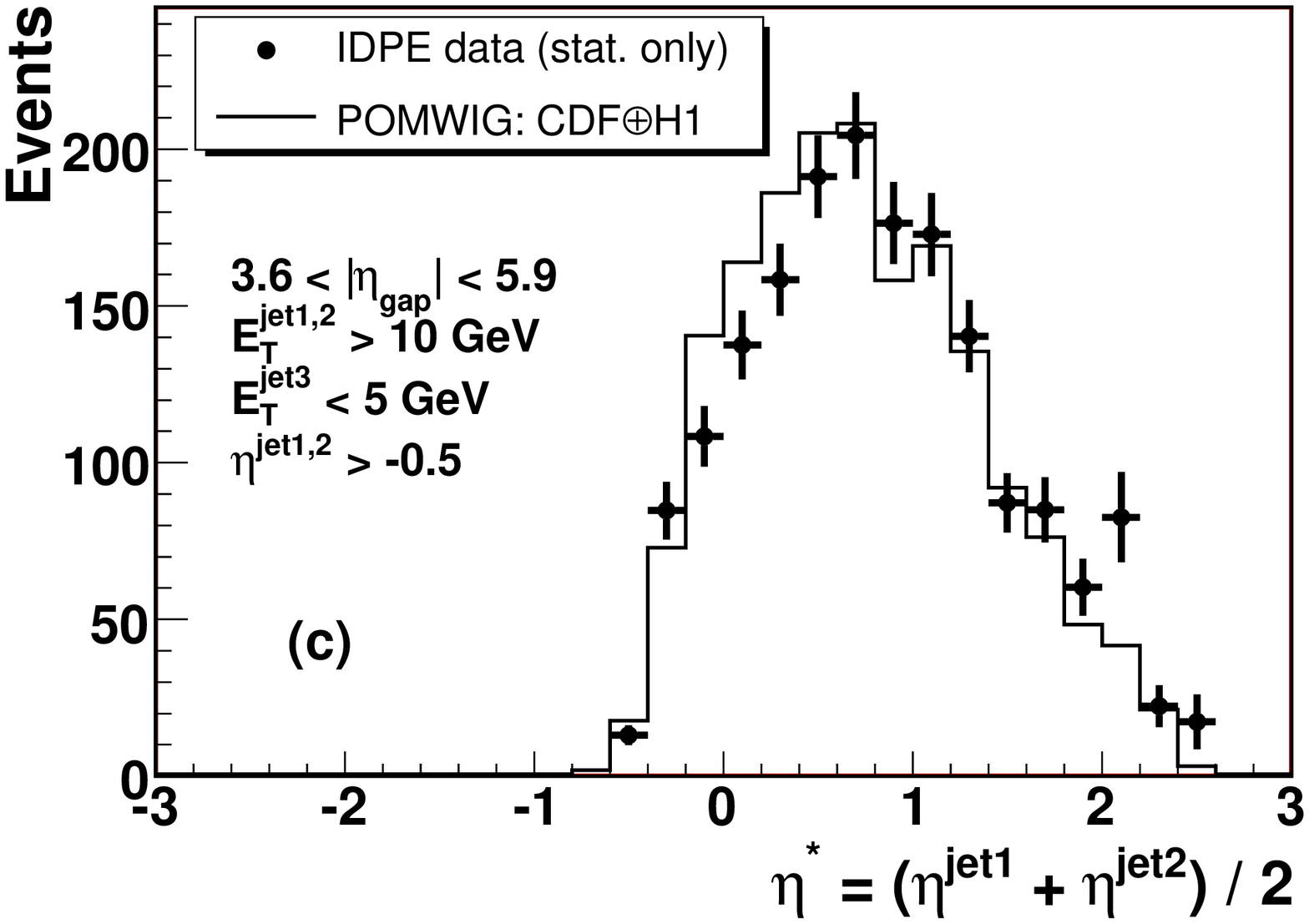}
   \includegraphics[width=4.2cm]{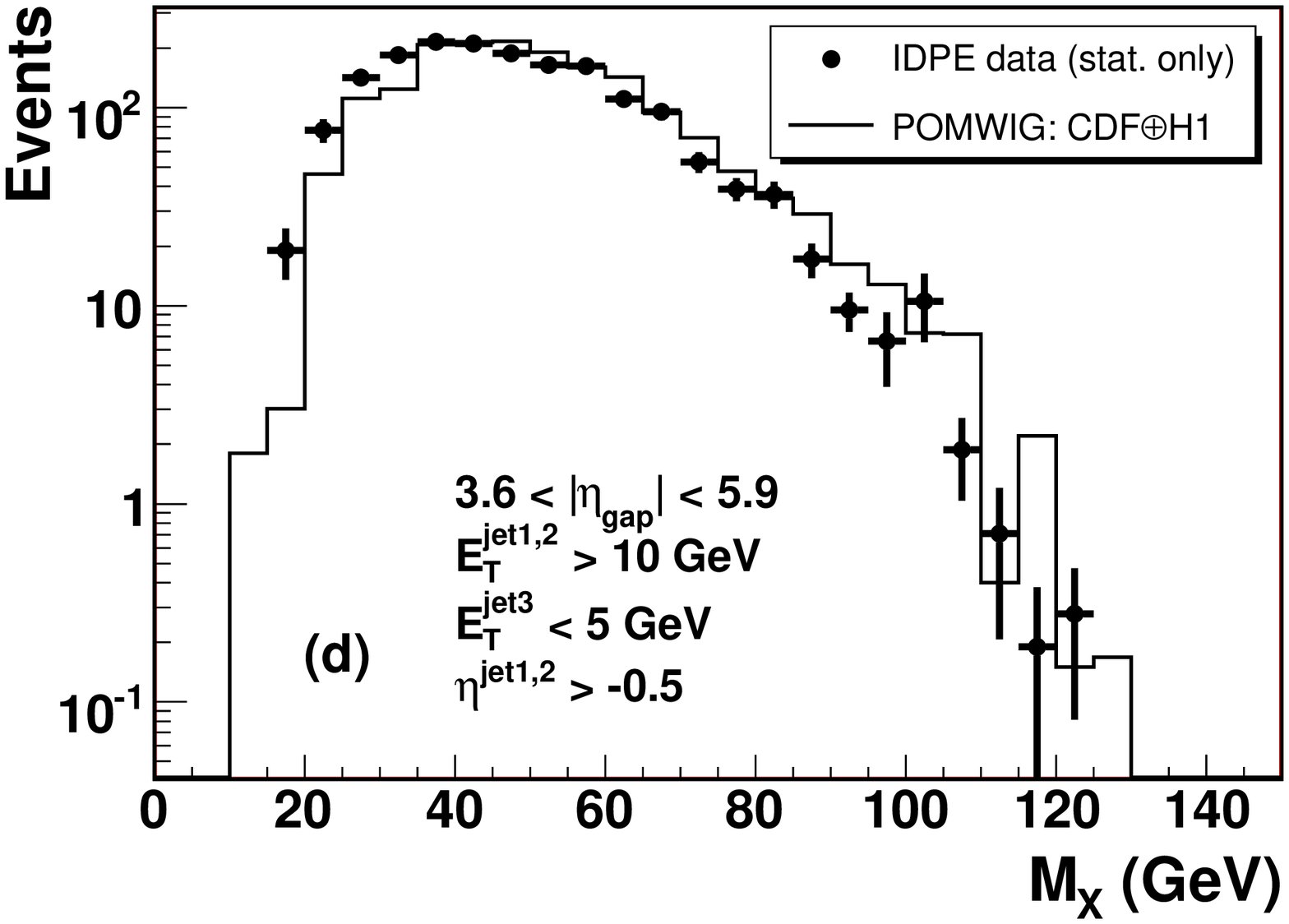}
   \caption{Comparison of IDPE data distributions of (a) $E_T^*$,  (b) $\eta^*$, (c) $M_{jj}$, and (d) $M_X$ for the background-rich region B with corresponding {\sc pomwig} distributions obtained using the CDF$\oplus$H1 DSF; the events plotted are those used in plots (a) and (b) of Fig.~\ref{fig:jet_eta_cuts}.   
\label{fig:regionB}}
 \end{center}
\end{figure}

Figure~\ref{fig:jet_eta_cuts} shows $\eta^{jet1}$ 
vs. $\eta^{jet2}$ distributions for DPE data satisfying all the above selection cuts, POMWIG dijet events generated with the CDF$\oplus$H1 DSF, and exclusive dijet events generated with two different exclusive dijet MC simulations, {\sc ExHuME} and {\sc ExclDPE}, which are described below in Sec.~\ref{subsec:exc_mc}. 
The {\sc pomwig} generated events, which do not contain an explicit exclusive contribution, and the data, which are dominated by non-exclusive events, are scattered symmetrically around $\eta^{jet1}=\eta^{jet2}=0$, as the requirements of $LRG_p$ and $LRG_{\bar p}$ accept the same range of momentum loss fractions $\xi_{\bar p}$ and $\xi_p$. Since, however, the recoil proton is not detected, the RPST trigger introduces a bias in the exclusive production case,  resulting in an asymmetric $\Pomeron_{\bar p}$-$\Pomeron_p$ collision with $\xi_{\bar p}>\xi_p$, boosting the dijet system toward negative $\eta$. 
We exploit this kinematic effect by splitting the data and the events generated by each MC simulation into two samples, A and B, defined in $\eta^{jet1}$-$\eta^{jet2}$ space as shown in Fig.~\ref{fig:jet_eta_cuts}.  The data samples A, for which at least one of the two leading jets has $\eta^{jet}<-0.5$, contains most of the exclusive signal events, while sample B, comprising all other events, has a  much reduced exclusive contribution. In Fig.~\ref{fig:regionB}, we compare IDPE data distributions of $E_T^*$, $\eta^*$, $M_{jj}$, and $M_X$ for the background-rich region B with the corresponding {\sc pomwig} distributions obtained using the CDF$\oplus$H1 DSF. Reasonable agreement between data and simulation is observed.

\subsection{Exclusive dijet Monte Carlo simulations}\label{subsec:exc_mc}
In the current analysis, we use two Monte Carlo event programs for generating exclusive dijet events: {\sc ExHuME} 1.3.1~\cite{ExHuME} and {\sc dpemc}~2.5~\cite{DPEMC}. {\sc ExHuME} is a LO matrix element event generator founded on the perturbative calculations presented in Ref.~\cite{KMRmethod}, while 
exclusive dijet production in DPEMC~2.5 ({\sc ExclDPE}) is based on the DPE non-perturbative Regge theory inspired model of Ref.~\cite{BL}. In  {\sc ExHuME}, we generate exclusive events using the MRST2002 next-to-leading order proton PDF~\cite{MRST}, and implement parton showering and hadronization using {\sc pythia}. In {\sc ExclDPE}, we generate exclusive dijet events with the default parameters, using {\sc herwig}~6.505~\cite{HERWIG} to simulate parton showering and hadronization. 

\begin{figure*}
 \begin{center}
 \includegraphics[width=5.8cm]{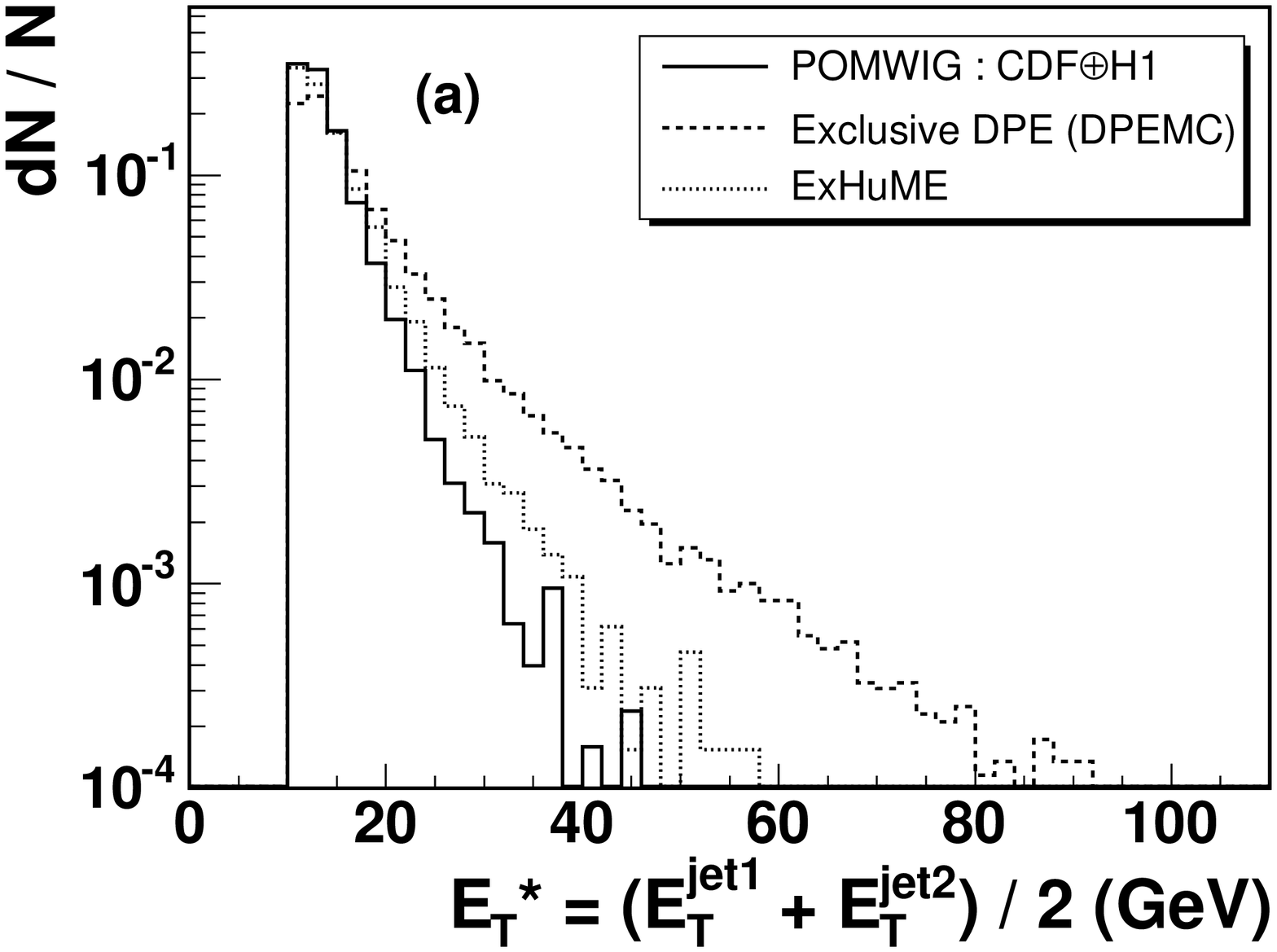}
 \includegraphics[width=5.8cm]{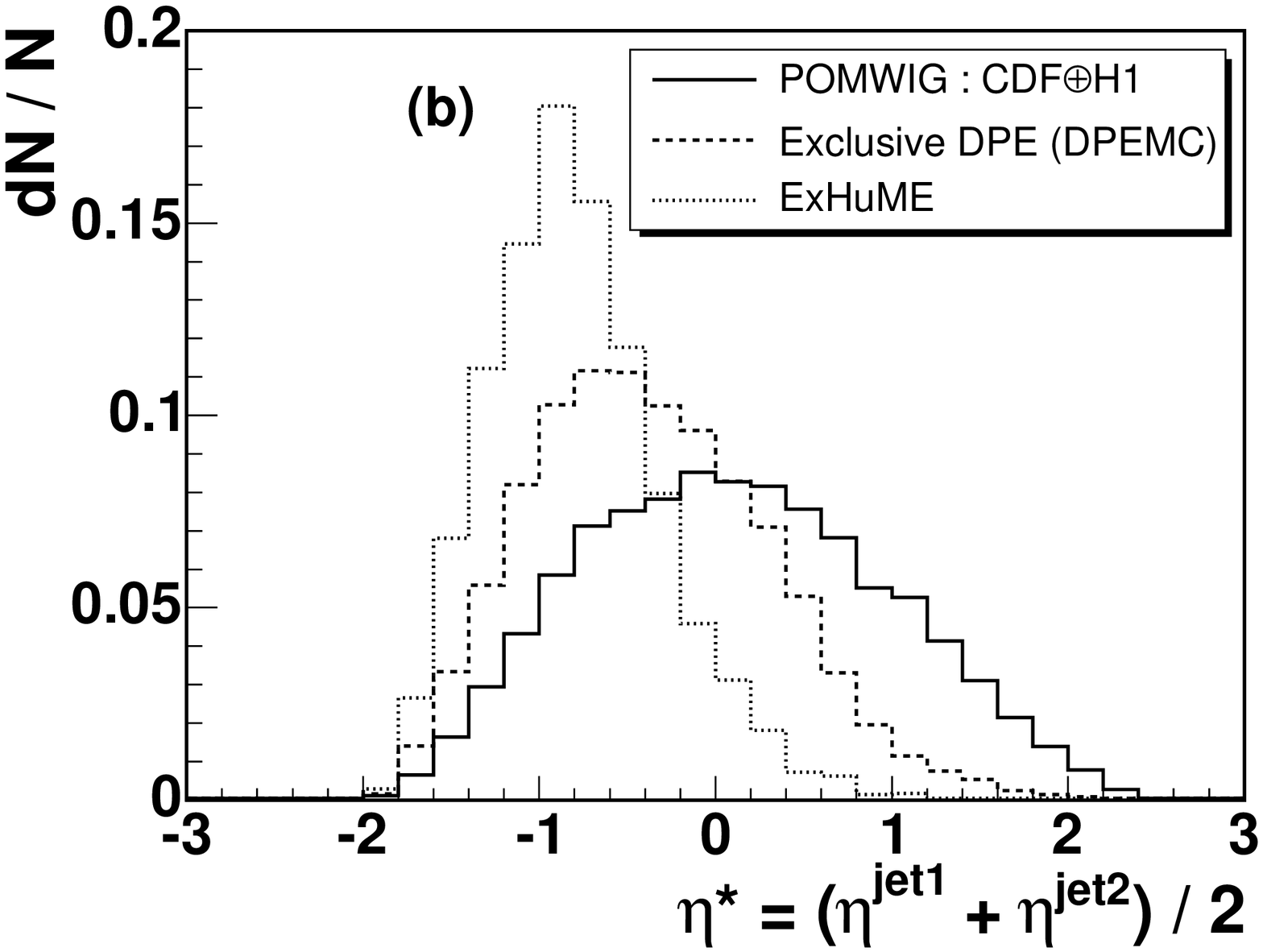}
 \includegraphics[width=5.8cm]{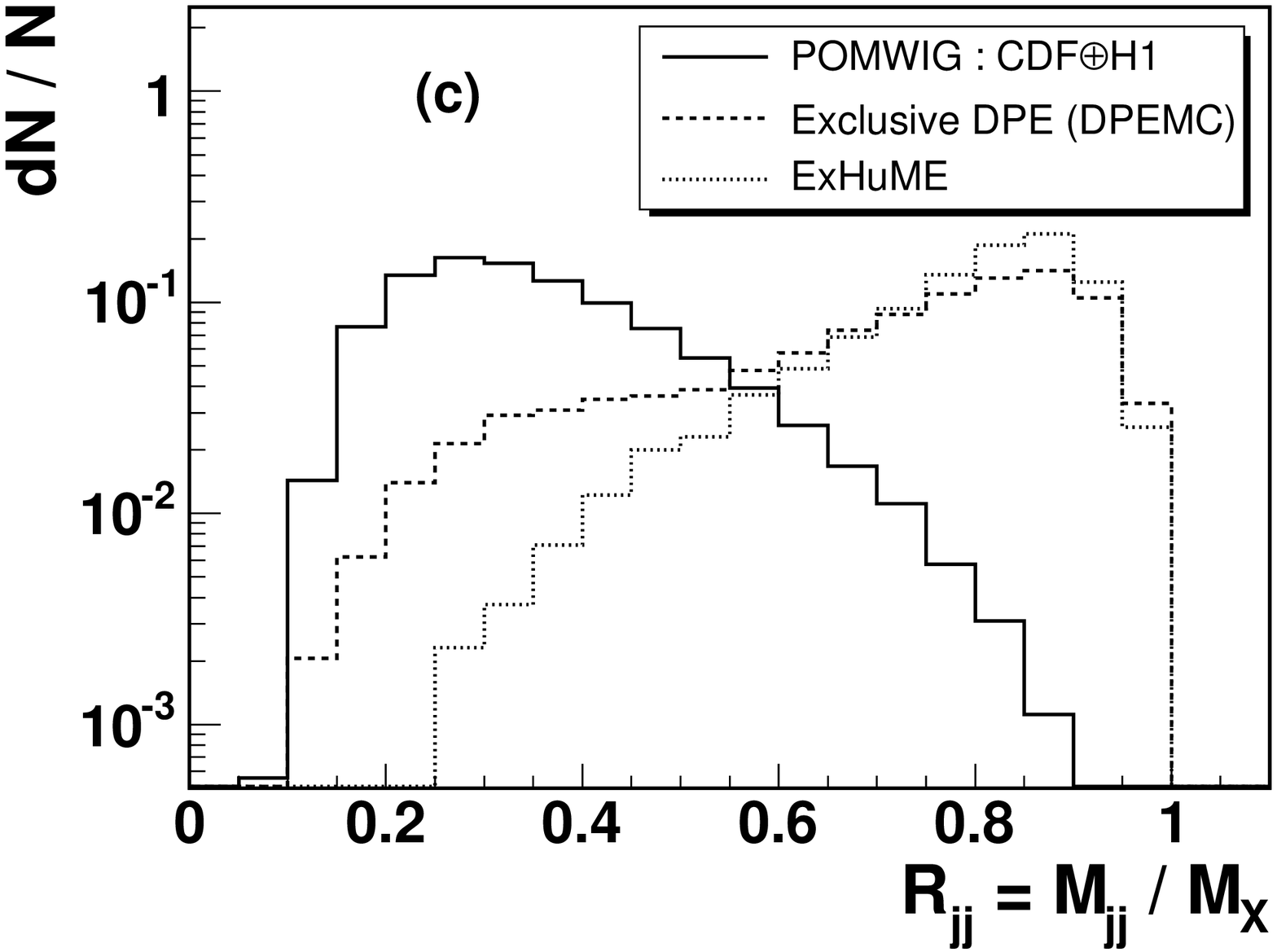}
 \caption{
(a) Mean $E_T^{jet}$ of the two leading jets, $E_T^*$ (b) mean $\eta^{jet}$,  $\eta^*$ and (c) dijet mass fraction, 
$R_{jj}$ for {\sc pomwig} 
DPE dijet events generated using the CDF$\oplus$H1 DSF diffractive/Pomeron structure function (solid histograms), 
and for exclusive dijet events generated with the {\sc ExclDPE} 
(dashed histograms) and {\sc ExHuME} MC simulations (dotted histograms). 
All distributions are normalized to unit area.
\label{fig:excl_jet}}
 \end{center}
\end{figure*}

Distribution shapes of $E_T^*$, $\eta^*$, and $R_{jj}$ for {\sc pomwig} DPE events and for events generated by {\sc ExHuME} and {\sc ExclDPE} are compared in Fig.~\ref{fig:excl_jet}. 
All distributions are produced using quantities reconstructed at the hadron level for events
selected with $E_T^{jet1,2}>10$~GeV, $|\eta^{jet1,2}|<2.5$, $0.03<\xi_{\bar{p}}<0.08$, and $3.6<|\eta^{gap}|<5.9$. The $E_T^*$ spectrum is harder (much harder) in 
{\sc ExHuME} ({\sc ExclDPE}) than in {\sc pomwig}, while the dijet system in both {\sc ExHuME} and {\sc ExclDPE} is boosted toward negative $\eta^*$ owing to the selected $\xi_{\bar p}$ range, as explained above in Sec.~\ref{subsec:search}. 
In the $R_{jj}$ distributions, the exclusive jets emerge around $R_{jj}\sim0.8$, while {\sc pomwig} events populate the low $R_{jj}$ region. Both {\sc ExHuME} and {\sc ExclDPE} MC distributions exhibit a long tail extending toward small $R_{jj}$ values due to gluon radiation. For comparison with data, the MC generated events are processed through a detector simulation.

\subsection{Comparison of data with combinations of inclusive and exclusive simulated events}\label{subsec:inc+exc}
\begin{figure*}
 \begin{center}
 \includegraphics[width=8.5cm]{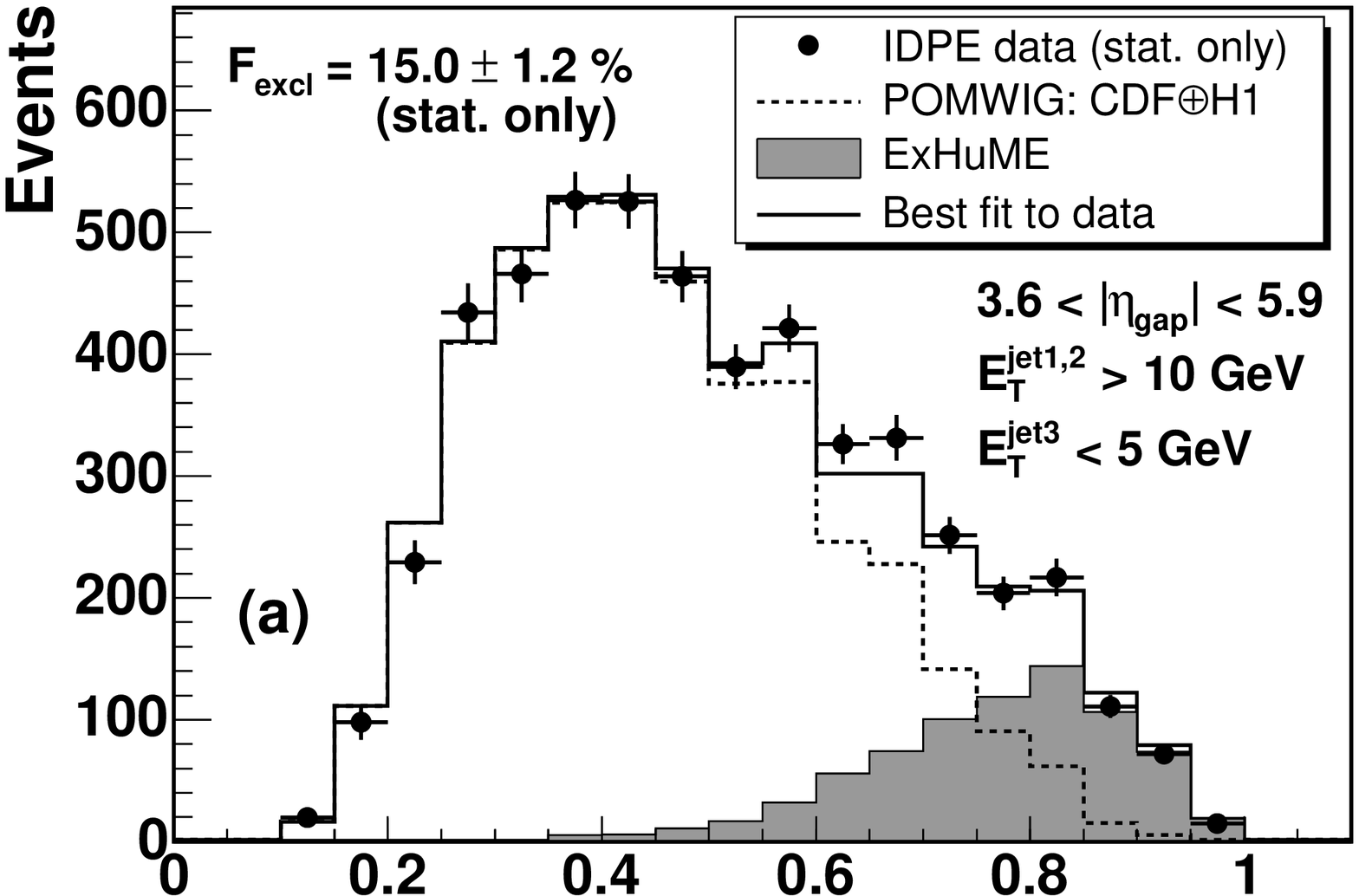}
 \includegraphics[width=8.5cm]{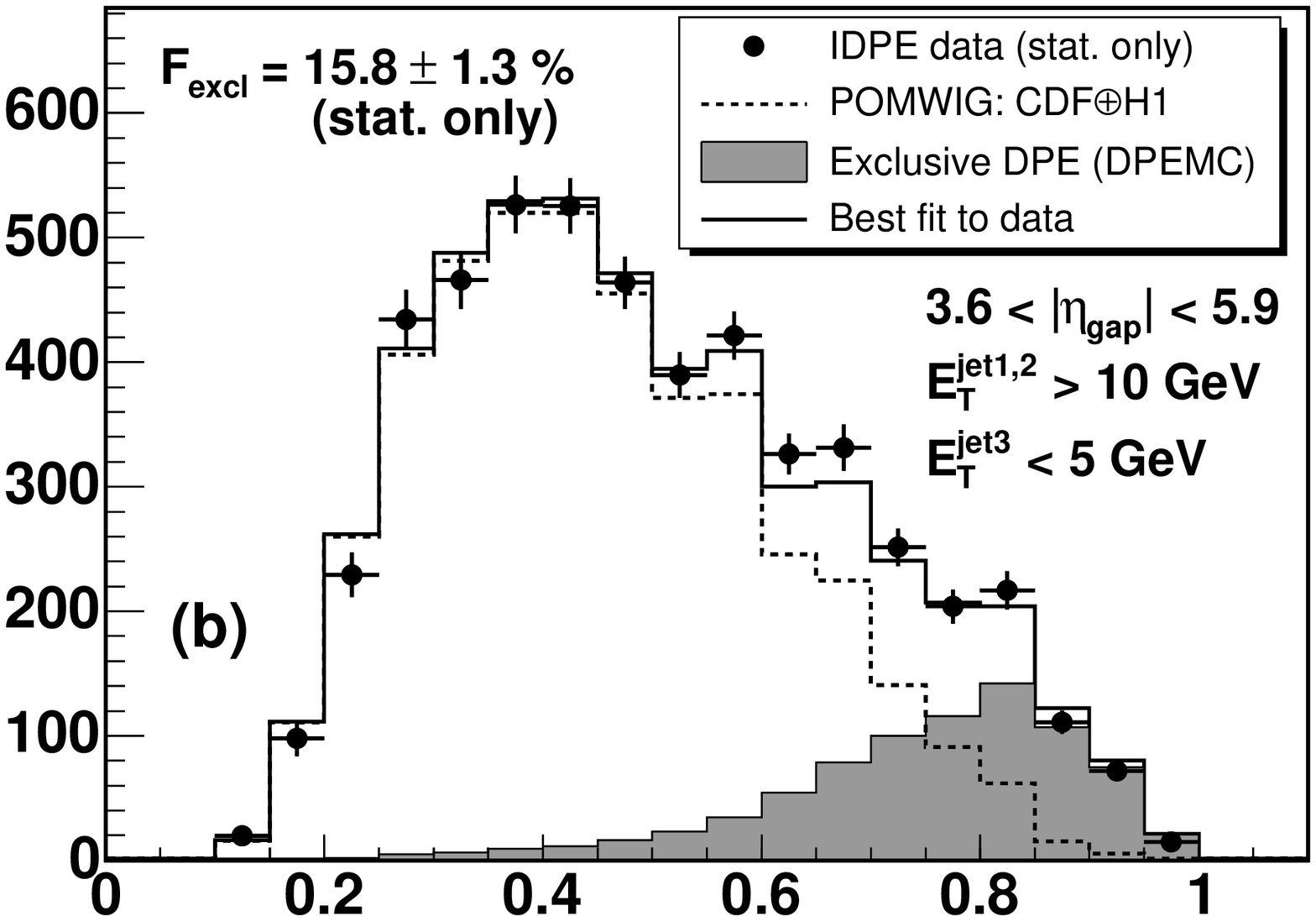}\\
 \includegraphics[width=8.5cm]{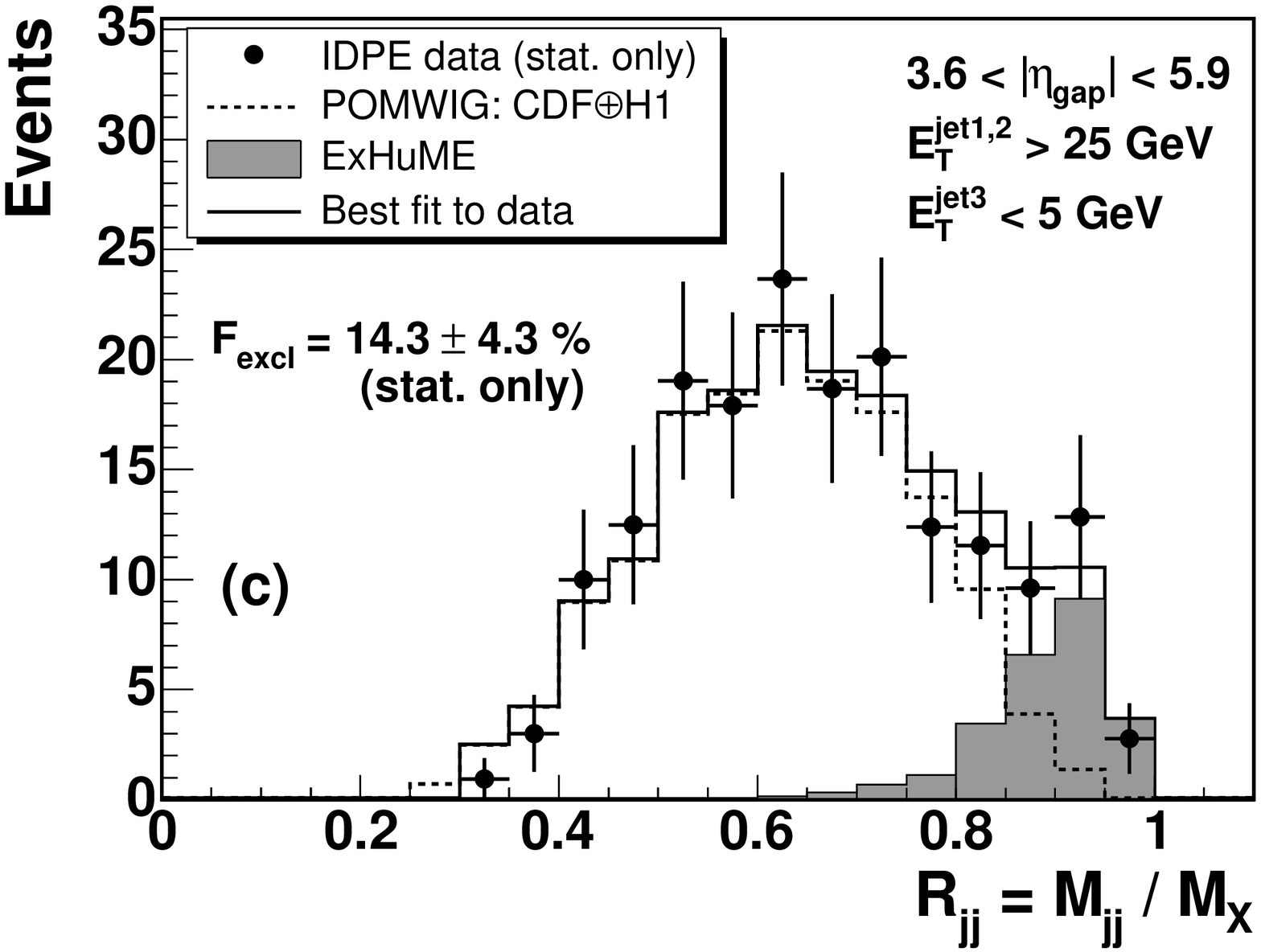}
 \includegraphics[width=8.5cm]{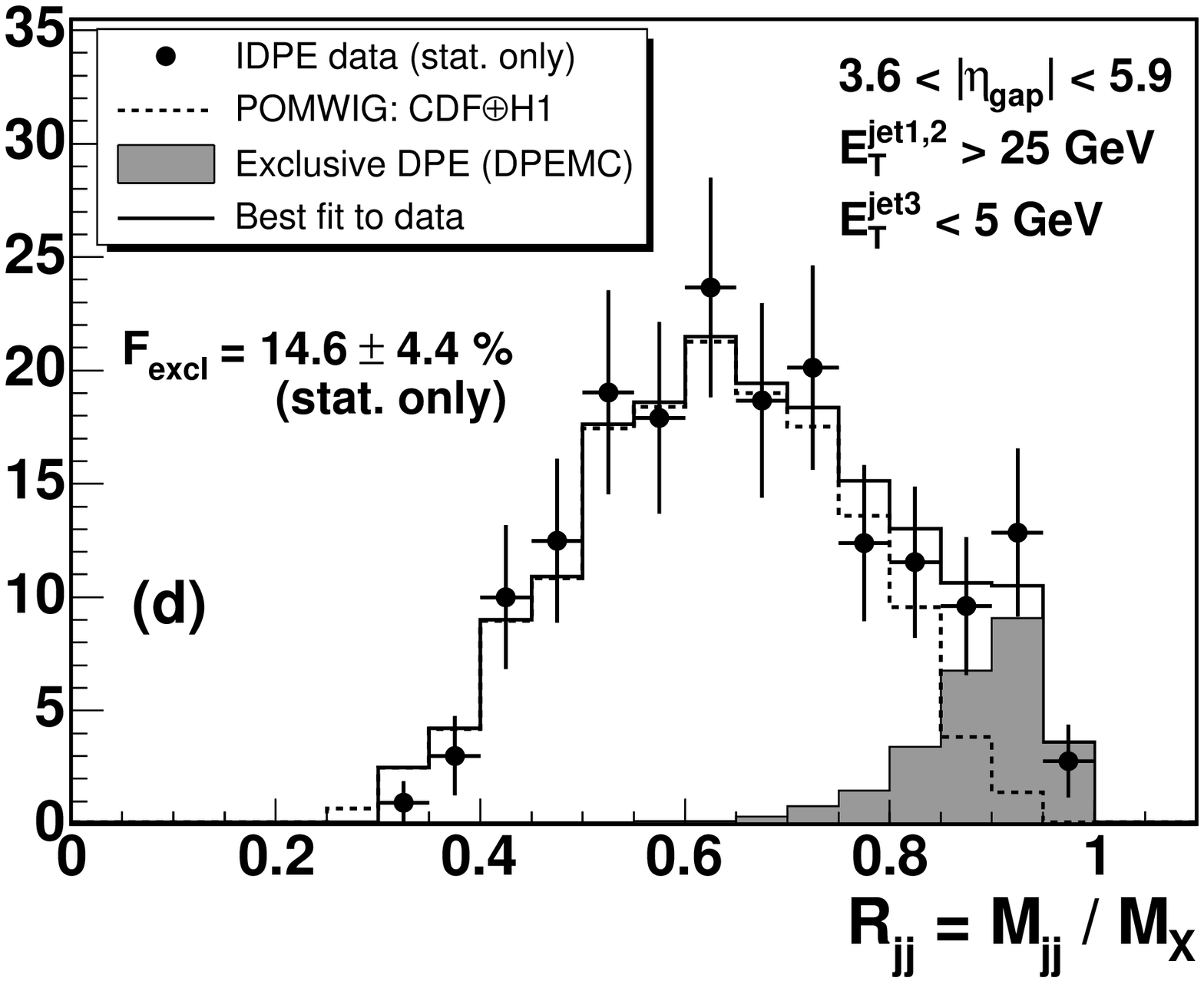}
   \caption{Dijet mass fraction in IDPE data (points)  and best fit (solid histograms) with a mixture of (i)
{\sc pomwig} generated events composed of {\sc pomwig} DPE signal and SD plus ND background events (dashed histogram),  
and (ii) exclusive dijet MC events (shaded histogram). The data and the MC events are selected from the respective IDPE samples after applying the additional veto cuts of $LRG_{\bar p}$ and $E_T^{jet3}<5$~GeV.
Plots (a) and (c) [(b) and (d)] present fits using {\sc ExHuME} [{\sc ExclDPE}] generated exclusive dijet events, while a requirement of  
$E_T^{jet2}>10$~GeV [25~GeV] is applied to plots (a) and (b) [(c) and (d)]. 
\label{fig:rjj_cut1_fit}}
 \end{center}
\end{figure*}

We first fit the $R_{jj}$ distribution of inclusive DPE dijet events satisfying the additional cuts (a) and (b) of Sec.~\ref{subsec:search}, but not requiring the $\eta^{jet}$-cut. Results are shown in Fig.~\ref{fig:rjj_cut1_fit}. In plots (a) and (b) the two highest $E_T$ jets in an event are required to have $E^{jet1,2}_T>10$~GeV and in plots (c) and (d) $>25$~GeV. The solid histogram in each plot is obtained from a binned maximum likelihood fit of the data with 
a combination of (i) {\sc pomwig} DPE plus SD and ND background events (dashed histograms) and (ii) exclusive signal events (shaded histograms) generated by {\sc ExHuME} for plots (a) and (c) or {\sc ExclDPE} for plots (b) and (d) satisfying the same cuts as the data. In each fit, the normalizations of the inclusive POMWIG and of the exclusive MC events are introduced as free parameters.
The $R_{jj}$ data distribution is well reproduced within statistical uncertainties with both {\sc ExHuME} and {\sc ExclDPE} based exclusive contributions, yielding exclusive fractions of $F_{excl}=15.0\pm1.2$~(stat.)\% and $F_{excl}=15.8\pm1.3$~(stat.)\%, respectively, for $E^{jet1,2}_T>10$~GeV. 
 
\begin{figure}
 \begin{center}
 \includegraphics[width=8.5cm]{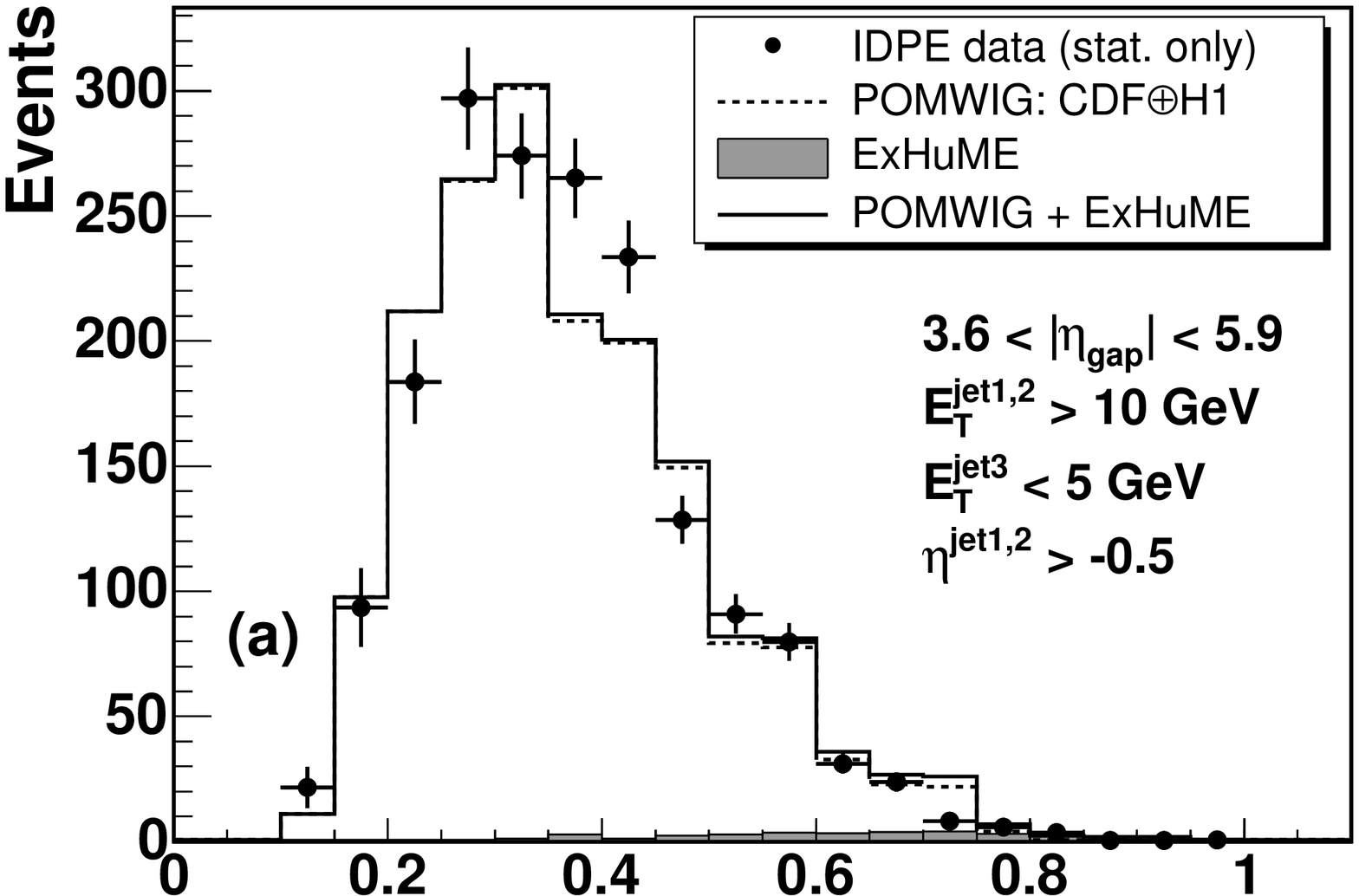}\\
 \includegraphics[width=8.5cm]{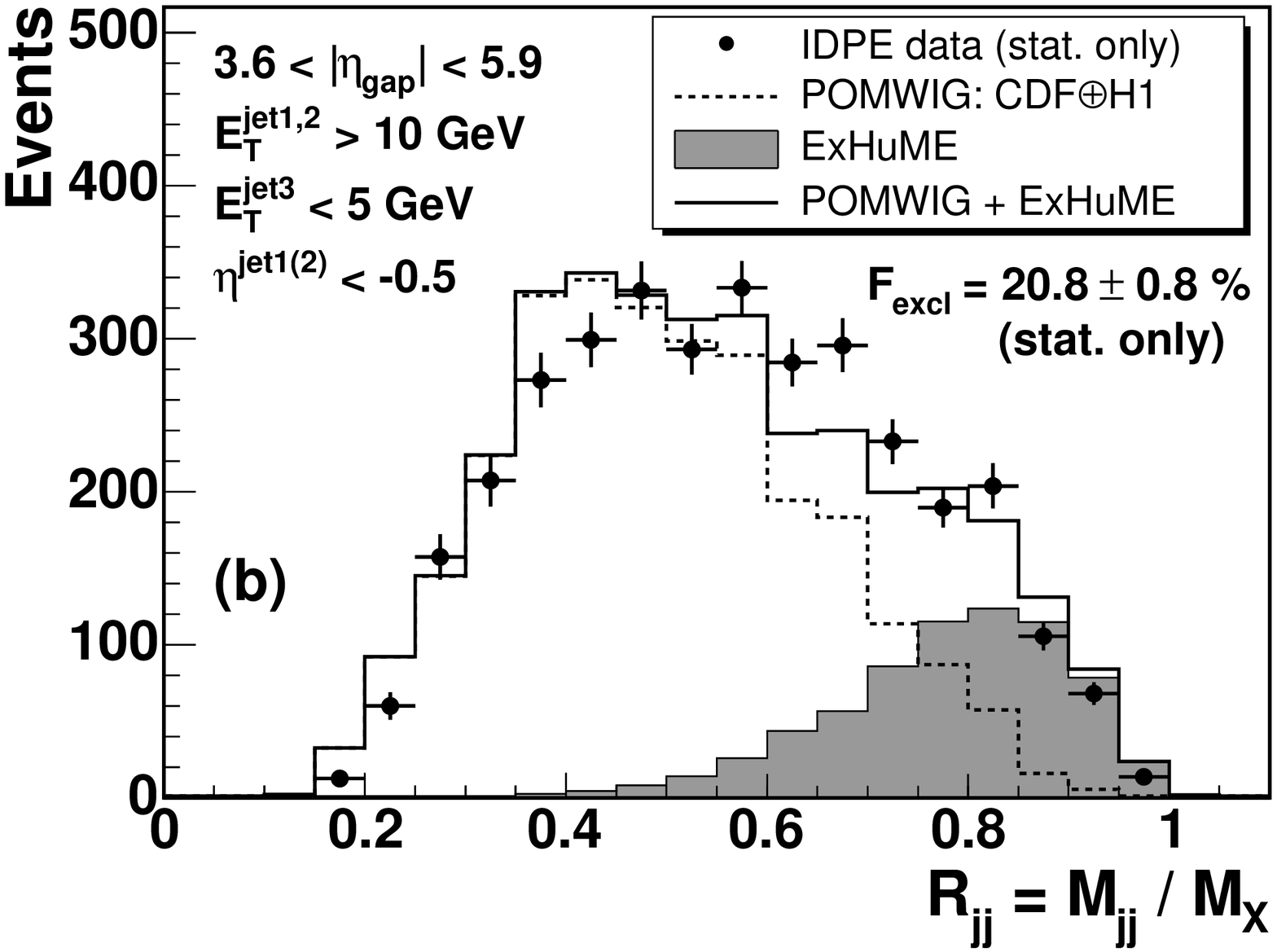}
 \caption{Dijet mass fraction for IDPE data (points) and for {\sc pomwig} generated events (dashed histogram) 
composed of {\sc pomwig} DPE plus SD and ND background events, and for {\sc ExHuME} generated exclusive 
dijet events (shaded histograms). The solid histogram is the sum of {\sc pomwig}$\oplus${\sc ExHuME} events. Plot (a)  shows distributions for event sample B, and plot (b) 
for event sample A. The events plotted pass all other selection cuts. The MC events are normalized using the results of the fits shown in 
Fig.~\ref{fig:rjj_cut1_fit} (a), scaled according to the actual number of events that pass the $\eta^{\rm jet}$-cut requirement. 
\label{fig:rjj_cut_nofit}}
 \end{center}
\end{figure}

As a control check, we add the requirement of the $\eta^{jet}$-cut and obtain distributions for data,  {\sc pomwig}, {\sc ExHuME}, and {\sc pomwig}+{\sc ExHuME} events separately for samples B and A, shown in Figs.~\ref{fig:rjj_cut_nofit}~(a) and (b), respectively. The relative normalizations of the total DPE data and MC 
 event samples are fixed to those obtained in the fits of Fig.~\ref{fig:rjj_cut1_fit} (a), and the $R_{jj}$ distributions are scaled by the number of events that pass the $\eta^{jet}$-cuts that define the data samples. As expected, no significant exclusive contribution is observed in the data sample B. In sample A, good agreement is observed between the data and the {\sc pomwig}+{\sc ExHuME} combination, indicating that the observed excess at high $R_{jj}$, whose fraction in the data has increased from $15.0\pm1.2$~(stat)\% to $20.8\pm0.8$~(stat)\% by the sample A selection cuts, is consistent with an exclusive signal in both shape and relative normalization. Similar agreement is observed using the {\sc ExclDPE} simulation.

\begin{figure}
 \begin{center}
   \includegraphics[width=8.5cm,clip=]{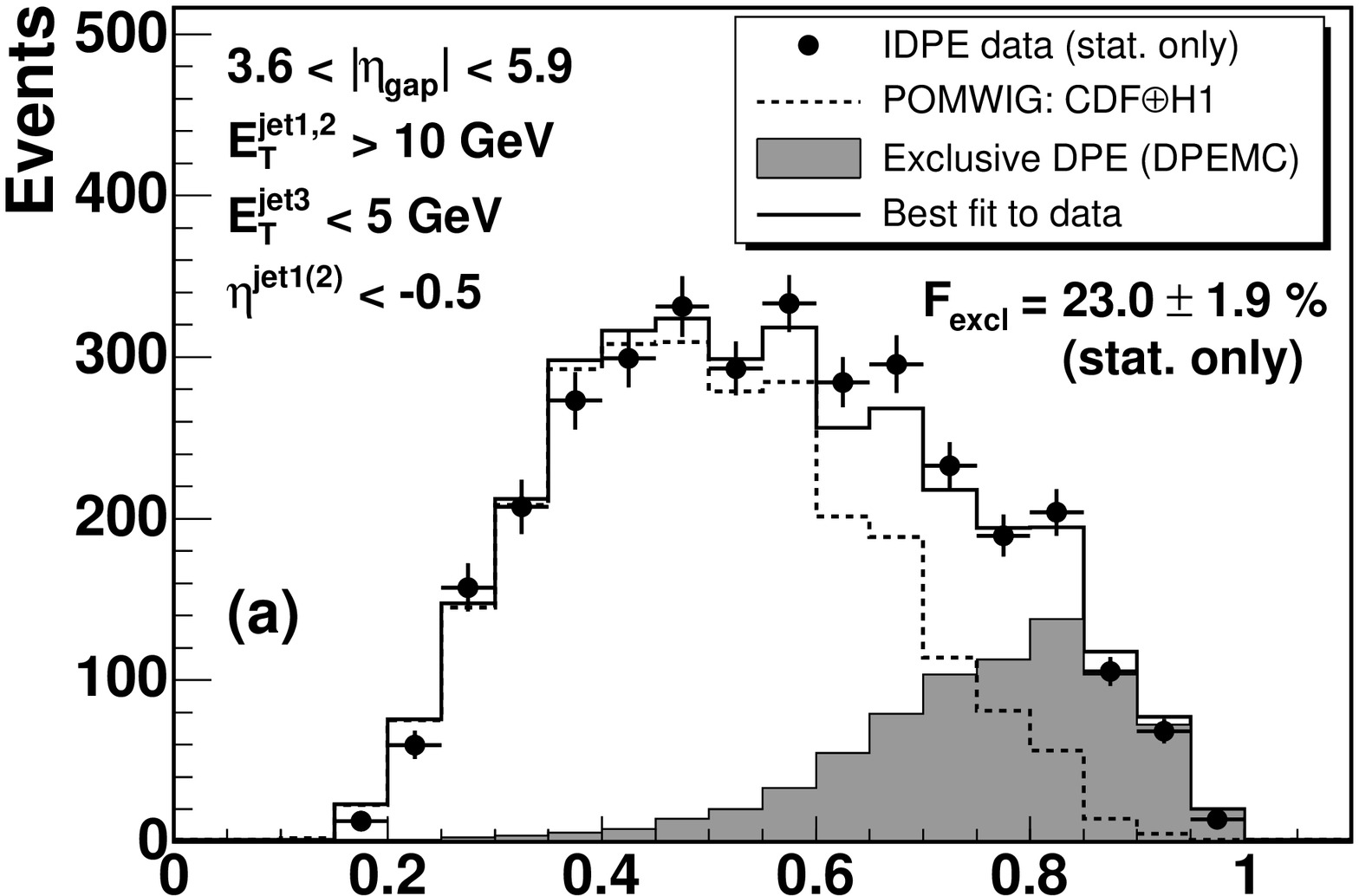}\\
   \includegraphics[width=8.5cm,clip=]{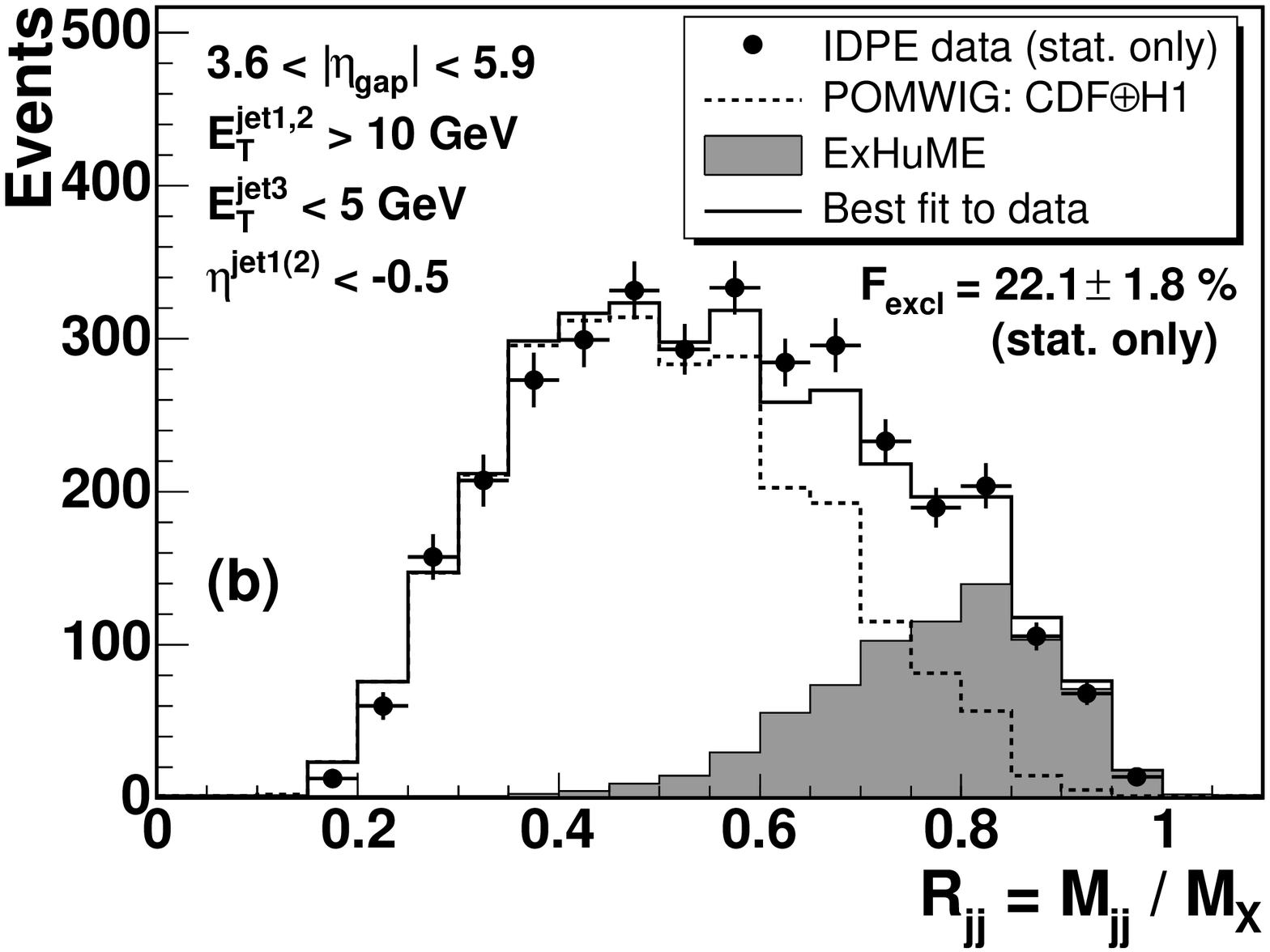}
   \caption{Dijet mass fraction for IDPE data (points) and best fit (solid histogram) to the data obtained from a combination of
{\sc pomwig} events (dashed histogram) composed of {\sc pomwig} DPE plus SD and ND background events, 
and exclusive dijet MC events (shaded histogram) generated using (a) {\sc ExclDPE} or (b) {\sc ExHuME}. The data and the MC events are from sample A and are required to pass all other selection cuts.
\label{fig:rjj_cut3_fit}}
 \end{center}
\end{figure}

The fit of the data with MC simulated events would be expected to improve if the normalizations were left free to be determined by the fit. Binned maximum likelihood fits to the data of sample A are shown in Fig.~\ref{fig:rjj_cut3_fit}. 
The fraction of exclusive dijet signal, $F_{excl}$, is found to be $23.0\pm1.9$~(stat)\% for {\sc ExclDPE} and 
$22.1\pm1.8$~(stat)\% for {\sc ExHuME}. 

\begin{figure}
 \begin{center}
   \includegraphics[width=8.5cm,clip=]{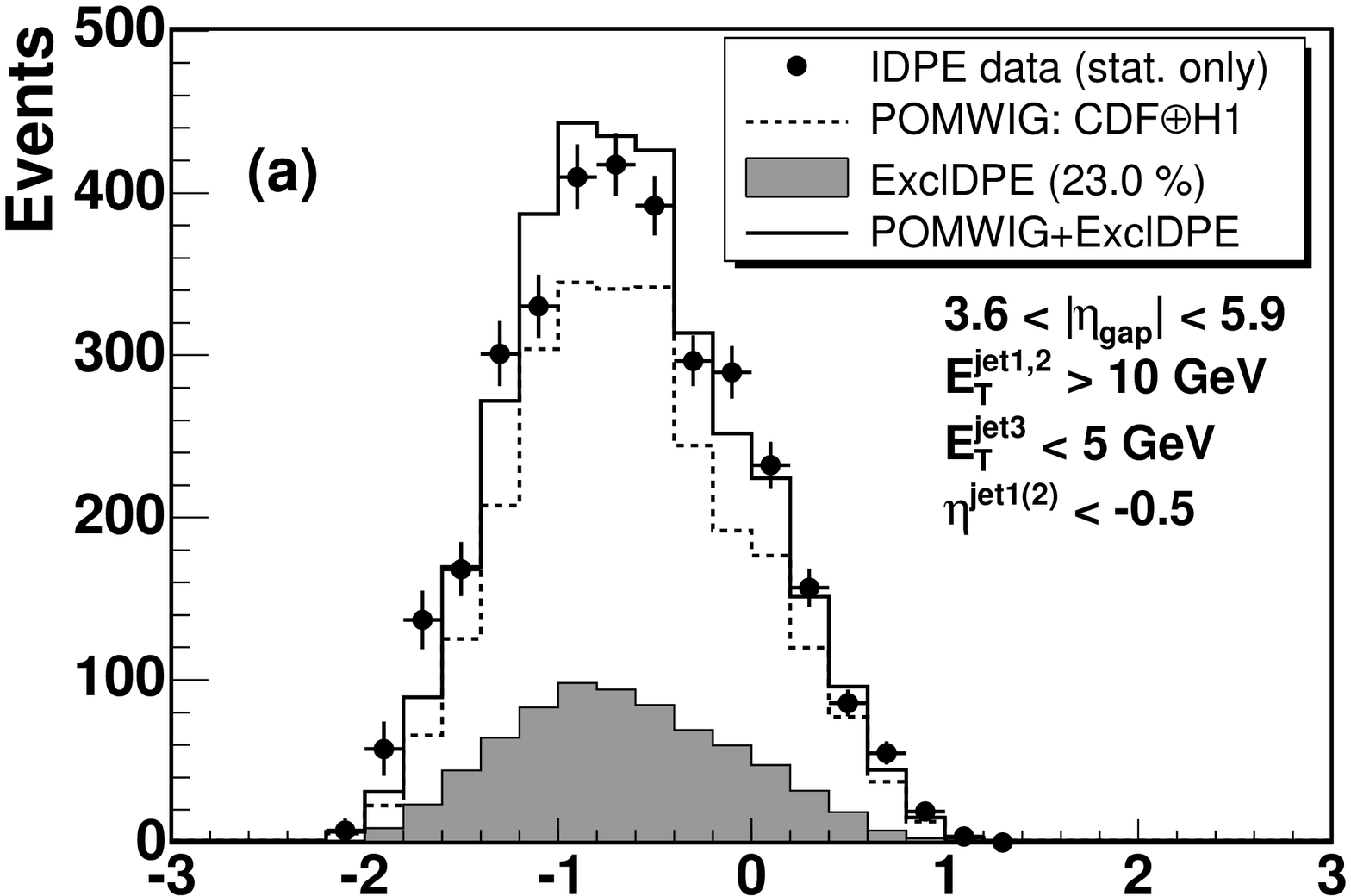}\\
   \includegraphics[width=8.5cm,clip=]{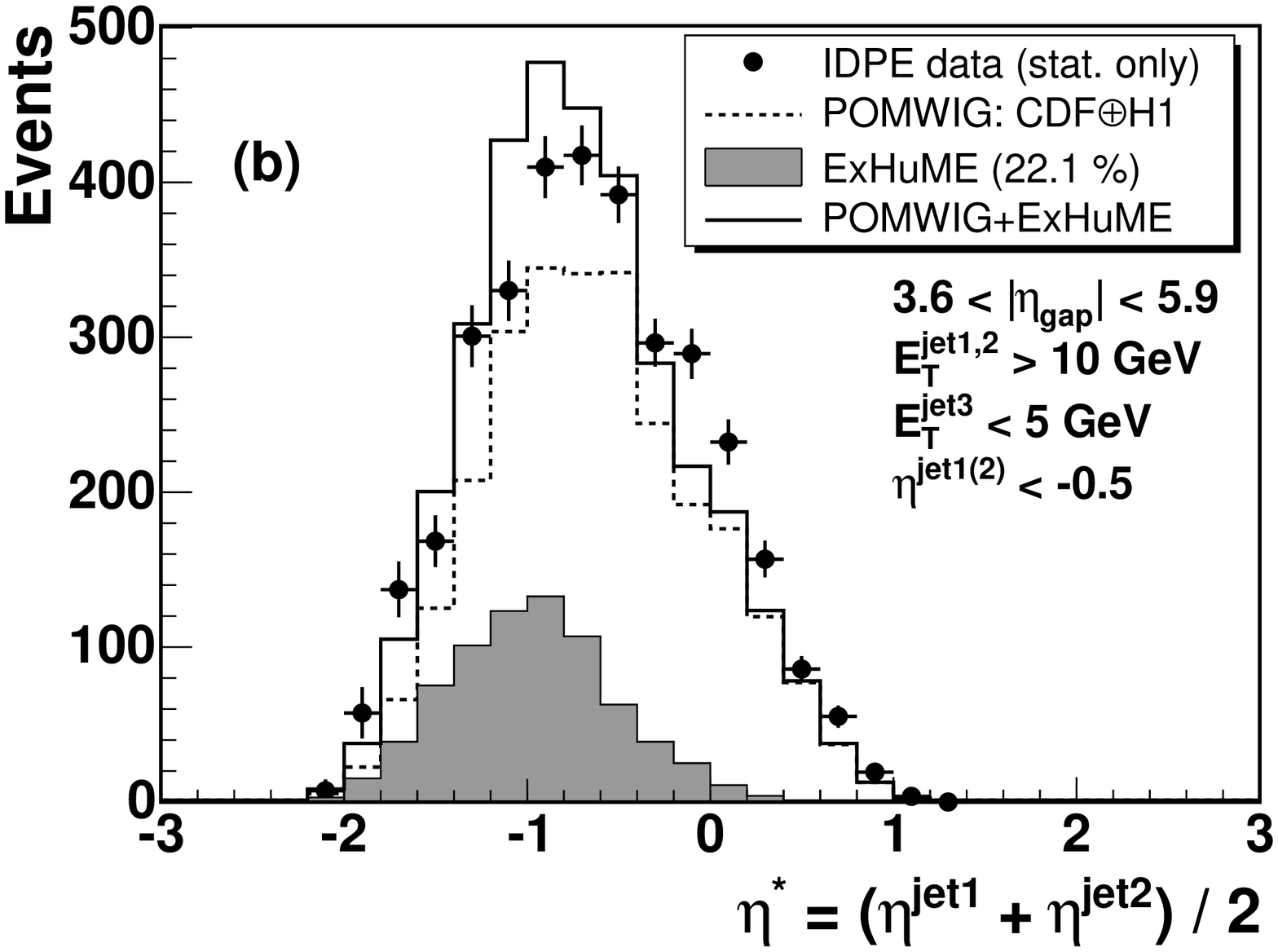}
   \caption{Comparison of mean pseudorapidity distributions $\eta^*$   between the IDPE data of sample A and the mixture of {\sc pomwig} and (a) {\sc ExclDPE} or (b) {\sc ExHuME} simulated events normalized by the fit presented in Fig.~\ref{fig:rjj_cut3_fit}. 
\label{fig:etastar}}
 \end{center}
\end{figure}

In Fig.~\ref{fig:etastar}, we compare the $\eta^*$ distribution of the data of sample A with a MC generated distribution using {\sc pomwig} with the CDF$\oplus$H1 diffractive structure function and an  admixture of an exclusive signal of (a) 23\% {\sc ExclDPE} or (b) 22\% {\sc ExHuME} generated events, where the normalization was fixed to that obtained from the fits in Fig.~\ref{fig:rjj_cut3_fit}. Considering that the normalization was not allowed to very in performing the fit, reasonable agreement between data and simulation is observed, confirming  our previous assertion that the broader data than {\sc pomwig} simulated distribution seen in Fig.~\ref{fig:pomwig_et_eta} is due to the exclusive contribution in the region around $\eta^*\sim -1$.  

To determine the sensitivity to the DSFs used in the simulations, we have extracted the fraction $F_{excl}$ using eight different combinations of DSFs, made up from each of the four DSFs used in {\sc pomwig} (CDF$\oplus$H1, CDF, H1-fit2 and ZEUS-LPS) with the DSF used in {\sc ExHuME} or in {\sc ExclDPE}. The eight $F_{excl}$ values obtained, listed in Table~\ref{tab:fexcl}, are mutually consistent within the quoted statistical uncertainties.

\begin{table}
 \caption{Fraction of exclusive dijet events in DPE dijet data, extracted from likelihood fits to data selected with the requirements of $E_T^{jet1,2}>10$~GeV, $E_T^{jet3}<5$~GeV,  $\eta^{jet1,2}>-2.5$, 
$\eta^{jet1(2)}<-0.5$, $0.03<\xi_{\bar{p}}<0.08$ and $3.6<|\eta^{gap}|<5.9$, using combinations of 
{\sc pomwig}+{\sc ExHuME} or {\sc pomwig}+{\sc ExclDPE} distribution shapes. 
Results are listed for four different DSFs used in {\sc pomwig}. 
The $\eta^{jet1(2)}<-0.5$ cut requires that at least one of the two highest $E_T$ jets be within  $\eta_{jet}<-0.5$.
Uncertainties are statistical only.\label{tab:fexcl}}
 \begin{ruledtabular}
 \begin{center}
  \begin{tabular}{lcc}
     DSF            & {\sc ExclDPE}     & {\sc ExHuME} \\\hline
     CDF$\oplus$H1  &  $23.0\pm1.9$~\%  &  $22.1\pm1.8$~\% \\
     CDF            &  $22.6\pm1.9$~\%  &  $21.7\pm1.8$~\% \\
     H1-fit2        &  $26.0\pm2.1$~\%  &  $24.7\pm2.0$~\% \\
     ZEUS-LPS       &  $25.4\pm2.1$~\%  &  $24.3\pm2.0$~\% \\
  \end{tabular}
 \end{center}
 \end{ruledtabular}
\end{table}

\section{Results}\label{sec:xsec}
In this section we present results for both inclusive DPE dijet cross sections, $\sigma_{DPE}^{incl}$, and for exclusive production, $\sigma_{jj}^{excl}$.  The inclusive cross sections are evaluated from the IDPE dijet data sample defined by the selection cuts listed in Eq.~(\ref{IDPEsample}). Although the exclusive events are expected to be concentrated in the region of $0.03<\xi_{\bar{p}}<0.08$, as determined from simulations and from a sub-sample of the data with recorded  RPS tracking information, the larger $\xi_{\bar{p}}^X$ range of $0.01<\xi_{\bar{p}}^X<0.12$ is used to ensure that there are no inefficiencies caused by resolution and/or possible systematic effects associated with the calorimeter based definition of $\xi_{\bar{p}}$~\cite{KG_d2006}. The results are corrected for backgrounds falling within this larger $\xi_{\bar{p}}^X$ region.  

Exclusive cross sections are obtained by scaling $\sigma_{DPE}^{incl}$ 
by the fraction of IDPE data that pass veto cuts (a), (b) and (c) of Sec.~\ref{subsec:search} and multiplying the result by $F_{excl} \cdot A_{excl}^{-1}$, 
where  $F_{excl}$ is the exclusive fraction and $A_{excl}$ the acceptance of the veto cut(s) for exclusive events.
As the veto cuts include a $LRG_{\bar p}$ requirement, we recalculate the correction for spoiled gaps due to multiple $\bar{p}p$ interactions with no vertices. We then scale $\sigma_{DPE}^{incl}$  
by the ratio of the correction for $p\oplus \bar p$  gaps to that for only a $p$-gap and apply it in evaluating $\sigma_{jj}^{excl}$. The acceptance of the exclusive cuts, $A_{excl}$, is obtained from the fraction of {\sc ExHuME} or {\sc ExclDPE} generated events passing the same cuts. 

In the following sections, we summarize the methods we used to calculate systematic uncertainties and their contributions to the total uncertainty.
\subsection{Jet $E_T$ smearing}\label{subsec:jetsmear_excl}
The large difference in the slope of the $E^{jet}_T$ distributions between inclusive {\sc pomwig} and exclusive MC generated events seen in Fig.~\ref{fig:excl_jet} results in different corrections for $E_T^{jet}$ smearing.
Corrections for {\sc ExHuME} and {\sc ExclDPE} generated events are derived using the method described in 
Sec.~\ref{subsec:jetsmear_incl}.
In obtaining the final results for $\sigma_{jj}^{excl}$ using {\sc ExHuME}, the cross sections extracted by the above procedure are multiplied by 
the ratio of the corrections obtained from the {\sc ExHuME} event sample
to the {\sc pomwig} based corrections to account for the difference in $E_T^{jet}$ spectra.
\subsection{Systematic uncertainties}\label{subsec:syst}
The systematic uncertainty in the exclusive fraction receives contributions from uncertainties in the jet energy scale, unclustered calorimeter energy determination, jet trigger efficiency, jet $E_T$ smearing, non-DPE background, RPS acceptance, luminosity determination, knowledge of the diffractive structure function, statistics of MC event samples, underlying event determination, and the modeling of the underlying event.  
\subsubsection{Jet energy scale}
The uncertainty in $E_T^{jet}$ associated with the jet energy scale (JES) is evaluated by varying the uncertainties
on the relative and absolute energy scale corrections by $\pm1\sigma$ in estimating the efficiency for triggering on a single calorimeter tower of $E_T>5$~GeV, while simultaneously monitoring the number of jets 
with $E_T^{jet}$ above the desired threshold. Due to the steeply falling $E^{jet}_T$ spectrum, 
the change in trigger efficiency increases with decreasing $E_T^{jet}$ from $^{-26}_{+34}$ \% 
for $10<E_T^{jet}<15$~GeV to $^{-11}_{+11}$ 
\% for $25<E_T^{jet}<35$~GeV, resulting in a variation of the number of IDPE dijet events accepted of $\pm21$~\% ($^{+32}_{-27}$~\%) for $E_T^{jet2}>10$~GeV ($E_T^{jet2}>25$~GeV). This is the dominant uncertainty in both the inclusive and exclusive dijet cross section measurements

\subsubsection{Unclustered calorimeter energy}
Uncertainties on the unclustered calorimeter energy scale affect the $\xi_{\bar{p}}^X$ measurement, which in turn leads to  
uncertainties not only on the number of observed events, but also potentially on $R_{jj}$ distribution 
shapes. However, the CCAL and PCAL energy scale uncertainties tend to cancel out in the $R_{jj}$ ratio.

Changing the energy scale of CCAL and PCAL by $\pm5$~\% in calculating $\xi_{\bar{p}}^X$ leads to a variation of $\pm1$~\% in the inclusive DPE dijet cross sections for $E_T^{jet1,2}>10$~GeV. 
Varying the MPCAL energy scale by $\pm30$~\%, independently of CCAL and PCAL, results in a cross section
change of $^{-7}_{+10}$~\%. This is due to more (less) events falling into the range of 
$0.01<\xi_{\bar{p}}^X<0.12$ when the MPCAL energy scale is lowered (raised). The number of data events 
at high $R_{jj}$ is less affected by these changes, since the MPCAL is less active for such events. 
After adjusting the jet energy scale, the fraction of exclusive dijet signal in the DPE data is re-evaluated by repeating the MC to data fits. The full difference between the number of inclusive events 
(exclusive signal fraction) obtained from the varied samples and that obtained from the
default sample is assigned as a systematic uncertainty on this correction. The uncertainties on the
exclusive cross sections propagated from these differences are $\pm13$~\% ($\pm21$~\%) for 
$E_T^{jet1,2}>10$ (25)~GeV.

\subsubsection{Jet trigger efficiency}
Jet trigger efficiencies have been discussed in Sec.~~\ref{subsec:eff}. The full difference of the efficiencies obtained from minimum-bias data and inclusive RPS triggered data
is taken as a systematic uncertainty and propagated to 
an uncertainty on the exclusive signal fraction.

\subsubsection{Jet $E_T$ smearing}
Corrections for inclusive DPE dijets are obtained from samples of {\sc pomwig} MC dijet events.
Statistical uncertainties on the correction factors are taken as systematic uncertainties and
propagated to uncertainties on $\sigma_{DPE}^{incl}$. The full difference between the exclusive dijet cross sections 
obtained using corrections derived from {\sc ExHuME} and {\sc ExclDPE} MC generated events is assigned as a systematic uncertainty on the exclusive cross sections. This uncertainty is $\pm4$~\% ($\pm6$~\%) for 
$E_T^{jet1,2}>10$ (25)~GeV.

\subsubsection{Non-DPE background}
The dominant non-DPE background uncertainty is associated with the SD background fraction of 
$F_{BG}^{SD}=14\pm3$~\%, which contributes an uncertainty on the cross sections of $\pm0.03/(1-0.14)=\pm3.5$~\%. The $\pm 3~\%$ uncertainty on 
$F_{BG}^{SD}$ has a negligible effect on the $R_{jj}$ shapes relative to other uncertainties.

\subsubsection{RPS acceptance}
The acceptance of the RPS trigger counters, $A_{RPS}$, could vary with beam conditions and changing counter efficiencies. During the data taking period, $A_{RPS}$ varied by at most $\pm6$~\%.
This value is assigned as a systematic uncertainty on  $A_{RPS}$ and propagated to
both inclusive and exclusive cross sections.

\subsubsection{Luminosity}
The luminosity uncertainty, which is applied to all cross sections, is 5.9~\%, with 4.4~\% due to the acceptance and operation of the 
luminosity monitor and 4.0~\% due to the uncertainty in normalization using the total $\bar{p}p$ cross section~\cite{LUM}. 

\subsubsection{Diffractive structure function}
We have examined the effect of the choice of DSF on comparisons between data and {\sc pomwig} generated event 
distributions. We find that the H1-fit2 and ZEUS-LPS DSFs yield similar 
kinematic distribution shapes, which are in reasonable agreement with the data, while the 
H1-fit3 and ZEUS-$M_X$ ones produce significantly different distributions and are clearly disfavored. Guided 
by these results, we use {\sc pomwig} 
events generated with the H1-fit2 DSF to perform the fits to the data and evaluate $F_{excl}$ and take the full difference between the exclusive cross sections obtained and those extracted using the default DSF of CDF$\oplus$H1 as a systematic uncertainty.

\subsubsection{Statistics of Monte Carlo event samples}
Statistical uncertainties in the MC generated events are taken into account in  MC to data likelihood fits performed to 
extract $F_{excl}$, so that the uncertainty in $F_{excl}$ is due to the uncertainties in both data and MC event samples.  The MC associated uncertainty is derived from the uncertainty in $F_{excl}$ by quadratically subtracting the data uncertainty, determined from the number of events in the extracted signal, and propagated to the MC contribution to the exclusive cross section uncertainty.

\subsubsection{Underlying event}
The observed excess of data over simulated events at high $R_{jj}$ in Fig.~\ref{fig:rjj_cut3_fit} could be due to an overestimate of the underlying event (UE) activity in the simulation.
We investigated this possibility by following the methodology previously developed by CDF in generic UE studies in $p\bar p$ collisions~\cite{Rfield}. The $\eta$-$\phi$ space is split into three regions with respect to the leading jet axis, the ``forward'' ($|\Delta\phi|<60^\circ$), the ``transverse'' ($60^\circ<\Delta\phi<120^\circ$ or $240^\circ<\Delta\phi<300^\circ$), and the ``away'' region ($120^\circ<\Delta\phi<240^\circ$), and the UE is evaluated in the transverse region, which is sensitive to the particles outside the jets. 

\begin{figure}
 \begin{center}
 \includegraphics[width=8.5cm]{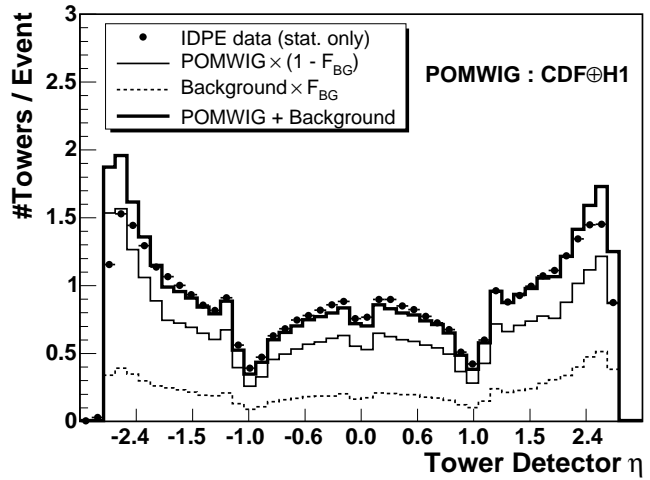}
 \caption{The number of calorimeter towers per event with $E_T>100$~MeV which are outside the jet cones versus tower detector $\eta$ for IDPE data (points) and {\sc pomwig} MC events (thick line) composed of {\sc pomwig} DPE signal (thin line) and the sum of SD and ND background events (dashed line) normalized to the background fraction per event. 
\label{fig:ntower_vs_eta_ue}}
 \end{center}
\end{figure}

Figure~\ref{fig:ntower_vs_eta_ue} shows the number of calorimeter towers outside the jet cones per event vs. tower detector $\eta$ for IDPE data and the simulation. 
Good agreement is observed, except for a discrepancy at the highest $|\eta|$ region, where the tower transverse size is smaller than that of hadron showers produced by particles interacting in the calorimeter. This results in several tower ``hits'' per particle at high $|\eta|$, which is difficult to accurately simulate. 

\begin{figure}
 \begin{center}
 \includegraphics[width=8.5cm]{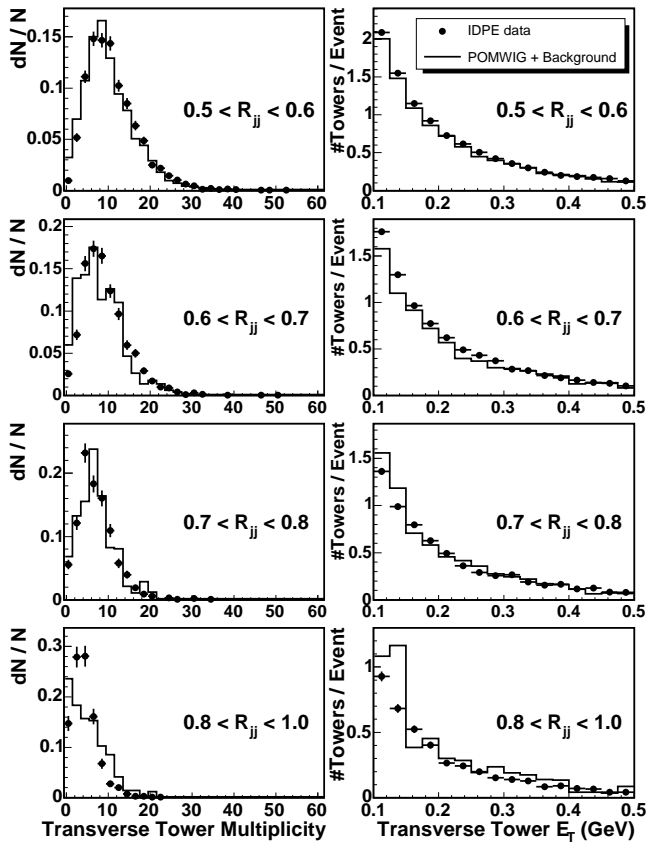}
 \caption{Calorimeter tower multiplicity and tower $E_T$ distributions in the transverse region with respect to the leading jet axis ($60^\circ<\Delta\phi<120^\circ$ or $240^\circ<\Delta\phi<300^\circ$) for four $R_{jj}$ bins in the region of $0.5<R_{jj}<0.9$. The tower multiplicity and tower $E_T$ distributions are normalized to unit area and to the number of transverse towers per event, respectively.
\label{fig:tower_multi_et}}
 \end{center}
\end{figure}

A more relevant UE comparison between data and simulation is that of transverse calorimeter tower multiplicity ($N_T$) and $E_T$ distributions between data and MC generated events, as shown in Fig.~\ref{fig:tower_multi_et} for different $R_{jj}$ bins in the range $0.5<R_{jj}<1.0$. Agreement between data and simulation is observed, except in the low multiplicity and low $E_T$ regions where the data points fall below the MC generated histograms. To quantify the effect of this discrepancy on the extracted exclusive signal fraction, we compare transverse tower $E_T$ distributions between data and MC in the region of $0.4<R_{jj}<0.7$, which is in the plateau of the distribution shown in Fig.~\ref{fig:rjj_cut3_fit} (to avoid edge effects). Then, we modify the UE in the simulation to minimize the $\chi^2/\mbox{d.o.f.}$ and re-evaluate the exclusive fraction. Using the  default MC simulation yields a  $\chi^2/\mbox{d.o.f}=1.4$. The UE is modified by scaling the tower $E_T$ of transverse calorimeter towers by a factor $F(E_T)=E_T\times (1.5-N_T/40)$ for $N_T<20$, yielding a $\chi^2/\mbox{d.o.f.}=0.4$, which is lower by 1 unit. The $E_T$ scaling has the effect of fewer calorimeter towers being rejected by the calorimeter threshold cuts, thereby resulting in a larger exclusive signal fraction, $F_{excl}^{scaled}=27.0\pm2.2$~(stat)\%. Since the discrepancy between the default and scaled MC distributions is only  one unit of $\chi^2/\mbox{d.o.f.}$, we retain the default value for the fraction and use the scaled result to assign a systematic uncertainty of $\pm (F_{excl}^{scaled}-F_{excl}^{default})/F_{excl}^{default})/2=\pm 9\%$      

\subsubsection{Exclusive dijet model}
The full difference between the exclusive cross section values obtained using the {\sc ExHuME} and {\sc ExclDPE} MC $R_{jj}$ shapes is assigned as a systematic uncertainty associated with the exclusive signal modeling (see Table~\ref{tab:xsec}). This difference is mainly due to different amounts of radiation emitted from the jets.

\begin{table*}[htp]
 \caption{Measured inclusive DPE and exclusive dijet cross sections, and the ratio of the exclusive to inclusive 
DPE dijet cross sections in the kinematic range $E_T^{jet1,2}>E_T^{min}$, $|\eta^{jet1,2}|<2.5$,
$0.03<\xi_{\bar{p}}<0.08$ (integrated over all $t_{\bar{p}}$), and $3.6<\eta_{gap}<5.9$.
\label{tab:xsec}}
 \begin{ruledtabular}
 \begin{center}
  \begin{tabular}{crrr}
   $E_T^{min}$  & $\sigma_{DPE}^{incl}\pm stat\pm syst$  & $\sigma_{jj}^{excl}\pm stat\pm syst$ & $R_{incl}^{excl}$ \\\hline
&&&\\
       10~GeV  & $14.5\pm0.1^{+9.8}_{-6.9}$ nb     & $1.10\pm0.04^{+1.29}_{-0.54}$ nb  &  $7.6\pm0.3^{+2.9}_{-1.2}$~\% \\
       15~GeV  & $1.43\pm0.02^{+0.89}_{-0.62}$ nb  & $112\pm7^{+84}_{-49}$ pb          &  $7.8\pm0.5^{+3.2}_{-1.2}$~\% \\
       20~GeV  & $267\pm6^{+166}_{-110}$ pb        & $15.7\pm2.0^{+15.5}_{-9.6}$ pb    &  $5.9\pm0.8^{+3.0}_{-2.1}$~\% \\ 
       25~GeV  & $76.0\pm2.7^{+37.0}_{-28.6}$ pb   & $4.84\pm0.96^{+4.11}_{-3.28}$ pb  &  $6.4\pm1.3^{+4.6}_{-3.9}$~\% \\
       35~GeV  & $14.6\pm1.2^{+5.3}_{-5.2}$ pb     & $1.37\pm0.49^{+1.08}_{-1.01}$ pb  &  $9.3\pm3.4^{+6.9}_{-6.6}$~\% \\
  \end{tabular}
 \end{center}
 \end{ruledtabular}
\end{table*}

\subsection{Cross sections}\label{subsec:xsec}
\begin{figure}
 \begin{center}
 \includegraphics[width=8.5cm]{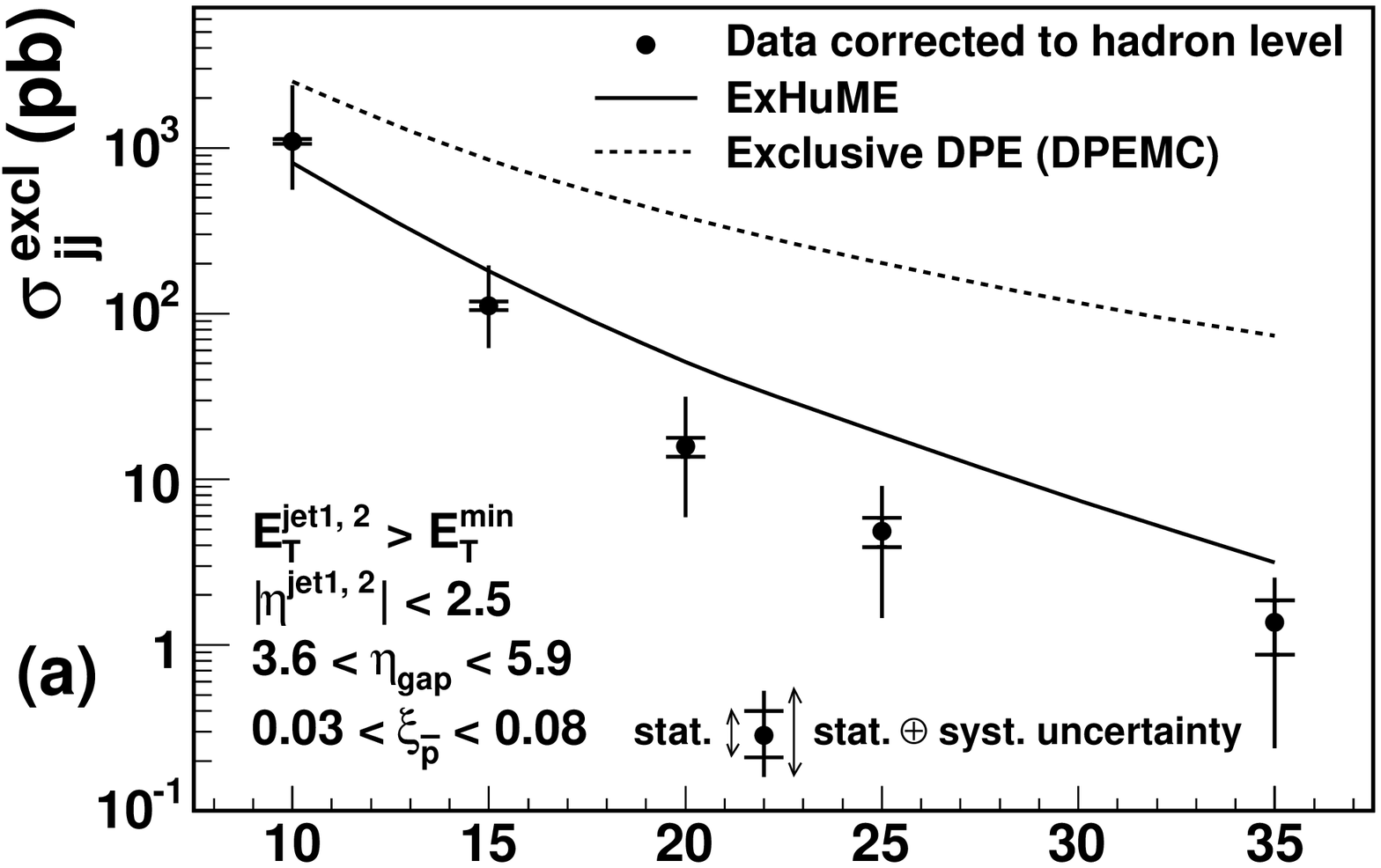}
 \includegraphics[width=8.5cm]{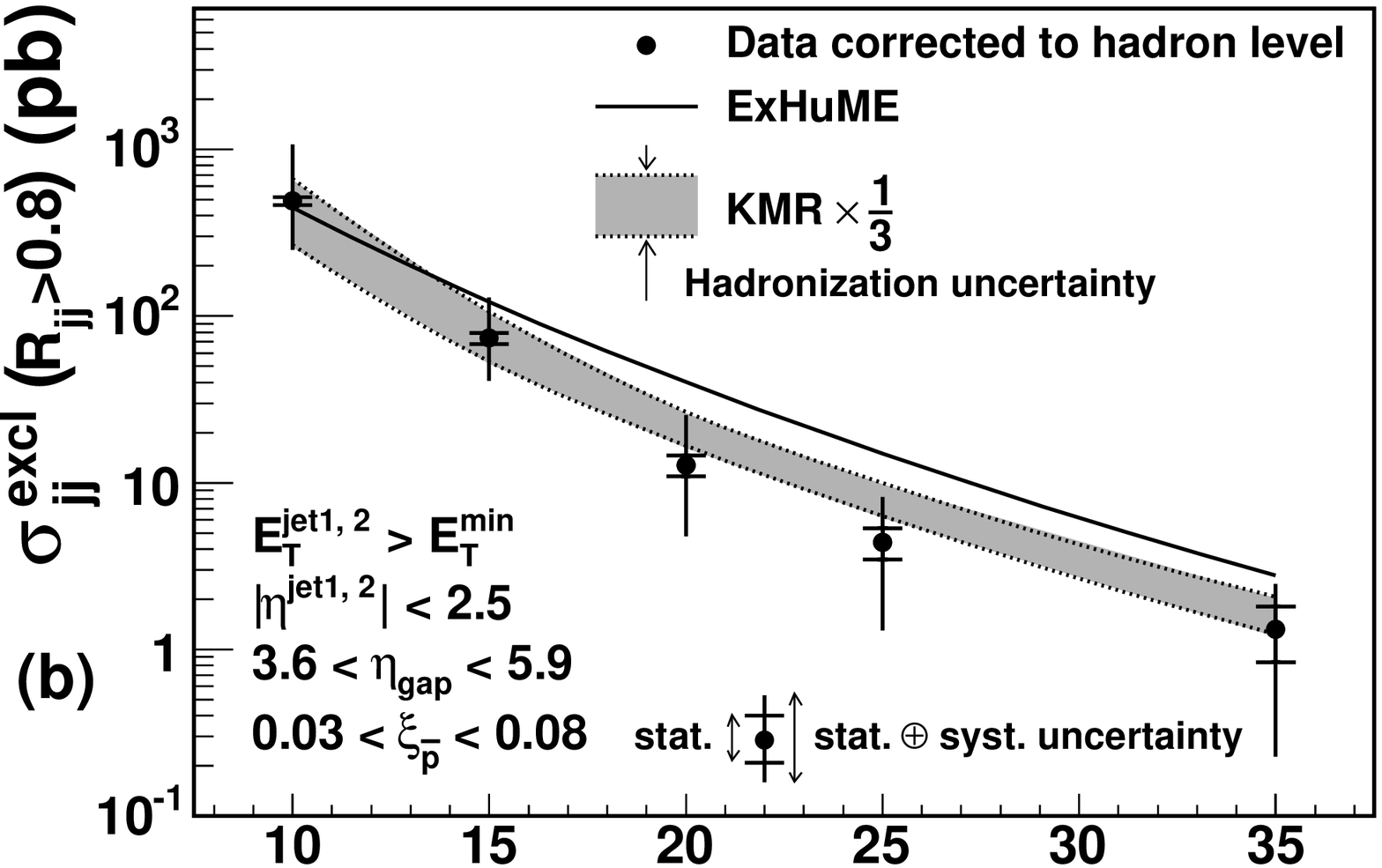}
 \includegraphics[width=8.5cm]{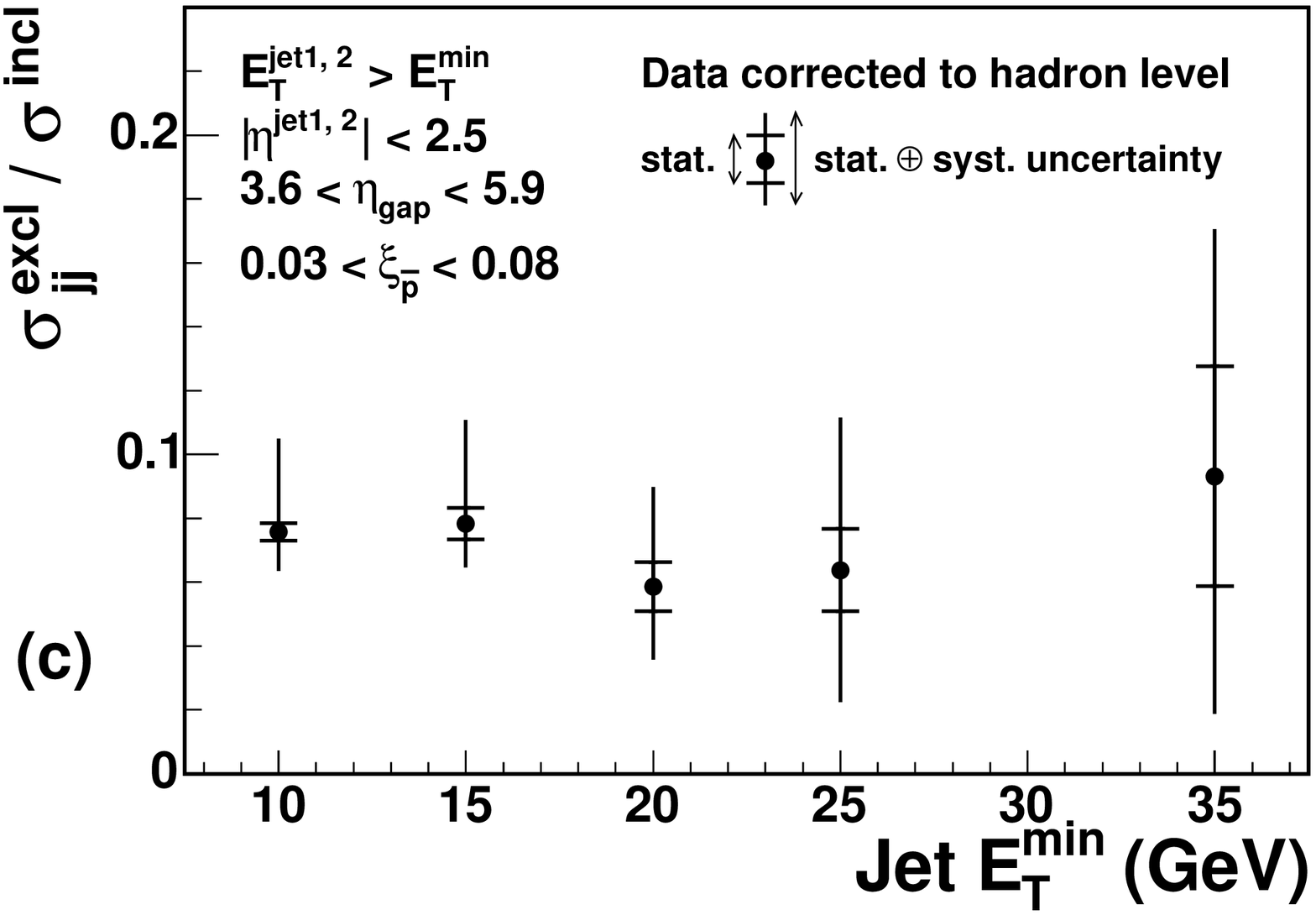}
 \caption{
Exclusive dijet cross sections for events with two jets of $E_T^{jet}>10$~GeV plotted vs. the minimum $E_T^{jet}$ of the two jets in the kinematic range denoted in the figures: (a) total exclusive cross sections compared with {\sc ExHuME} and {\sc ExclDPE} predictions; (b) exclusive cross sections for events with $R_{jj}>0.8$ compared with {\sc ExHuME} (solid curve) and with the LO analytical calculation from Ref.~\cite{KMRmethod} (see also Ref.~\cite{KMRrecent}) scaled down by a factor of three (dashed lines) - the shaded area represents  uncertainties in the calculation due to hadronization effects; and (c)
the ratio of total exclusive to inclusive DPE cross sections.
 \label{fig:xsec_excl_vs_et}}
 \end{center}
\end{figure}

Measured cross sections
for inclusive DPE and exclusive dijet production, and the ratio of exclusive to inclusive DPE dijet cross 
sections for different $E_T^{jet1,2}$ thresholds, are presented in Table~\ref{tab:xsec}. The listed systematic uncertainties consist of all those discussed above added in quadrature.  
The exclusive cross sections are plotted in Fig.~\ref{fig:xsec_excl_vs_et}~(a), where they are compared with hadron-level predictions of {\sc ExHuME} and {\sc ExclDPE} Monte Carlo simulations. The {\sc ExHuME} predictions are favored by the data in both normalization and shape. The exclusive signal for $E_T^{jet1,2}>10$~GeV is established at a significance level of 6.1~$\sigma$, determined from the value of $R^{excl}_{incl}=7.6\pm 0.3^{+2.9}_{-1.2}$~\% presented in Table~\ref{tab:xsec}.  The value of 6.1~$\sigma$ is obtained as the ratio of the central value of $R^{excl}_{incl}=7.6$ divided by an uncertainty composed of the statistical uncertainty of 0.3\% and the ``downward'' systematic uncertainty of 1.2\%, combined in quadrature and yielding $(0.3^2+1.2^2)^{1/2}=1.24$~\%, as the systematic uncertainty of -1.2\% comes from an upward fluctuation of the background.

In Fig.~\ref{fig:xsec_excl_vs_et}~(b), we compare the data exclusive cross section for events with $R_{jj}>0.8$ plotted vs. jet $E_T^{min}$ with the {\sc ExHuME} prediction and with the analytical calculation of exclusive dijet cross sections from Ref.~\cite{KMR} (KMR). The $\sigma_{jj}^{excl}$ is recalculated using the observed exclusive signal in the region of $R_{jj}>0.8$. The KMR cross sections, which are based on a LO parton level calculation of the process $gg \rightarrow gg$, have an ${\cal{O}}(3)$ systematic uncertainty. The kinematic  cuts used in KMR are slightly different from those used in the present analysis, which could lead to an effect of $\sim\pm20$~\% on the predicted cross section values~\cite{ref:KR_private}. The good agreement seen in Fig.~\ref{fig:xsec_excl_vs_et} between the measured $\sigma_{jj}^{excl}$ and the KMR predictions multiplied by a factor of $1/3$ suggests that the data are consistent with the KMR predictions within the quoted uncertainties. An even better agreement is reached by rescaling the parton transverse momentum in the KMR calculation to the measured jet transverse energy~\cite{KMRrecent}.

The ratio of exclusive to inclusive DPE dijet cross sections measured from the data as a function of jet $E_T^{min}$ is shown in Fig.~\ref{fig:xsec_excl_vs_et}~(c).

\section{Heavy Flavor Quark Jets}\label{sec:hf}
One of the most characteristic features of exclusive dijet production is that at high dijet mass fraction it is dominated by the parton level process $gg \rightarrow gg$, as contributions from $gg \rightarrow q\bar{q}$ are suppressed. Born level cross sections for exclusive production of a color-singlet dijet system of mass $M$ are given by~\cite{KMR2002}
\begin{eqnarray}
  \frac{d\hat{\sigma}_{excl}}{dt}(gg \rightarrow gg)       &=& \frac94\:\frac{\pi\alpha_s^2}{E_T^4}\label{eq:excl_gg}\\
  \frac{d\hat{\sigma}_{excl}}{dt}(gg \rightarrow q\bar{q}) &=& \frac{\pi\alpha_s^2}{6E_T^4}\:\frac{m_q^2}{M^2}\:\left(1-\frac{4m_q^2}{M^2}\right),\label{eq:excl_qq}
\end{eqnarray}
where $E_T$ is the transverse energy of the final state parton 
and $m_q$ is the quark mass. 
The suppression of $gg\rightarrow q\bar  q$ is due to the factor  $(m_q^2/M^2)(1-4m_q^2/M^2)$, which vanishes as $m_q^2/M^2 \rightarrow 0$ ($J_z=0$ selection rule~\cite{KMR2001}). Exclusive $gg \rightarrow q\bar{q}$ contributions are also strongly suppressed in NLO and NNLO QCD, and in certain higher orders~\cite{KRS}.   

The predicted exclusive $q\bar q$-dijet suppression offers the opportunity of searching for an exclusive signal in IDPE data by comparing the inclusive dijet $R_{jj}$ shape with that of data containing identified $q\bar q$ dijets. 
The presence of an exclusive dijet signal in the IDPE event sample would be expected to appear as a suppression in the ratio of $q\bar q$ to inclusive events at high $R_{jj}$. This data driven method avoids the use of MC simulations and can be used to corroborate the MC-based extraction of the exclusive signal from the inclusive data sample. As many systematic effects cancel in measuring the ratio, a relatively small $q\bar q$ event sample can provide valuable information. 

To ensure quark origin, we select jets from heavy flavor (HF) $b$- or $c$-quarks, identified from secondary vertices produced from the decay of intermediate $B$ or $D$ mesons using the SVX~II detector. Both $b$- and $c$-quark jets are used, since the suppression mechanism holds for all quark flavors. 

Below,  in Sec.~\ref{HFsample} we describe the HF data sample and event selection requirements, in Sec.~\ref{HFeff} we evaluate the HF selection efficiencies and backgrounds, and in Sec.~\ref{HFresults} we present the HF jet fraction results. 
\subsection{Data sample and event selection\label{HFsample}}
The data used in this analysis were collected at a full rate (no pre-scaling) with a trigger satisfying the same requirements as the DPE trigger, Jet5+RPS+$\overline{\rm BSC1_p}$, plus an additional one designed to enhance the HF jet content. 
The latter required the presence of at least one track with transverse momentum $p_T>2$~GeV/c displaced from the IP by a distance $d$ of $0.1<d<1.0$~mm, where $d$ is the distance of closest approach of the track to the IP~\cite{SVT}. 
The total integrated luminosity of this data sample is $200\pm12$ pb$^{-1}$.

Jets are reconstructed using a CDF Run~I based iterative cone algorithm~\cite{JetClu} with an $\eta$-$\phi$ cone of radius 0.4. 
The  {\sc secvtx} tagging algorithm is used to search for a displaced secondary vertex due to a $B$ or $D$ meson decay within a jet cone.  This algorithm seeks tracks with hits in the SVX~II within the jet cone, and reconstructs the secondary vertex from those which are significantly displaced  from the primary vertex. A jet is considered {\sc secvtx} tagged if it has a secondary
vertex consisting of at least two (or three) such tracks with $p_T>1$ (0.5) GeV/c. 
Events are further required to pass the IDPE selection criteria listed in Eq.~(\ref{IDPEsample}). 
This selected ``pretag'' event sample contains 34~187 jets  with at least two tracks in the SVX~II. 
Applying the {\sc secvtx} tagging algorithm to the jets in the pretag sample yields 1,118 tagged jets with $E_T^{jet}>10$~GeV and $|\eta^{jet}|<1.5$. 

\subsection{Heavy flavor selection efficiencies\label{HFeff}}
\begin{figure}
 \begin{center}
 \includegraphics[width=8.5cm]{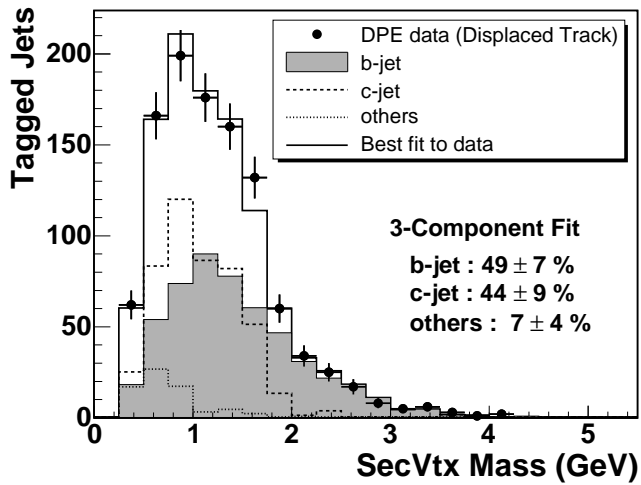}
 \caption{Displaced secondary vertex mass distribution for jets tagged by the {\sc secvtx} algorithm in DPE data (points).
 The solid histogram shows the result of a three-component binned maximum likelihood fit to the data, using Monte 
Carlo templates of distribution shapes, consisting of $b$-jets (shaded histogram), $c$-jets (dashed histogram), 
and other jets (dotted histogram) obtained from {\sc pythia} dijet events.
\label{fig:svmass_3fit}}
 \end{center}
\end{figure}

A {\sc secvtx} tag in a jet without a HF quark is labeled as a ``mistag.'' The mistag probability per jet, measured from 
inclusive jet data, is parameterized as a function of the number of tracks, $E_T$, $\eta$, 
and $\phi$ of the jet,
and the sum over the  $E_T$ values of all jets with $E_T^{jet}>10$~GeV and $|\eta^{jet}|<2.4$ in the event. The mistag background in
tagged jets is estimated by weighting each jet in the pretag sample by the mistag probability and summing up the
weights over all jets in the sample. The total number of mistag jets, ${\rm N}_{mistag}$, is measured to be 
$104\pm 15$. This number is consistent with a  background estimate obtained from studies of the distribution of the invariant mass distribution $M_{svtx}$ of charged particles associated with a displaced secondary vertex 
(Fig.~\ref{fig:svmass_3fit}). 
The mistag background for a given $R_{jj}$ interval is evaluated by applying the mistag probability to the jets in that interval of the pretag sample.

\begin{figure*}
 \begin{center}
 \includegraphics[width=8.5cm]{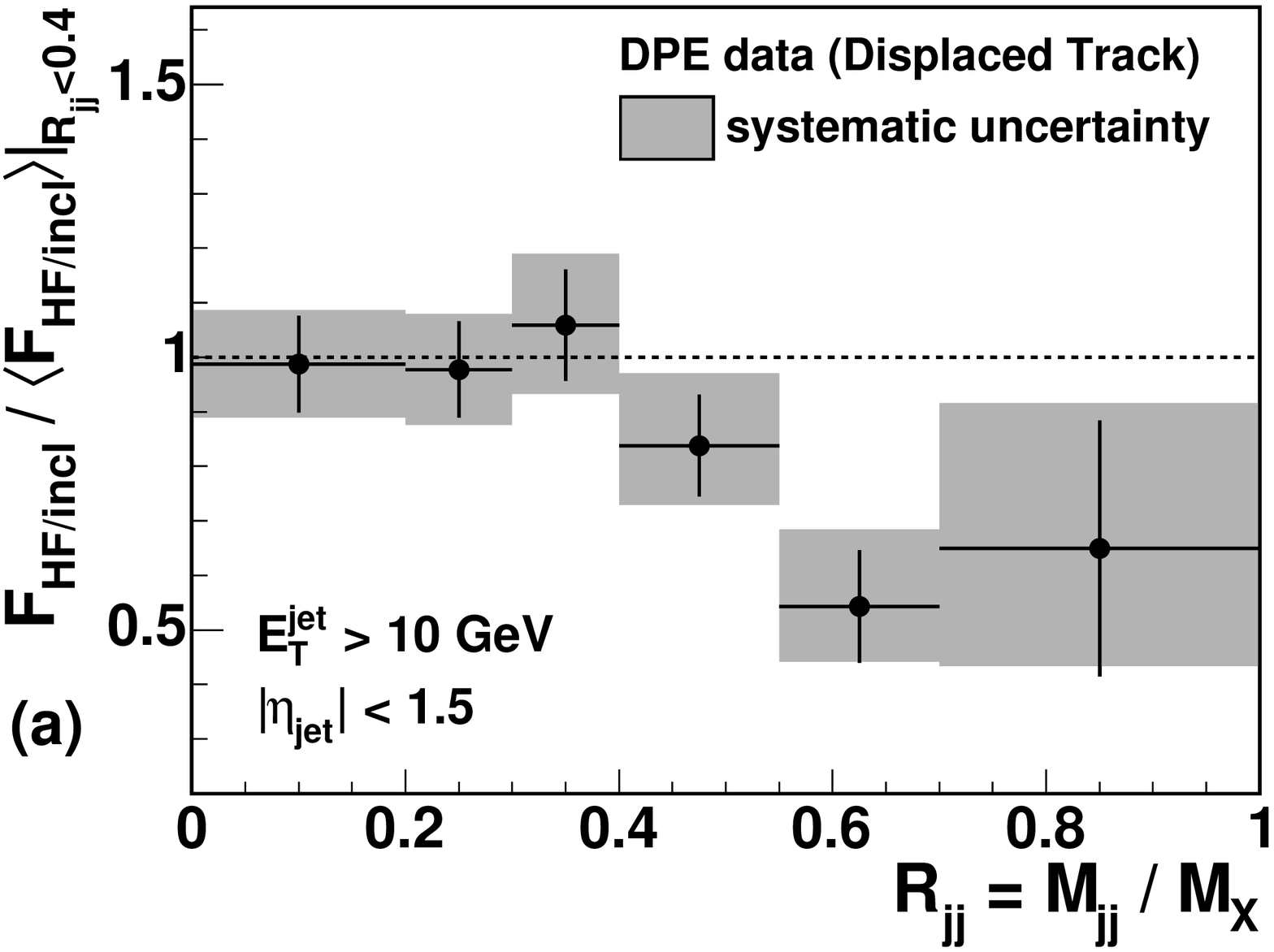}
 \includegraphics[width=8.5cm]{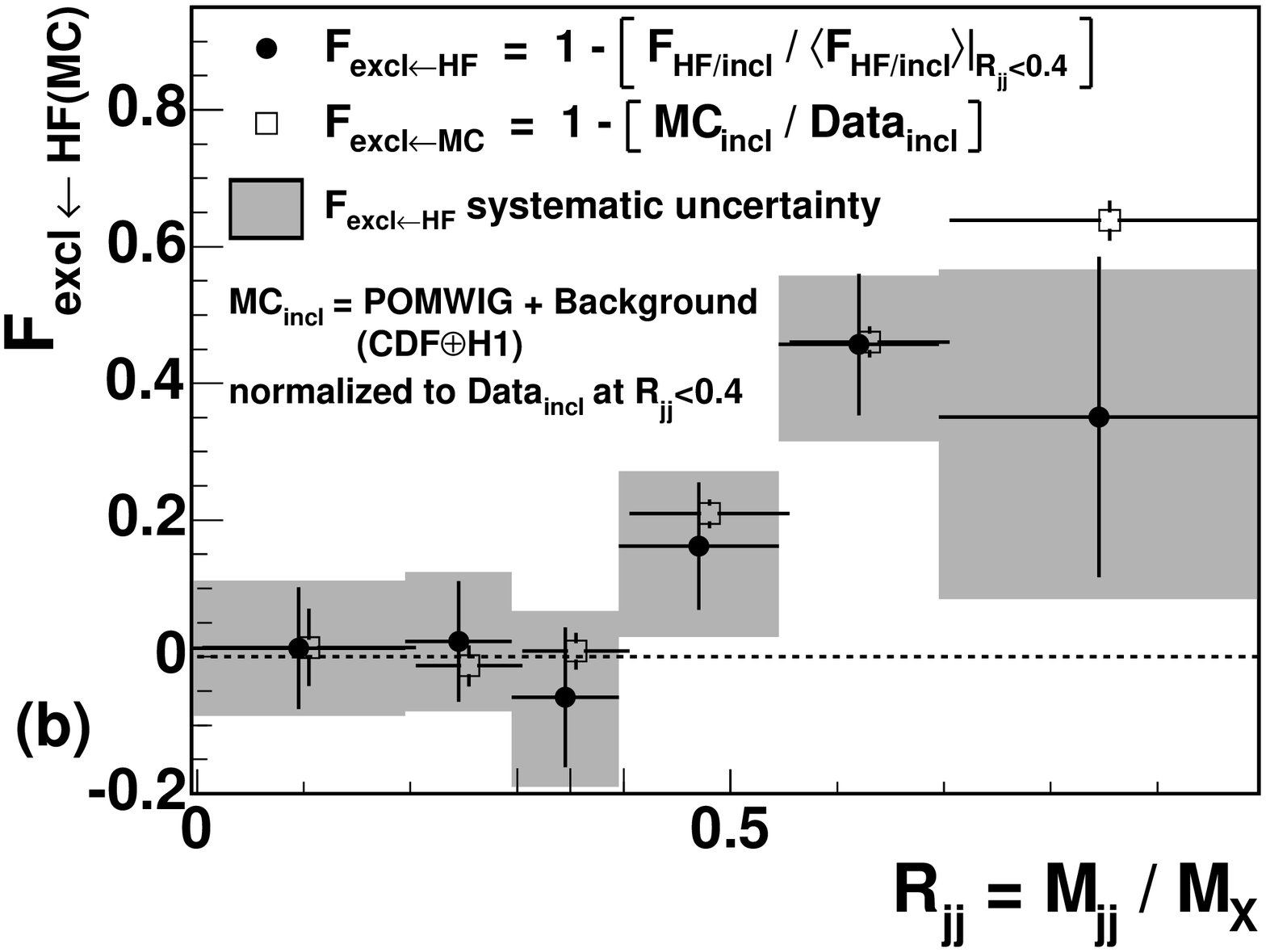}
 \caption{ (a) 
Measured ratio $F_{HF/incl}$ of heavy flavor jets to all inclusive jets of $E_T^{jet}>10$~GeV and $|\eta^{jet}|<1.5$ as a function of dijet mass fraction $R_{jj}$, normalized to the weighted average value in the region of $R_{jj}<0.4$, with systematic uncertainties represented by the shaded band; 
(b) 
values of $F_{excl \leftarrow HF}=1-F_1$ (filled circles) and $F_{excl \leftarrow MC}=1-F_2$ (open squares) as a function of $R_{jj}$, where $F_1=F_{HF/incl}/<F_{HF/incl}>|_{R_{jj}<0.4}$, plotted on left vs. $R_{jj}$, and $F_2$ is the ratio of {\sc pomwig} MC to inclusive dijet events obtained from the studies presented in Sec.~\ref{sec:excl_search} - the error bars (shaded band) represent statistical ($F_{excl \leftarrow HF}$ systematic) uncertainties.
\label{fig:fex_vs_rjj}}
 \end{center}
\end{figure*}

The efficiency for tagging HF jets by the {\sc secvtx} algorithm depends on the composition of $b$- and $c$-jets of the data sample to which the algorithm is applied, and is therefore evaluated using a combination of a Monte Carlo simulation and HF jet fractions obtained from the data. The tagging 
efficiency for a $b\,(c)$-jet, $\epsilon_b$($\epsilon_c$), is obtained using {\sc pythia} MC dijet events passed through a 
detector simulation, and is corrected for discrepancies observed between MC events and data. The fraction of $b\,(c)$-jets in the DPE data sample, $F_b$($F_c$), is obtained from  the fit to the $M_{svtx}$ distribution shape (Fig.~\ref{fig:svmass_3fit}).
The tagging efficiency for a HF jet, $\epsilon_{tag}^{HF} = F_{HF}/[F_b/\epsilon_b + F_c/\epsilon_c]$,
is found to be $7.9\pm1.4$~\% for $E^{jet}_T\sim15$~GeV, where $F_{HF}=F_b+F_c$.

The data are corrected for efficiencies associated with the displaced track (DT) requirement in the trigger. These efficiencies are obtained from a data sample collected without requiring a displaced track. 
Selecting from this sample events that pass the DT requirement, we obtain the DT trigger efficiency for inclusive jets as the ratio  $\epsilon_{all}^{DT}=N_{all}^{DT}/N_{all}$, where
$N_{all}^{DT}$ is the number of jets in the selected events and $N_{all}$ is the total number of jets in the sample. 
Similarly, the efficiency for HF jets is obtained as 
$\epsilon_{HF}^{DT}=N_{HF}^{DT}/N_{HF}=[N_{tag}^{DT}(1-F_{mistag}^{DT})]/[N_{tag}(1-F_{mistag})]$, 
where $N_{tag}$ ($N_{tag}^{DT}$) is the number of tagged jets and $F_{mistag}$ ($F_{mistag}^{DT}$) the mistag background fraction in events without (with) the DT requirement.
$F_{mistag}^{DT}$ is evaluated from a 3-component MC template fit to the $M_{svtx}$ data distribution, consisting of $b$-, $c$-, and $other$-jets at experimentally measured proportions (Fig.~\ref{fig:svmass_3fit}), while $F_{mistag}$ is measured from the corresponding $M_{svtx}$ fit 
to the DPE data collected without the DT requirement.

\subsection{Heavy flavor jet fraction results\label{HFresults}}
Results for the fraction $F_{HF/incl}$ of HF jets  to all inclusive jets of $E_T^{jet}>10$~GeV and $|\eta^{jet}|<1.5$ for the IDPE event sample
 are shown in Fig.~\ref{fig:fex_vs_rjj}~(a) as a function of dijet mass fraction $R_{jj}$. The fraction is normalized to the mean value of the ratio
of the HF to inclusive events over the four $R_{jj}$ bins in the region of $R_{jj}<0.4$, so that systematic uncertainties correlated among $R_{jj}$ bins cancel out, as e.g. the uncertainties from corrections for data to MC tagging efficiency discrepancies or from  the mistag background fraction estimate before and after the DT trigger requirement. Thus,
\begin{equation}
 F_{HF/incl} = {\langle F_{HF/incl} \rangle|_{R_{jj}<0.4}} \cdot \frac
{\sigma^{incl}}{\sigma^{incl}+\sigma^{excl}},
\label{eq:HFincl}
\end{equation}
where $\sigma^{incl}$ is the inclusive DPE jet production cross section only,  $\sigma^{excl}$ is the  exclusive cross section, and $\langle F_{HF/incl} \rangle|_{R_{jj}<0.4}$ is the mean value of $F_{HF/incl}$ in the range $R_{jj}<0.4$~\cite{HFfootnote}. An exclusive dijet production rate contributing to the total rate but which is suppressed in HF dijet production would be expected to appear as a suppression in  the fraction $F_{HF/incl}$ at high $R_{jj}$~\cite{KMR2002}. The suppression seen in Fig.~\ref{fig:fex_vs_rjj}~(a) is examined here for consistency with this hypothesis.

\paragraph*{\bf The statistical uncertainties} shown in Fig.~\ref{fig:fex_vs_rjj} are dominated by the low statistics HF event sample. Measuring the fraction of HF to inclusive dijet events has the advantage of reducing systematic uncertainties common to both event samples. The fraction is corrected for mistag 
backgrounds, tagging efficiency for HF jets, and displaced track trigger efficiencies for inclusive and HF jets.

\paragraph*{\bf The systematic uncertainties} on the ratio $F_{HF/incl}$, shown in Fig.~\ref{fig:fex_vs_rjj} as shaded areas, are due to the uncertainties associated with the background and efficiency estimates.
The mistag background uncertainty is evaluated from the uncertainties associated with the 
determination of the mistag probability and propagated to an uncertainty in the fraction.
The HF-jet tagging efficiency, derived from combinations of {\sc pythia} MC generated events and data, has an  uncertainty propagated from the statistical uncertainty of the MC generated event sample, uncertainties from the correction for discrepancies between the tagging efficiencies derived from MC and data, and uncertainties from the $b$- and $c$-jet fractions in the tagged jet data sample. The displaced track trigger efficiency 
has two sources of systematic uncertainty: the statistical uncertainty of the IDPE sample collected 
without the displaced track requirement, and the uncertainty on the mistag background fractions before and after applying the mistag requirement to the sample.
In addition to the above uncertainties, we assign a systematic uncertainty associated with an increasing trend of $c$- to $b$-jet fraction found in {\sc pythia} generated events, which could contribute to a decreasing HF-jet fraction with $R_{jj}$ due to the tagging efficiency $\epsilon_c$ being lower than $\epsilon_b$. 

To examine the consistency between the MC based extracted exclusive dijet fraction and the data based suppression of the exclusive HF to inclusive dijet production rates we compare in Fig.~\ref{fig:fex_vs_rjj}~(b) the $R_{jj}$ residual distributions defined as $F_{excl \leftarrow MC}\equiv 1-\left[{\rm MC}_{\rm incl}/{\rm Data}_{\rm incl}\right]$, for which  the excess is defined as the inclusive DPE events observed above the inclusive $R_{jj}$ distribution (composed of
{\sc pomwig} MC events with ND and SD backgrounds) normalized to the DPE data at $R_{jj}<0.4$ (open squares), and $F_{excl \leftarrow HF}\equiv 1-\left[{\rm F}_{\rm HF/incl}/\langle{\rm  F}_{\rm HF/incl} \rangle|_{R_{jj}<0.4}\right]$ (filled circles). 
The absolute values and $R_{jj}$ dependence of the $F_{excl \leftarrow HF}$ points in the region of $0.4<R_{jj}<1.0$ are consistent with those of $F_{excl \leftarrow MC}$, 
supporting an interpretation of the observed $F_{HF/incl}$ distribution as a manifestation of the suppression of HF quark jets
in exclusive production. 

\section{Exclusive Dijets and Diffractive Higgs Production}\label{sec:higgs}
The search for Higgs bosons is one of the top priorities of the LHC experiments. While the main effort of both the ATLAS and CMS experimental plans is directed toward searches for inclusively produced Higgs bosons, an intense interest has developed toward  exclusive Higgs boson production, $p+p\rightarrow p+H+p$~\cite{FP420}. The exclusive Higgs production channel presents several advantages, including: (i) it can provide events in an environment of suppressed QCD backgrounds for the main Higgs decay mode of $H\rightarrow b_{jet}+\bar b_{jet}$, due to the $J_z=0$ selection rule discussed in Sec.~\ref{sec:hf}, (ii) the Higgs mass can be measured accurately with the missing mass technique by detecting and measuring the momentum of the outgoing protons~\cite{albrow}, (iii) the spin-parity of the Higgs boson can be determined from the azimuthal angular correlations between the two outgoing protons, and (iv) the method is universally sensitive to all exclusive Higgs production mechanisms. 

Theoretical predictions for exclusive Higgs boson production cross sections range from $\sim 200$~fb~\cite{BL} to 2-6 fb~\cite{KKMR} for a Higgs boson mass of $\sim 120$~ GeV at $\sqrt s$=14 TeV at the LHC.  However, since  exclusive Higgs boson and exclusive dijet production proceed through similar diagrams, as illustrated in Figs.~\ref{fig:SD_DPE}b (with no Pomeron remnants) and Fig.~\ref{fig:excl_diagram},
the models can be calibrated by comparing their predictions for exclusive dijet cross sections with measured values at the Tevatron.
Furthermore, measured exclusive dijet cross sections at the Tevatron may also be used to evaluate backgrounds to the process $H \rightarrow b\bar{b}$ from exclusive $gg$ dijet production  with 
gluons misidentified as $b$-quarks in $b$-tagging, or from $b$-quarks produced by 
gluon splitting, $g \rightarrow b\bar{b}$.

\subsection{M$_{jj}$ distribution}
\begin{figure}
 \begin{center} 
 \includegraphics[width=8.5cm]{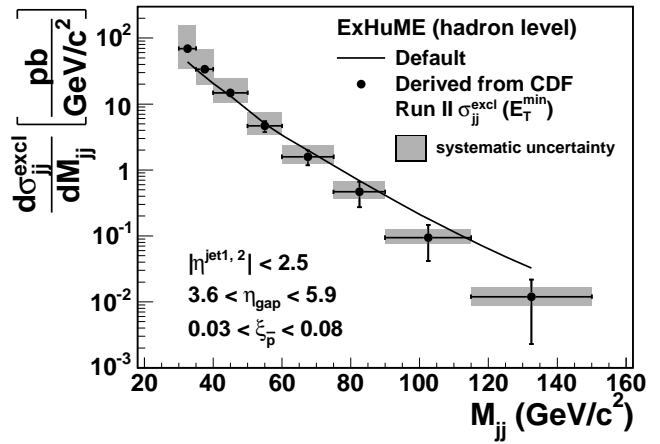}
 \caption{{\sc ExHuME} exclusive dijet differential cross section at the hadron level vs. dijet mass $M_{jj}$. The filled points show
cross sections derived from the measured $\sigma_{jj}^{excl}$ values shown in Fig.~\ref{fig:xsec_excl_vs_et}~(top) using the procedure described in the text. The vertical error bars on the points and the shaded band represent statistical and systematic uncertainties, respectively, obtained by propagating the
corresponding uncertainties to the measured values of $\sigma_{jj}^{excl}$. The solid curve is the cross section predicted by {\sc ExHuME} using the default settings.
\label{fig:xsec_vs_mjj}}
 \end{center} 
\end{figure}

The measured exclusive dijet cross section presented in Fig.~\ref{fig:xsec_excl_vs_et} vs. jet $E_T^{min}$ is converted to a cross section vs. dijet mass $M_{jj}$ using the {\sc ExHuME} Monte Carlo simulation with $M_{jj}$ reconstructed at the hadron level.
From the measured values of $\sigma_{jj}^{excl}$ for the
$E_T^{jet1,2}$ thresholds given in Table.~\ref{tab:xsec}, we obtain the cross section for each of the following $E_T^{jet2}$ intervals;
10-15 GeV, 15-20 GeV, 20-25 GeV, 25-35 GeV, and 35 GeV or higher. After applying a  hadron level $E_T^{jet2}$ cut, the {\sc ExHuME} $M_{jj}$ distribution for each 
$E_T^{jet2}$ interval is normalized to the cross section
for that interval. Summing up over all the normalized $M_{jj}$ distributions yields the {\sc ExHuME}-based exclusive dijet differential cross section
as a function of $M_{jj}$, $d\sigma_{jj}^{excl}/dM_{jj}$. The values obtained are corrected for
a possible bias caused by the minimum threshold requirement of $E_T^{jet2}>10$ GeV by comparing the $M_{jj}$ distributions with and without the 
$E_T^{jet2}$ cut. 
The derived $d\sigma_{jj}^{excl}/dM_{jj}$ distribution is shown for $M_{jj}>30$ GeV/c$^2$ in Fig.~\ref{fig:xsec_vs_mjj} (solid circles).
This distribution falls slightly faster than the 
default {\sc ExHuME} prediction (solid curve), as one would expect from the fact that the measured  $\sigma_{jj}^{excl}(E_T^{min})$ falls somewhat more steeply with $E_T^{min}$ than that of {\sc ExHuME} (Fig.\ref{fig:xsec_excl_vs_et}), but overall there is reasonable agreement.
This result supports the {\sc ExHuME} prediction, and thereby the perturbative QCD calculation of Ref.~\cite{KMRmethod} on which {\sc ExHuME} is based.

\subsection{Higgs boson cross section}
From the {\sc ExHuME} resulting values of $d\sigma_{jj}^{excl}/dM_{jj}$, we obtain $\sigma_{jj}^{excl}\approx 360$~fb for the range $115<M_{jj}<145$~GeV/c$^2$, which corresponds to a $\pm12$~\% mass window around $M_{jj}=130$~GeV/c$^2$ for jets within the kinematic region defined by the cuts denoted in Fig.~\ref{fig:xsec_vs_mjj}.
For SM Higgs boson production at the Tevatron, perturbative calculations~\cite{KKMR} predict $\sigma_{H}^{excl}\sim0.2$~fb with a factor of 2-3 uncertainty for a Higgs boson mass of $m_H=120$~GeV/c$^2$, which leads to a ratio of exclusive Higgs signal to dijet background of $R_{H/jj}\sim 6\times10^{-4}$. This value is in agreement with the  estimate of $R_{H/jj}=6\times10^{-4}$ given in Ref.~\cite{KKMR} for $m_H=120$~GeV/c$^2$ 
using an experimental missing mass resolution of $\Delta M_{missing}=3$~GeV/c$^2$ at the LHC, rendering support to the prediction of the SM Higgs exclusive production cross section of 3~fb (with a factor of 3 uncertainty) presented in Ref.~\cite{KMRmethod}. Measurements of exclusive dijet production rates in the Higgs
mass range at the LHC could  further constrain $\sigma_{H}^{excl}$ through  $R_{H/jj}$.

Models of exclusive Higgs production may also be tested using measured cross sections for exclusive $\gamma\gamma$ production, $p+p\rightarrow p+\gamma\gamma+p$, a process similar to exclusive dijet production. In the model of Ref.~\cite{KMRgammagamma}, the $\gamma\gamma$ production is represented by the diagrams of Fig.~\ref{fig:excl_diagram} in which ``jet'' is replaced by ``$\gamma$''. A recent CDF measurement ~\cite{CDFgammagamma} yielded a cross section upper limit close to the predicted value, providing further support for this exclusive production model.   

\section{Summary and Conclusion}\label{sec:conclusion}
We have presented results from studies of dijet production in $\bar{p}p$ collisions at $\sqrt{s}=1.96$~TeV using events with a leading antiproton detected in 
a Roman Pot Spectrometer and a forward rapidity gap on the outgoing proton side, collected by the CDF II detector during Fermilab Tevatron Run II. These events, presumed to be produced by double Pomeron exchange (DPE), were extracted from a data sample of integrated luminosity 310 pb$^{-1}$.
In particular, we have demonstrated the presence of exclusively produced dijets, $\bar{p}+p \rightarrow \bar{p}+\mbox{dijet}+p$, by means of 
detailed studies of distributions of the dijet mass fraction $R_{jj}$, defined as the dijet mass 
divided by the DPE system mass. In comparisons of data $R_{jj}$ distributions with inclusive {\sc pomwig}~\cite{POMWIG} Monte Carlo simulations, we observe an excess of events 
in the data over the Monte Carlo predictions at high $R_{jj}$, which is consistent in terms of kinematic distribution shapes with the presence of an exclusive dijet signal as modeled by the {\sc ExHuME}~\cite{ExHuME} and exclusive DPE in {\sc dpemc}~\cite{DPEMC} Monte Carlo simulations. To facilitate comparison with theoretical predictions, the exclusive dijet cross section, $\sigma_{jj}^{excl}$, and the ratio of exclusive dijet to inclusive DPE dijet cross sections have been measured as a function of minimum $E_T$ threshold of the two leading jets in an event. The measured values of $\sigma_{jj}^{excl}$ favor the {\sc ExHuME} over the {\sc dpemc} predictions, and are found to be consistent with predictions from perturbative calculations presented in Ref.~\cite{KMRmethod}.

The Monte Carlo based extraction of the exclusive dijet signal is checked experimentally using a largely independent sample of heavy flavor $b$-tagged jet events extracted from  200 pb$^{-1}$ of DPE data collected with a special trigger requiring a track displaced from the interaction point. As exclusive dijet production from $gg \rightarrow q\bar{q}$ is predicted to be suppressed by the $J_Z=0$ selection rule relative to production through $gg \rightarrow gg$, the ratio of identified heavy flavor quark jets to inclusive jets is expected to decrease at high $R_{jj}$.  For jets of $E_T^{jet}>10$~GeV, we observe a suppression of the ratio of heavy flavor jets to inclusive jets in the region of $R_{jj}>0.4$, 
which is consistent in shape and magnitude with the expectation from the exclusive signal extracted by the MC based method based on Ref.~\cite{KMRmethod}.

The present results, representing the first observation of exclusive dijet production in high energy $\bar pp$ collisions, provide a benchmark template against which to calibrate theoretical calculations of exclusive Higgs boson production. The prospects for an observation of exclusive Higgs boson production at the LHC have been briefly discussed in light of our measured exclusive dijet cross sections.         
\begin{acknowledgments}
We thank the Fermilab staff and the technical staffs of the participating institutions for their vital contributions. This work was supported by the U.S. Department of Energy and National Science Foundation; the Italian Istituto Nazionale di Fisica Nucleare; the Ministry of Education, Culture, Sports, Science and Technology of Japan; the Natural Sciences and Engineering Research Council of Canada; the National Science Council of the Republic of China; the Swiss National Science Foundation; the A.P. Sloan Foundation; the Bundesministerium f\"ur Bildung und Forschung, Germany; the Korean Science and Engineering Foundation and the Korean Research Foundation; the Science and Technology Facilities Council and the Royal Society, UK; the Institut National de Physique Nucleaire et Physique des Particules/CNRS; the Russian Foundation for Basic Research; the Comisi\'on Interministerial de Ciencia y Tecnolog\'{\i}a, Spain; the European Community's Human Potential Programme; the Slovak R\&D Agency; and the Academy of Finland.
\end{acknowledgments}


\begin{thebibliography}{99}
\bibitem{rapidity}
Rapidity, $y=\frac{1}{2}\ln\frac{E+p_L}{E-p_L}$, and pseudorapidity, $\eta=-\ln\tan\frac{\theta}{2}$, are used interchangeably for particles detected in the calorimeters, 
since in the kinematic range of interest in this analysis they are approximately equal.
\bibitem{excl_define}While the exclusive DPE process is part of the inclusive, it has become traditional to refer to the processes with and without Pomeron remnants as ``inclusive'' and ``exclusive,'' respectively. In this paper we will follow this tradition.
\bibitem{COORD}    
  We use a coordinate system with origin at the center of the CDF detector, $z$ coordinate along the proton beam direction, and $y$ coordinate pointing up; the $x$ coordinate points away from the center of the accelerator ring.
\bibitem{H1Q2max} The maximum value of $Q^2$ in the H1 x2
fit3 Pomeron PDF is 75 GeV$^2$ and is used as a default setting in the {\sc POMWIG} generator.
\bibitem{HFfootnote} We assume that in the region of $R_{jj}<0.4$ there is no HF contribution from exclusive jet production (see Fig.~\ref{fig:rjj_cut1_fit}), so that $\sigma_{HF}/\sigma^{incl}=\sigma_{HF}^{incl}/\sigma^{incl}=\langle F_{HF/incl} \rangle|_{R_{jj}<0.4}$ , where $\sigma_{HF}^{incl}$ is the  HF jet cross section obtained from inclusive DPE events.
\end{thebibliography}
\end{document}